\documentclass[useAMS,usenatbib]{mn2e}

\usepackage{graphicx}
\usepackage{times}
\usepackage{natbib}
\usepackage{amsmath}

\renewcommand{\d}{\mathrm{d}}

\newcommand{\gtrsim}{\ga}
\newcommand{\lesssim}{\la}

\usepackage[usenames,dvipsnames]{color}

\def\Omegal{{\Omega_{0,\rm \Lambda}}}

\newcommand{\HI}{\textrm{H}\,\textsc{i}}
\newcommand{\HII}{\textrm{H}\,\textsc{ii}}
\newcommand{\Lya}{\textrm{Ly}$\alpha$}

\title[Origin of cosmic abundances]{Origin of Cosmic Chemical Abundances}
\author[U.~Maio \& E.~Tescari]{
Umberto~Maio$^{1,2}$ \thanks{E-mail: umaio@aip.de, maio@oats.inaf.it},
Edoardo~Tescari$^{3,4}$
\\
${}^1$Leibniz-Institut f\"ur Astrophysik, An der Sternwarte 16, 14482 Potsdam, Germany\\
${}^2$INAF -- Osservatorio Astronomico di Trieste, via G. Tiepolo, 11, 34131 Trieste, Italy\\
${}^3$School of Physics, University of Melbourne, Parkville, VIC 3010, Australia\\
${}^4$ARC Centre of Excellence for All-Sky Astrophysics (CAASTRO)\\
}

\begin{document}

\date{Accepted 2015 July 27. Received 2015 July 09; in original form 2014
  December 15.}
\pagerange{\pageref{firstpage}--\pageref{lastpage}}\pubyear{0}
\maketitle

\label{firstpage}

\begin{abstract}
Cosmological N-body hydrodynamic computations following atomic and molecular chemistry (e$^-$, H, H$^+$, H$^-$, He, He$^+$, He$^{++}$, D, D$^+$, H$_2$, H$_2^+$, HD, HeH$^+$), gas cooling, star formation and production of heavy elements (C, N, O, Ne, Mg, Si, S, Ca, Fe, etc.) from stars covering a range of mass and metallicity are used to explore the origin of several chemical abundance patterns and to study both the metal and molecular content during simulated galaxy assembly.
The resulting trends show a remarkable similarity to up-to-date observations of the most metal-poor damped Lyman-$\alpha$ absorbers at redshift $z\gtrsim 2$.
These exhibit a transient nature and represent collapsing gaseous structures captured while cooling is becoming effective in lowering the temperature below $\sim 10^4\,\rm K$, before they are disrupted by episodes of star formation or tidal effects.
Our theoretical results agree with the available data for typical elemental ratios, such as [C/O], [Si/Fe], [O/Fe], [Si/O], [Fe/H], [O/H] at redshifts $z\sim 2-7$.
Correlations between \HI\ and H$_2$ abundances show temporal and local variations and large spreads as a result of the increasing cosmic star formation activity from $z\sim 6$ to $z\sim 3$.
The scatter we find in the abundance ratios is compatible with the observational data and is explained by simultaneous enrichment by sources from different stellar phases or belonging to different stellar populations.
Simulated synthetic spectra support the existence of metal-poor cold clumps with large optical depth at $z\sim 6$ that could be potential population~III sites at low or intermediate redshift.
The expected dust content is in line with recent determinations.

\end{abstract}



\section{Introduction}\label{Sect:introduction}


The current understanding of cosmic structure formation relies on gas cooling and collapse within growing dark-matter potential wells \cite[e.g.][]{SaslawZipoy1967,GunnGott1972}.
The resulting gaseous fragments represent the sites where star formation takes place and subsequent stellar evolution processes enrich the medium with heavy elements.
Metal pollution is determined by the final fate of stars which synthesise metals during their lives and deliver them into the surrounding medium via supernova (SN) explosions or stellar winds. The relative production of each chemical element is tightly linked to the mass of the progenitor star: massive short-lived stars synthesise and predominantly expel oxygen and the $\alpha$-capture elements via type II SNe explosions
\cite[SN~II; e.g.][]{WW1995, ChieffiLimongi2004, HegerWoosley2010},
whereas less massive long-lived stars can enrich the surrounding medium through type Ia SNe (SN~Ia) which produce substantial amounts of iron-peak elements.
Therefore, to understand the observed abundance patterns at different epochs via theoretical models of structure formation it is important to follow cosmic chemical evolution of individual elements with appropriate delay-times.
\\
Several works \cite[see e.g.][for recent reviews]{Ciardi2005, Bromm2011} have shown that primordial structure formation can take place via catastrophic molecular-driven runaway collapse in rare high-density peaks already at redshift $z > 10$, when the Universe was less than 0.5~Gyr old.
At such early times metal spreading by primordial sources would rapidly pollute the local medium up to metallicities\footnote{
  Metallicities are indicated by $Z$ and are defined as the mass fraction of all the heavy elements in the gas phase.
}
$Z \sim 10^{-2} Z_\odot$ \cite[][]{Maio2010, Maio2011b, Wise2012, Wise2014, BiffiMaio2013} and could reach values of $Z \sim 0.1 Z_\odot $ \cite[][]{Campisi2011,Salvaterra2013, Pallottini2014,Ma2015} by $z\sim 6$, when the age of the Universe was roughly 1 Gyr.
At more recent epochs larger $Z$ values could be attained \cite[][]{Tornatore2007,Dave2011,Dave2013, Wuyts2014}.
This process induced a rapid transition from a primordial metal-free star formation regime led by non-equilibrium H$_2$ gas cooling \cite[][]{SaslawZipoy1967} to an enriched one characterised by efficient fine-structure metal line cooling \cite[e.g.][]{BrommLoeb2003, SantoroShull2006, Maio2007} and the subsequent formation of long-lived stars with masses $\lesssim 1 M_{\odot}$.
At lower redshifts, the Universe became increasingly dominated by cosmic star formation down to $z\sim 2$.
This epoch is responsible for converting a significant amount of gas into stars and for a general flattening or decrement of the specific star formation rate, as observed at redshifts $2\lesssim z \lesssim 6$ \cite[e.g.][]{Daddi2007,Michalowski2010,Reddy2012,Stark2013, Bouwens2014, Duncan2014}.
During this period the first intermediate- and low-mass stars evolved and locally contributed heavy elements, thereby changing the composition of the cosmic medium.
\\
Although the global picture seems qualitatively clear, there are still many quantitative uncertainties due to our incomplete knowledge of nucleosynthesis and abundance evolution over cosmological times.
Additional unknowns are linked to the effects of the environment at various $z$, which might influence the chemical patterns of cosmic objects.
Despite the utility of simple stellar population models in testing typical trends of abundance ratios, they have limited predictive power in explaining inhomogeneous metal spreading that requires 3D numerical calculations.
\\
In this respect, a valuable tool for testing chemical abundances in the high-redshift Universe is offered by damped Ly-$\alpha$ systems (DLAs) \cite[e.g.][]{Wolfe1986,Foltz1986,Smith1986}, both from an observational point of view \cite[see e.g.][for a review]{WolfeGawiserProchaska2005} and from the theoretical one.
Numerical investigations available in the literature have mainly focused on
\HI\ properties
\cite[e.g.][]{Katz1996,HaehneltSteinmetzRauch1998,Potzen2008,Popping2009,Erkal2012,Cen2012,Duffy2012,Yajima2012,Rahmati2013}
and neglect heavy-element abundance ratios, which offer a powerful tool to study the origin and evolution of cosmic chemical enrichment. This is one of the primary goals of our paper.
\\
DLAs are defined as clouds of predominantly neutral gas with \HI\ column
densities in excess of $2\times10^{20}$ atoms cm$^{-2}$
\cite[e.g.][]{Moller1998, Prochaska2003},
which appear to be associated with galaxies at the faint end of the galaxy
luminosity function and represent neutral proto-galactic objects that are just building up their stellar mass
\cite[][]{Fynbo2010, Krogager2012, Bouche2012}.
They are usually cold or ``warm'' structures with typical kinetic temperatures between a few $10^3\,\rm K$ 
\cite[][]{Petitjean2000, Srianand2012}
and $\sim 10^4\,\rm K$ 
\cite[][]{PettiniCooke2012, Carswell2012, Noterdaeme2012} 
constituted by largely neutral or molecular hydrogen
\cite[e.g.][]{Vladilo2001, Cui2005}
and with metal content that can be measured precisely
\cite[][]{Pettini2008, Battisti2012}.
Owing to their thermodynamical properties, DLAs are the preferential sites for the initial stages of gas cooling and star formation and are useful probes of metal absorption lines at redshift $z > 6$
\cite[][]{DodoricoMolaro2004, Simcoe2012, Becker2012}.
These systems are also valuable tests for stellar evolution parameters 
\cite[e.g.][]{Salvadori2012, Kulkarni2013}
and chemical evolution models in low-metallicity environments or in the early Universe
\cite[e.g.][]{Bolton2011, Maio2013, Finlator2013, Keating2014}.
They are precious probes of the abundance of nuclides formed during primordial nucleosynthesis, such as deuterium, D \cite[e.g.][]{Cooke2014}, and have also been employed to investigate possible variations in the fundamental physical constants (e.g. the ratio between the proton and electron mass, $\mu$) at different cosmic epochs \cite[e.g.][]{WendtMolaro2012, King2011, Rahmani2013}.
Observationally, DLAs do not appear to be preferentially associated with bright actively star forming galaxies \cite[e.g.][]{Fumagalli2015}, however, both indirect experiments \cite[][]{WolfeChen2006,Rafelski2011} and direct searches for host galaxies \cite[][]{Moller2004, Fynbo2008} suggest a physical connection between DLAs and star formation.\\
Considering the availability of DLA spectra covering a wide range of redshifts, we use them in our work as observational tools for theoretical predictions of chemical abundance ratios.
\\
In this paper, we present results for atomic and molecular abundances of cold, gas-rich environments at $ 2\lesssim z \lesssim 7$ as expected from 3D numerical simulations of cosmological structures including non-equilibrium chemistry, detailed stellar evolution and chemical enrichment.
We compare our theoretical predictions for cosmic chemical abundances to the
current suite of observational data on intermediate- and high-$z$ observations
of metal-poor DLA systems.
We study the effects of cosmic pollution on abundance ratios for different heavy elements and investigate the typical trends as a function of cosmic time. We find that typical ratios are usually scattered due to simultaneous enrichment by different stellar phases or by different stellar populations and are consistent with a general decrease of metal content at high $z$.
Molecular fractions show temporal and local variations as a result of star formation activity and related feedback processes.
\\
The paper is organized as follows.
Details on the numerical implementation and the data sample are given in Sect.~\ref{Sect:method}; chemical abundances are discussed in Sect.~\ref{Sect:results}; and the theoretical expectations from synthetic spectral observations are presented in Sect.~\ref{Sect:synt_qso}.
We summarise and conclude in Sect.~\ref{Sect:conclusions}.


\section{Method}\label{Sect:method}


In the following subsections, we briefly describe the most important features of the cosmological calculations we have performed (Sect.~\ref{Sect:simulations}), the selection criteria for our analyses (Sect.~\ref{Sect:selection}) and the data samples we have used (Sect.~\ref{Sect:data}).

\subsection{Simulations}\label{Sect:simulations}

\begin{table}
\begin{center}
\caption{ Reaction network.}
\label{tab:reactions}
\begin{tabular}{lr}
\hline
\hline
Reactions & References for the rate coefficients\\
\hline
 	H    + e$^-$   $\rightarrow$ H$^{+}$  + 2e$^-$ & A97 / Y07 / M07\\
	H$^+$   + e$^-$  $\rightarrow$ H     + $\gamma$    & A97 / Y07 / M07\\
        H + $\gamma$ $\rightarrow$  H$^+$ + e$^-$  & A97 / Y07 / M07 \\
	He   + e$^-$   $\rightarrow$ He$^+$  + 2e$^-$    & A97 / Y07 / M07\\
	He$^+$  + e$^-$   $\rightarrow$ He   + $\gamma$     & A97 / Y07 / M07\\
        He + $\gamma$ $\rightarrow$  He$^{+}$ + e$^-$   & A97 / Y07 / M07 \\
	He$^+$  + e$^-$   $\rightarrow$ He$^{++}$ + 2e$^-$    & A97 / Y07 / M07\\
	He$^{++}$ + e$^-$   $\rightarrow$ He$^+$  + $\gamma$ & A97 / Y07 / M07\\
        He$^+$ + $\gamma$ $\rightarrow$  He$^{++}$ + e$^-$  & A97 / Y07 / M07 \\
	H    + e$^-$   $\rightarrow$ H$^-$    + $\gamma$  & GP98 / Y07 / M07\\
        H$^-$ + $\gamma$ $\rightarrow$  H + e$^-$  & A97 / Y07 / M07 \\
	H$^-$    + H  $\rightarrow$ H$_2$  + e$^-$       & GP98 / Y07 / M07\\
        H    + H$^+$ $\rightarrow$ H$_2$$^+$  + $\gamma$ & GP98 / Y07 / M07\\
        H$_2^+$ + $\gamma$ $\rightarrow$ 2 H$^+$ + e$^-$ & A97 / Y07 / M07\\
        H$_2^+$ + $\gamma$ $\rightarrow$  H + H$^+$  & A97 / Y07 / M07\\
        H$_2$$^+$  + H  $\rightarrow$ H$_2$  + H$^+$     & A97 / Y07 / M07\\
	H$_2$   + H   $\rightarrow$ 3H            & A97 / M07\\
	H$_2$   + H$^+$ $\rightarrow$ H$_2$$^+$  + H     & S04 / Y07 / M07\\
  	H$_2$   + e$^-$   $\rightarrow$ 2H   + e$^-$ & ST99 / GB03 / Y07 / M07\\
      	H$^-$    + e$^-$   $\rightarrow$ H    + 2e$^-$   &A97 / Y07 / M07\\
       	H$^-$    + H   $\rightarrow$ 2H    + e$^-$       & A97 / Y07 / M07\\
       	H$^-$    + H$^+$ $\rightarrow$ 2H             &P71 / GP98 / Y07 / M07\\
       	H$^-$    + H$^+$ $\rightarrow$ H$_2$$^+$  + e$^-$& SK87 / Y07 / M07\\
        H$_2$$^+$  + e$^-$   $\rightarrow$ 2H            &GP98 / Y07 / M07\\
        H$_2$$^+$  + H$^-$  $\rightarrow$ H + H$_2$   &A97 / GP98 / Y07 / M07\\
        H$_2$ + $\gamma$ $\rightarrow$  H$_2^+$ +  e$^-$  & A97 / Y07 / M07 \\
        H$_2$ + $\gamma$ $\rightarrow$ 2 H & A97 / R01 / Y03 / M07 \\
        D    + H$_2$   $\rightarrow$   HD   + H     & WS02 / M07\\
        D$^+$  + H$_2$   $\rightarrow$   HD   + H$^+$  & WS02 / M07\\
        HD   + H   $\rightarrow$   D   + H$_2$     & SLP98 / M07\\
        HD   + H$^+$  $\rightarrow$   D$^+$  + H$_2$   & SLP98 / M07\\
        H$^+$  + D   $\rightarrow$   H    + D$^+$   & S02 / M07\\
        H    + D$^+$  $\rightarrow$   H$^+$  + D    & S02 / M07\\
        D$^+$  + e$^-$  $\rightarrow$   D + $\gamma$    & GP98 \\
        D  + $\gamma$  $\rightarrow$ D$^+$  + e$^-$  & GP98 \\
        He    + H$^+$  $\rightarrow$ HeH$^+$ + $\gamma$  & RD82/ GP98 / M07\\
        HeH$^+$    + H $\rightarrow$ He  + H$_2^+$    & KAH79 / GP98 / M07\\
        HeH$^+$    + $\gamma$ $\rightarrow$ He  + H$^+$ & RD82 / GP98 / M07\\
\hline
\hline
\end{tabular}
\end{center}
Notes:
$\gamma$ stands for photons;
P71~=~\cite{Peterson1971};
KAH79~=~\cite{KAH1979};
RD82~=~\cite{RD1982};
SK87~=~\cite{SK1987};
A97~=~\cite{Abel_et_al1997};
GP98~=~\cite{GP98};
SLP98~=~\cite{SLD_1998};
ST99~=~\cite{ST99};
R01~=~\cite{Ricotti2001};
WS02~=~\cite{Wang_Stancil_2002};
S02~=~\cite{Savin_2002};
GB03~=~\cite{GB03};
Y03~=~\cite{Yoshida2003};
S04~=~\cite{Savin_et_al2004};
Y07~=~\cite{Yoshida2007early};
M07~=~\cite{Maio2007}.
\end{table}
The numerical calculations performed in this work extend our previous studies that investigate the interplay between galaxies and the intergalactic medium from low redshift ($z<2$) \cite[][\textsc{Angus} project]{Tescari2014,Katsianis2013arXiv} to the epoch of reionization and above \cite[][]{Maio2010, Maio2011, Salvaterra2013}.
Our hydrodynamical calculations are carried out via the parallel numerical code P-Gadget3, an updated version of P-Gadget2 \cite[][]{Springel2005}.

\subsubsection{Implementation}

The improved implementation used here combines several physical processes and, in particular, contains a self-consistent treatment of low-temperature cooling by molecules and metals, as described in \cite{Maio2007, Maio2010}.
We include detailed non-equilibrium atomic and molecular chemistry evolution for e$^-$, H, H$^+$, H$^-$, He, He$^+$, He$^{++}$, H$_2$, H$^+_2$, D, D$^+$, HD and HeH$^+$
\cite[][]{Yoshida2003, Maio2006, Maio2007,PM2012},
determined according to the reaction network in Tab.~\ref{tab:reactions}.
Temporal evolution is computed by solving the following equation for each chemical species with number density $n_i$ at temperature $T$:
\begin{equation}
\label{noneq_eq}
\frac{ {\rm d} n_i}{{\rm d} t}= \sum_p\sum_q k_{pq,i}(T) n_p n_q - \sum_l k_{li}(T) n_l  n_i,
\end{equation}
where ${\d t}$ is the time interval, $ k_{pq,i}(T) $ is the (temperature-dependent) creation coefficient of species $i$ as a result of interactions of species $p$ and $q$, while $k_{li}(T)$ is the destruction coefficient due to interactions of species $i$ and $l$ (see references in Tab.~\ref{tab:reactions}).
The non-equilibrium treatment imposes additional constraints on the simulation timestep, hence abundances are computed each $\sim 1/10$ the hydrodynamical timestep.
This choice assures a robust convergence of our calculations, as previously checked in literature \cite[e.g.][and references therein]{Abel_et_al1997,Abel2002,Yoshida2003,Maio2007,Maio2010,PM2012}, and allows us to have reliable estimates for the gas molecular content even at zero metallicities and in environments where simple collisional equilibrium is questionable.
\HI\ self-shielding corrections are accounted for on post-processing as in \cite{Rahmati2013}, taking into account density, temperature and the evolving UV background\footnote{
We note that similar fitting formulas can bear uncertainties due to unknown ionisation corrections.
}
\cite[][]{HaardtMadau2012}.
We additionally include cooling, star formation, stellar evolution and metal spreading \cite[][]{Tornatore2007, Maio2009, Maio2010, Maio2011b, Tescari2014} by tracing individual elements (He, C, N, O, Ne, Mg, Si, S, Ca, Fe) according to available stellar yields \cite[][]{WW1995, vandenHoek1997, Woosley2002, Thielemann2003}\footnote{
  We remind the reader that metallicity-dependent iron yields by \cite{WW1995} are overproduced by a factor of about 2.
See e.g. \cite{Timmes1995oct}, \cite{Francois2004} and references therein.
We take into account this correction.
},
UV radiation \cite[][]{HaardtMadau2001} and feedback effects \cite[see more details in e.g.][]{Springel2005, Tescari2014}.
The UV background can heat the gas and dissociate molecules in the inter-galactic medium (IGM), but cooling and star formation in dense collapsing sites are little affected.
Feedback from black holes although included is usually negligible at the redshift and masses considered here \cite[][]{Bird2014}.
The stellar lifetimes are given by \cite{Matteucci1986} and \cite{Renzini1986}.
The assumed stellar initial mass function (IMF) covers the range [0.1, 100] $\rm M_\odot$ and follows the log-normal shape proposed by \cite{Chabrier2003}, numerically approximated as in \cite{Tescari2014}.
Stars with masses above $8\, \rm M_\odot$ explode as SNe~II and inject an energy of $10^{51}\,\rm erg$ which heats the surrounding medium.
Lower-mass stars evolve through AGB or SNe~Ia phase.
Gas and heavy elements can be ejected by star forming regions via winds (at $450\,\rm km/s$). We mimic metal diffusion in the interstellar medium by smoothing individual metallicities over the neighbouring particles in the SPH kernel.
We refer to \cite{Maio2013b} for extended tests through simulations of isolated objects.
\\
We adopt a $\Lambda$CDM background cosmological model with present-day expansion parameter normalised to 100 km/s/Mpc of $h=0.704$, and baryon, matter and cosmological-constant parameters equal to $\Omega_{0,b}=0.0456$, $\Omega_{0,m}=0.272$, $\Omegal=0.728$ \cite[][]{Komatsu2011}.
Spectral parameters are assumed to be $\sigma_8=0.8$ and $n=0.96$.
We consider a box of 12 Mpc$/h$ (comoving) a side, where gas and dark-matter fields are sampled by $384^3$ particles for each species, resulting in a gas and dark-matter resolution of
$\sim 3.86 \times 10^5\,\rm M_\odot/{\it h}$ and 
$\sim 1.92 \times 10^6\,\rm M_\odot/{\it h}$,
respectively, and softening length of 1.5 kpc$/h$.
\\
We note that for a precise picture hydrodynamical simulations should resolve both rare objects on large scales and very small structures on tiny scales.
Because of numerical limitations, a fair trade-off to assess properly the luminosity functions and halo statistics at high redshift is normally obtained with boxes of $\sim 10-30$ Mpc/{\it h} a side \cite[e.g.][]{Campisi2011, Salvaterra2013, MaioViel2014}.
In terms of space resolution, 1.5~kpc/{\it h} softening length used in this work is enough to provide a realistic description of galaxies at high and low redshift \cite[][]{Tescari2014, Katsianis2013arXiv}, within the limits of the modeling.
In terms of star formation and metal evolution, our cosmic star formation rate densities converge at $z\lesssim 15$ as well as metal distributions and filling factors \cite[e.g.][]{Maio2010, Maio2011b, Maio2012}.
These considerations hold even for different cosmologies and model parameters \cite[][]{MaioIannuzzi2011,Maio2011,Maio2011cqg,Maio2012,MaioKhochfar2012,MaioViel2014}.
\\
It is worth stressing that limitations of stellar evolution results are often related to severe unknowns in stellar evolution features.
{\it Rectius}, in order to have accurate metal ratios at different epochs it is fundamental to follow temporal evolution and mass-dependent yields from SN~II, AGB and SN~Ia stars.
While massive SN~II explosions dominate O and $\alpha$ content in early star forming phases, lower-mass AGB stars are particularly important at intermediate epochs, because their ejecta contain significant amounts of C, N and possibly O.
Hence, they can alter the pre-existing chemical patterns.
Dwarf stars ending their life as SN~Ia are expected to cause, instead, a remarkable boost in Fe content.
Since we follow stellar evolution from all the aforementioned stellar phases and consistently with mass-dependent yields and lifetimes, our results should be very robust.
In this work we focus on regular population~II stellar generations, neglecting the primordial population~III (popIII) ones, since their contribution to the cosmic star formation is believed to be minor at these epochs \cite[e.g.][]{Maio2010, Maio2011b, Wise2014}.
\\
In addition, when assessing gas molecular content in different environments it is important to follow detailed molecular evolution with a proper non-equilibrium scheme, as the one adopted here.
Simple phenomenological prescriptions for H$_2$ and HD can lead to misleading conclusions.
Effects from dissociating Lyman-Werner radiation coming from the first stars can have some impacts on the local gas, but the following collapse phases are usually dominated by resonant or fine-structure transition cooling from newly ejected heavy elements \cite[e.g.][]{Maio2007}.
These have been included as detailed in the previous paragraphs.
\\
Cosmic structures are identified by means of friends-of-friends and Subfind algorithms \cite[e.g.][and references therein]{Dolag2009} and are post-processed to trace masses, positions, velocities, star formation rates, H, He, H$_2$, HD, C, N, O, Ne, Mg, Si, S, Ca, Fe, etc., and all the relevant physical properties in each object.
To avoid spurious fragmentation or numerical artifacts \cite[][]{BateBurkert1997, Maio2009, deSouza2013}, we consider only those structures that are resolved with at least 200 gas particles or, equivalently, $\sim 10^3$ total particles.
This ensures that behaviour of the gas within bound structures is properly
treated.
\\
The cosmological redshifts we consider in this work are $z = $ 6.564, 4.992, 4.186, 2.995, corresponding to cosmic times of $\sim$ 0.85, 1.2, 1.5, 2.2 Gyr after the Big Bang, respectively (for our adopted cosmology).

\subsubsection{Basic properties}

\begin{figure}
\includegraphics[width=0.45\textwidth]{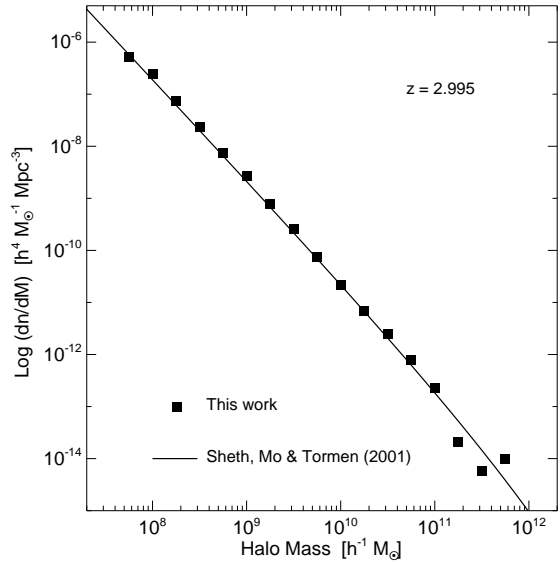}\\
\vspace{-0.5cm}
\caption[]{\small
  Simulated mass function at redshift $z = 2.995$ (points) compared to theoretical predictions at the same redshift according to \cite{Sheth2001} (solid line).
}
\label{fig:mf}
\end{figure}
\begin{figure}
\includegraphics[width=0.45\textwidth]{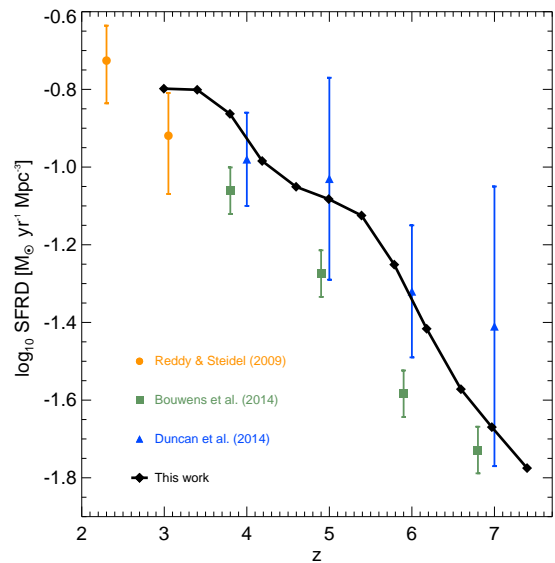}\\
\vspace{-0.5cm}
\caption[]{\small
  Cosmic star formation rate density, SFRD, as expected from our work (solid line with diamonds) and from observational determinations with 1$\sigma$ error bars by \cite{ReddySteidel2009} (circles), \cite{Bouwens2014} (squares) and \cite{Duncan2014} (triangles).
}
\label{fig:sfr}
\end{figure}
\begin{figure}
\includegraphics[width=0.5\textwidth]{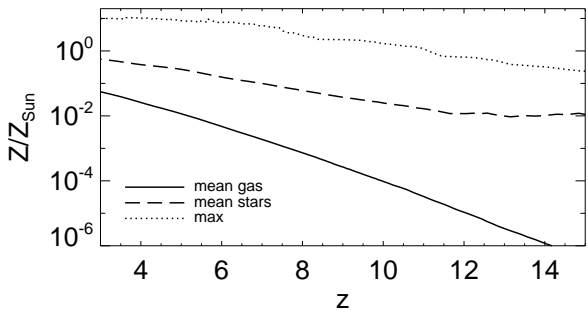}\\
\vspace{-0.5cm}
\caption[]{\small
  Metallicity evolution, $Z$, as a function of redshift, $z$. Lines refer to mean gas metallicity (solid), mean stellar metallicity (dashed) and maximum metallicity (dotted).
}
\label{fig:Z}
\end{figure}
\begin{figure*}
\includegraphics[width=0.3\textwidth]{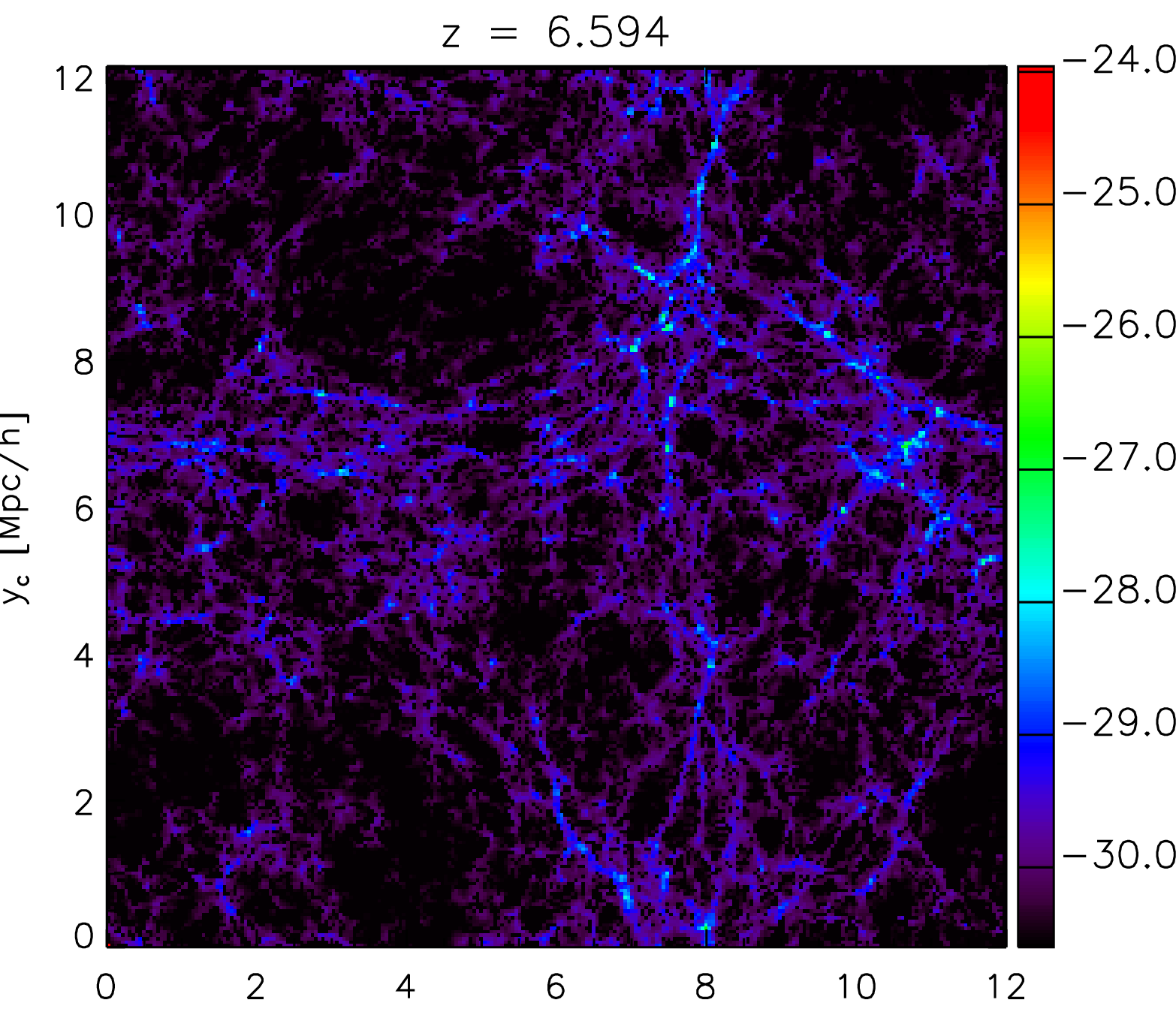}
\hspace{0.2cm}
\includegraphics[width=0.3\textwidth]{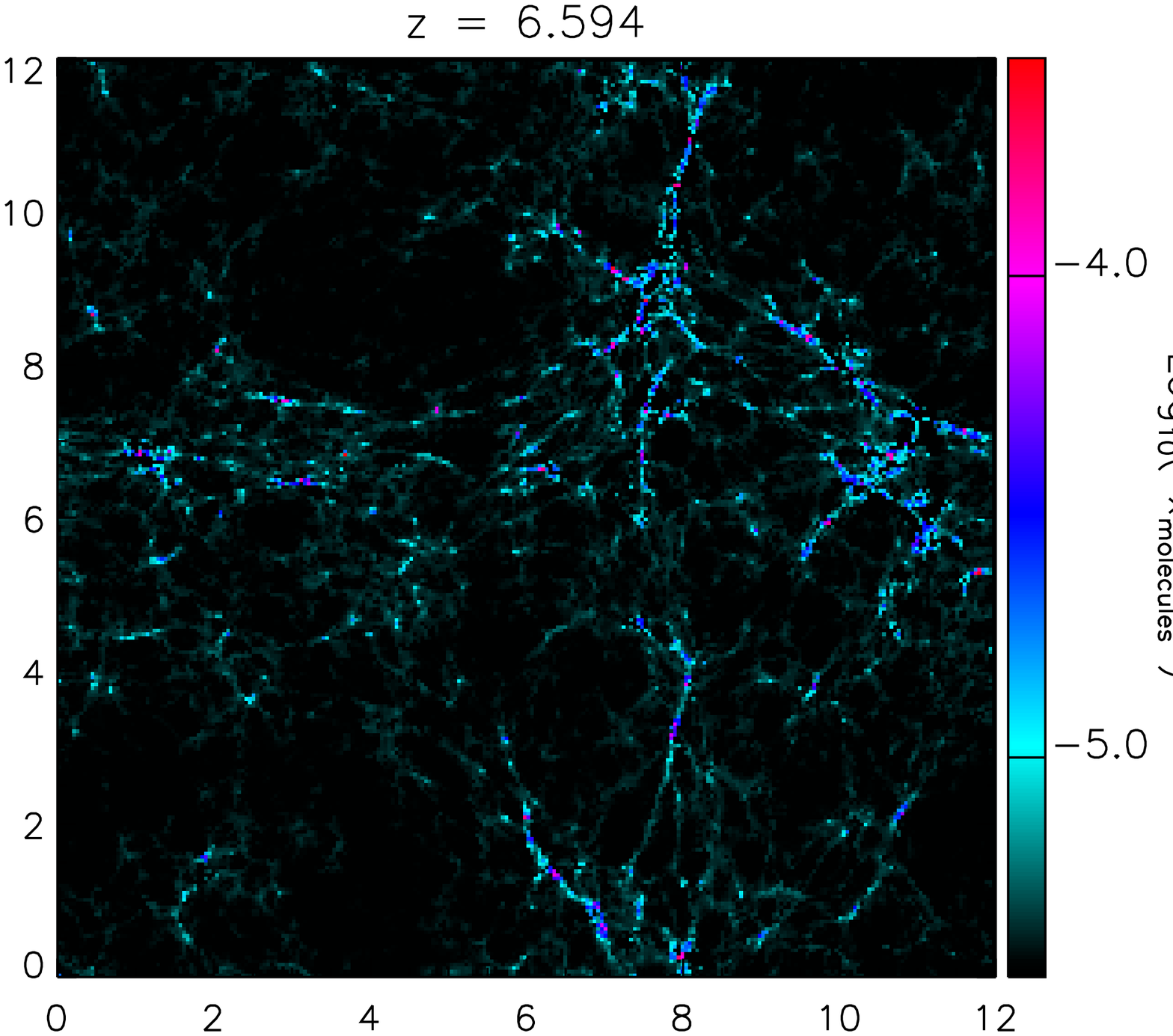}
\includegraphics[width=0.3\textwidth]{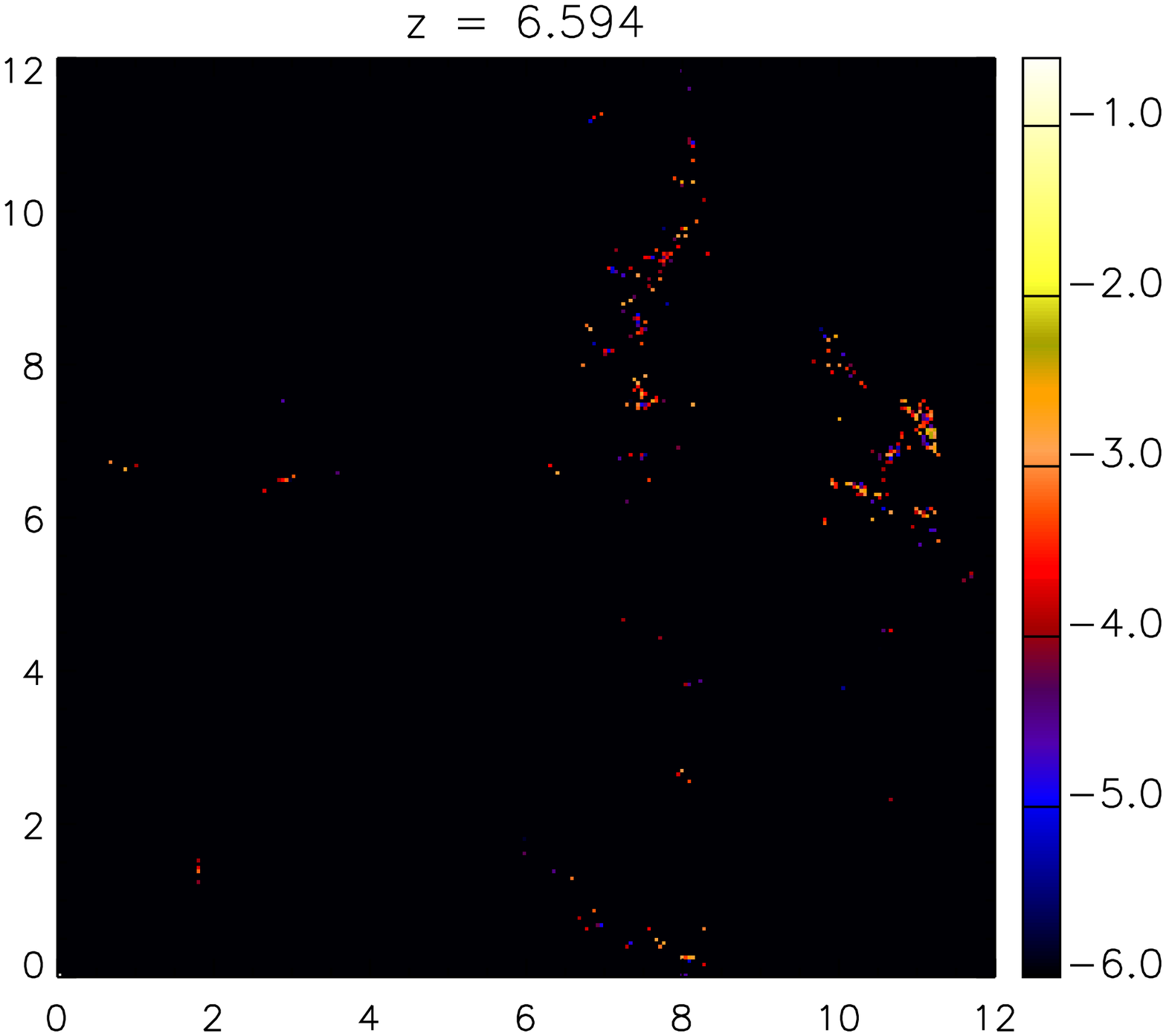}\\
\vspace{0.2cm}
\includegraphics[width=0.3\textwidth]{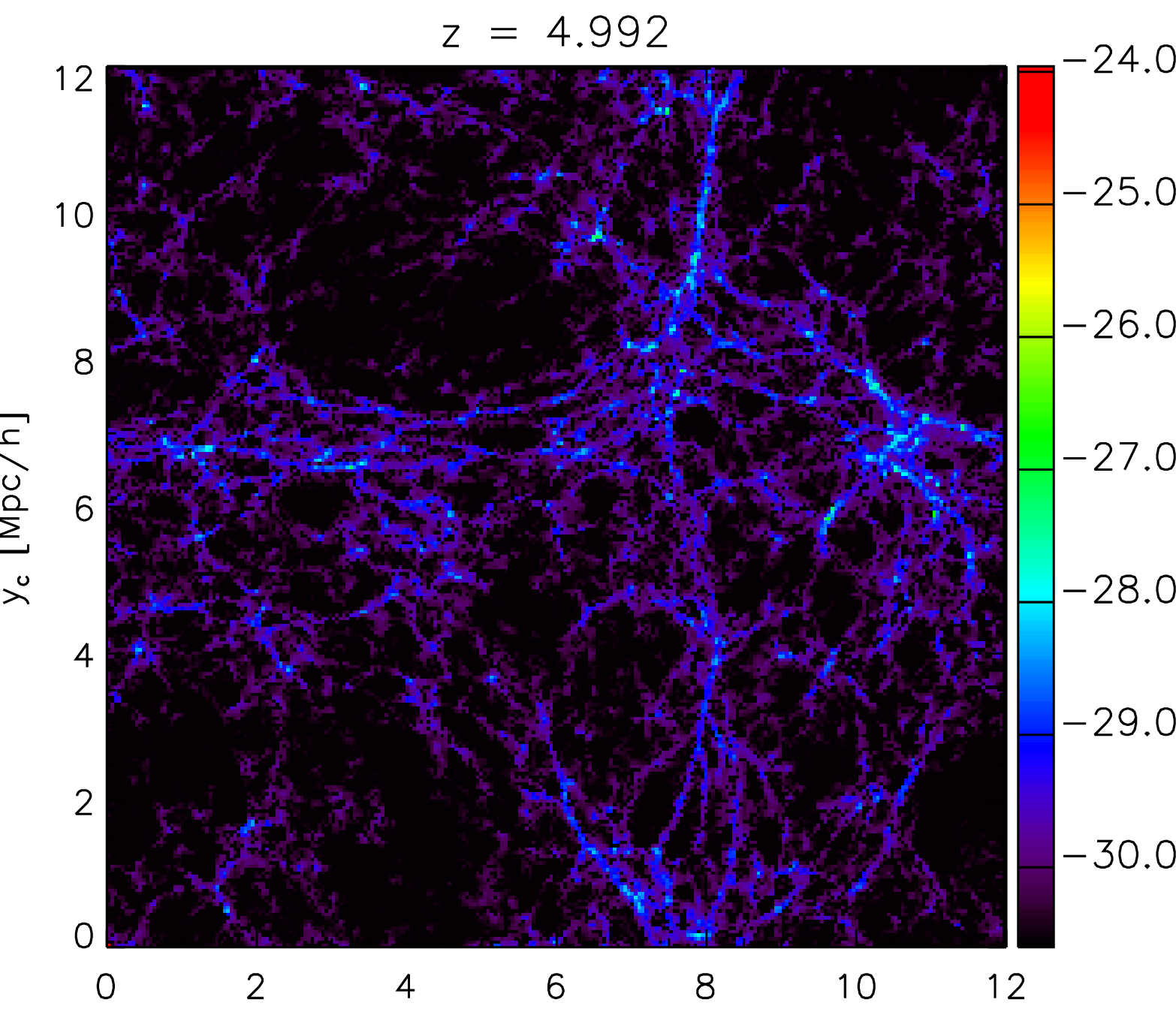}
\hspace{0.2cm}
\includegraphics[width=0.3\textwidth]{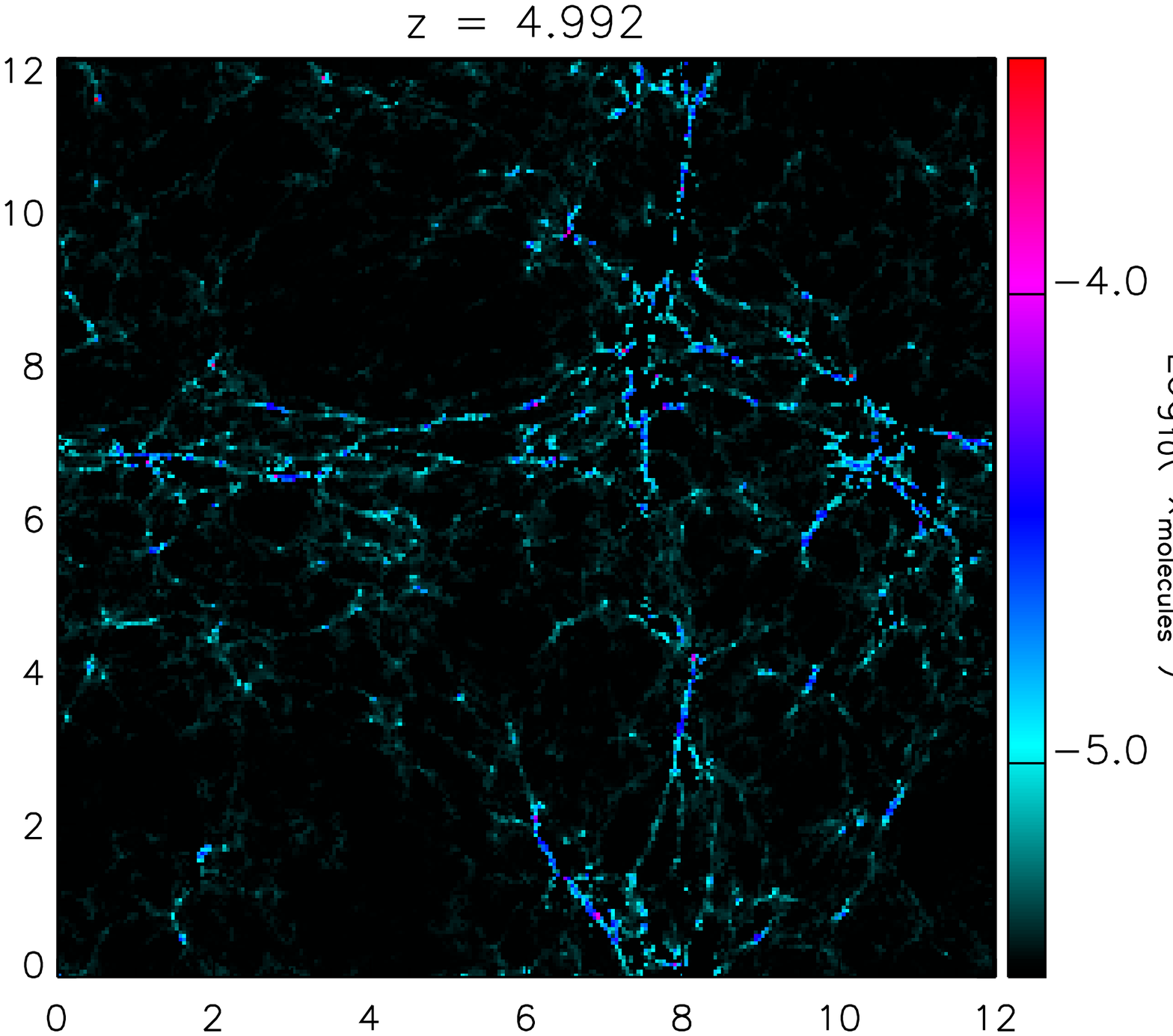}
\includegraphics[width=0.3\textwidth]{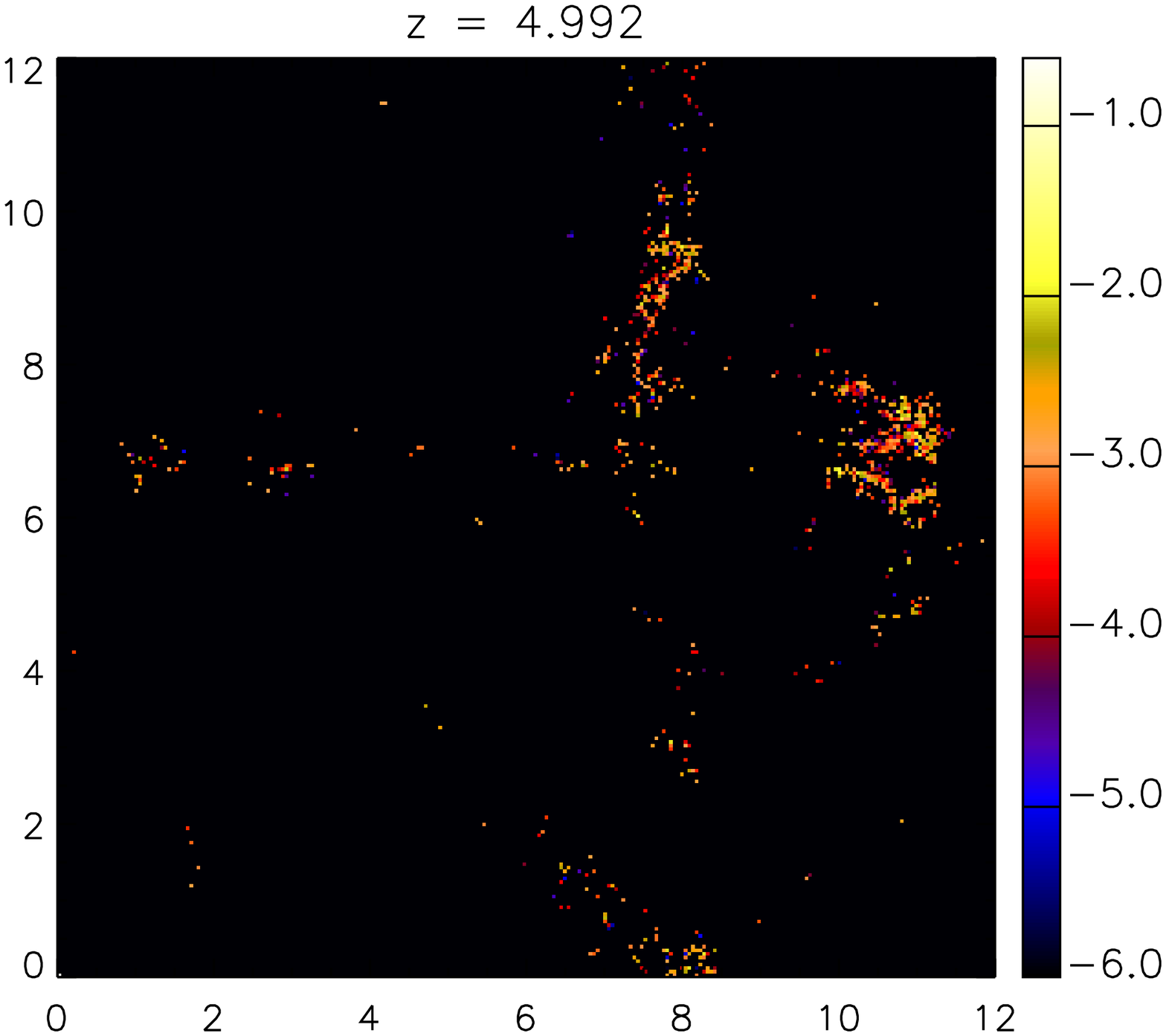}\\
\vspace{0.2cm}
\includegraphics[width=0.3\textwidth]{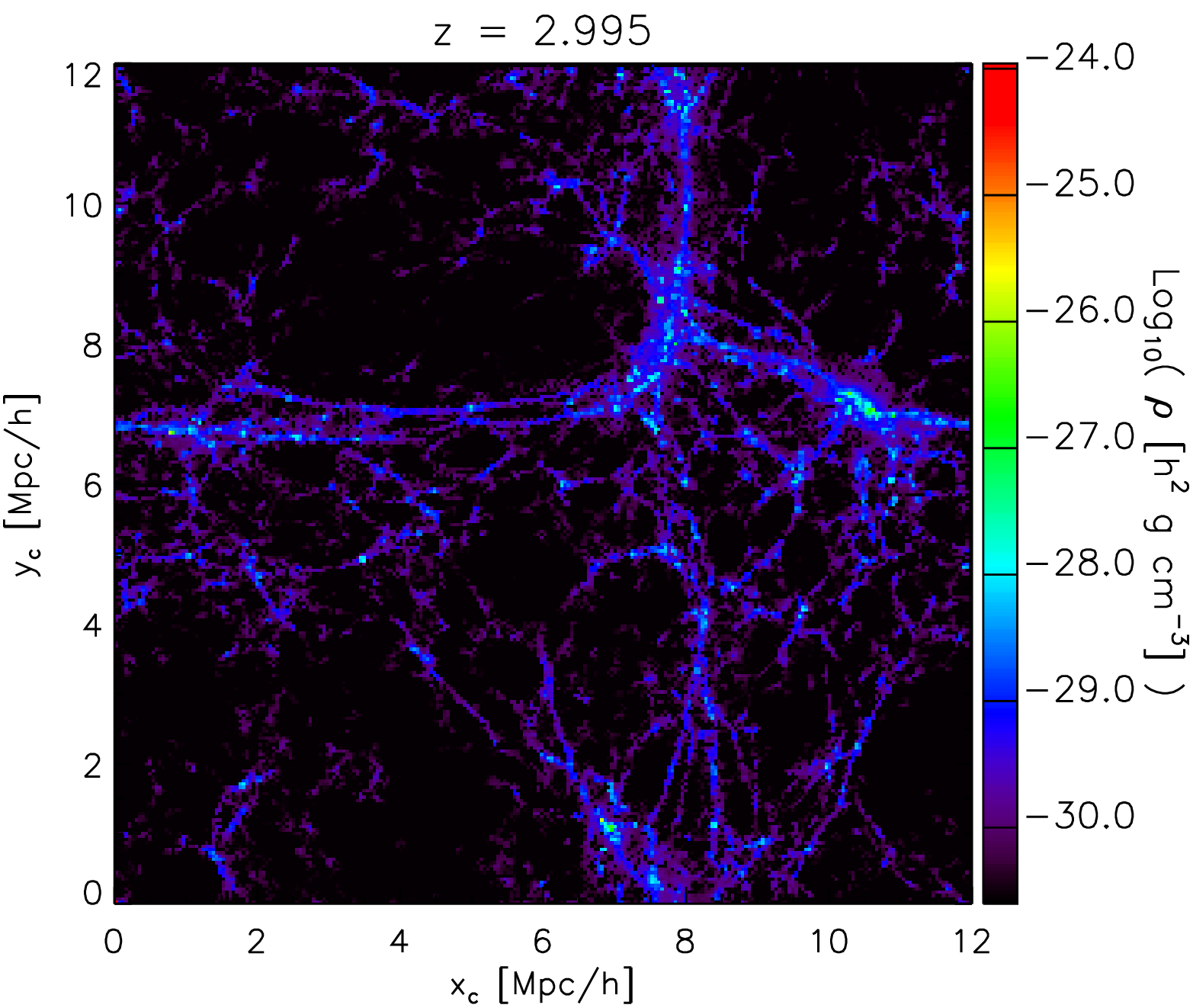}
\hspace{0.2cm}
\includegraphics[width=0.3\textwidth]{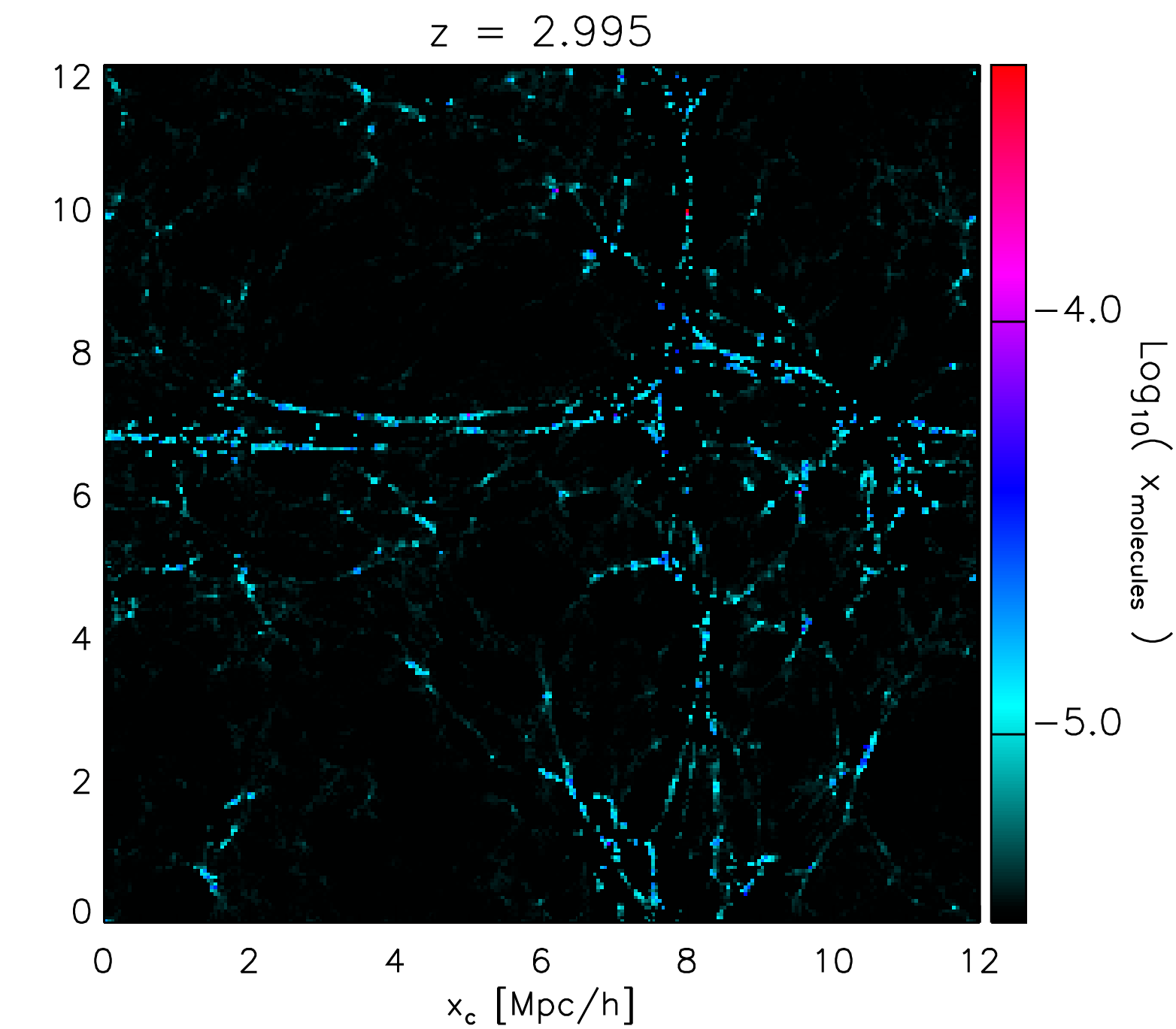}
\includegraphics[width=0.3\textwidth]{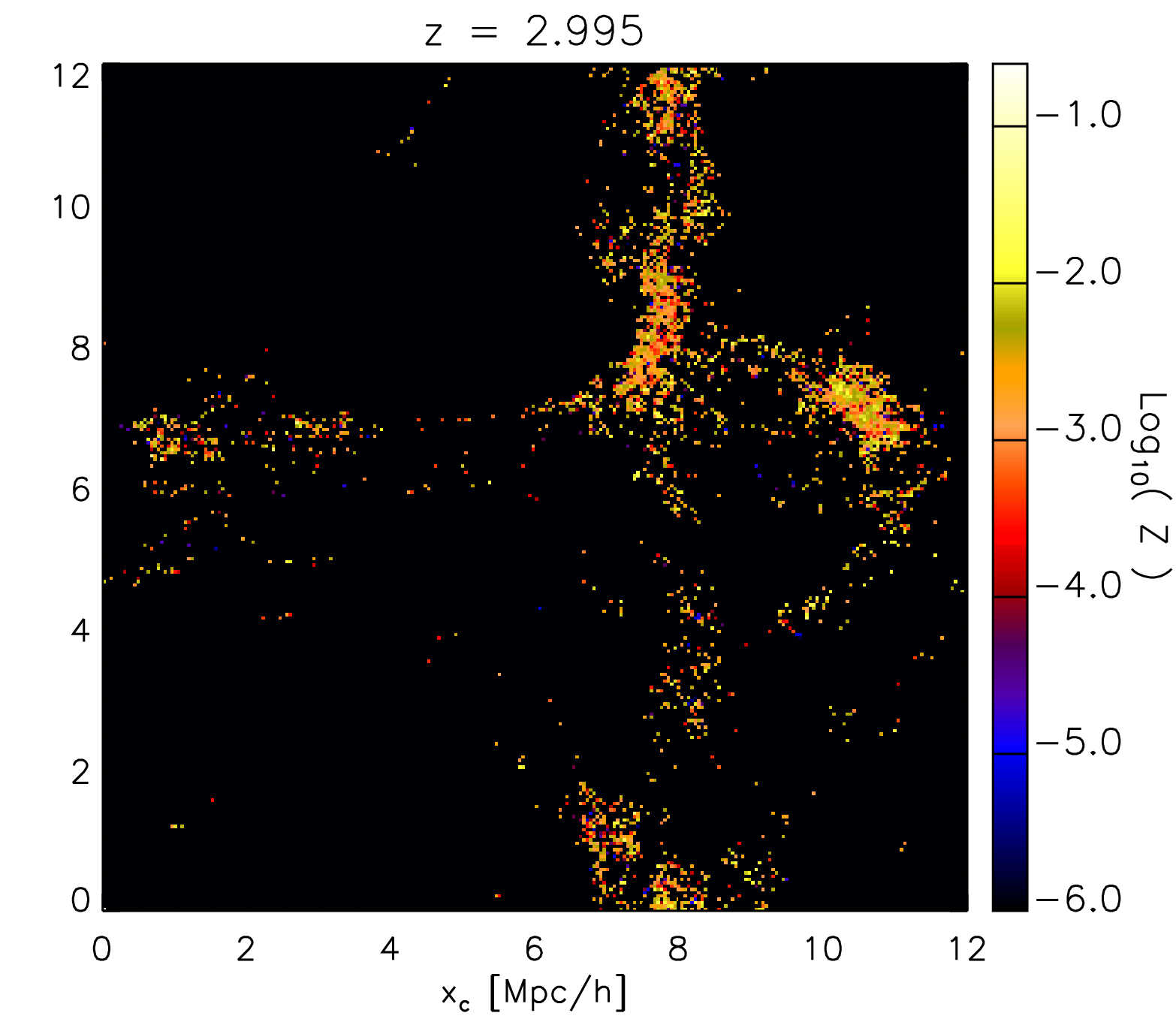}\\
\caption[]{\small
  Projected physical quantities at various redshift for a slice passing through the center of the box and 0.01~Mpc/{\it h} (comoving) thick. The values of $x_{\rm c}$ and $y_{\rm c}$ refer to the comoving coordinates in Mpc/{\it h} along the $x$ and $y$ box sides, respectively.
Left, central and right columns refer to mean gas density, molecular fraction and metallicity, respectively.
}
\label{fig:maps}
\end{figure*}

Numerical results are tested against analytical calculations in Fig.~\ref{fig:mf}, where the mass function extracted from the halo catalogue at $z=2.995$ (points) is displayed against corresponding theoretical expectations (solid line).
The general agreement between numerical and analytical calculations at all masses is satisfying and supports the reliability of our results. We refer the interested reader to our previous works \cite[e.g.][and references therein]{MaioViel2014} for more focused discussions on the effects of different parameters, resolution dependence and power spectra in similar simulations.
\\
In Fig.~\ref{fig:sfr} we also show the evolution of the cosmic star formation rate density, SFRD, which is consistent with observational determinations by \cite{Duncan2014} and features an increasing trend from $\sim 0.02$ up to $\sim 0.2\,\rm M_\odot yr^{-1} Mpc^{-3}$, as expected at $z\sim 2-7$, with a flattening around $z\sim 3$.
\\
The corresponding metallicity evolution is shown in Fig.~\ref{fig:Z}, where maximum, mean stellar and mean gas metallicities are plotted.
In this figure, $Z$ is computed as the mass fraction derived by C, Ca, O, N, Ne, Mg, S, Si, Fe, etc. contributions.
The increasing behaviour of the mean metal content in cosmic gas is a clear feature of metal production from star forming regions.
These star forming regions can pollute the surrounding gas through feedback effects and can boost gas mean metallicities up to $\sim 0.1\, Z_\odot$ by $z\lesssim 4$.
Enrichment can reach maximum values that exceed the solar abundance, locally.
The large spread in metallicity at any $z$ is a consequence of two main effects.
The first is due to spreading events starting from the high-density sites and polluting pristine low-density environments through wind and SN feedback.
The second is due to zero-metallicity particles that are particularly abundant at early epochs and bring down the mean values when averaging over all the particles
(see further discussion in Sect.~\ref{Sect:synt_qso}).
The metallicity of the stellar populations (mean stars in the legend) show some temporal variations, growing from 
$Z\simeq 10^{-2}\,Z_\odot$ to $Z\simeq Z_\odot$ 
between $z\sim 15$ and $z\sim 3$,
but featuring little evolution already at $z\lesssim 7$.
\\
To have a clear hint on the origin of such metals we will investigate in the next sections their abundance patterns, which vary mainly with the mass-dependent stellar yields and lifetimes of the actual stellar population.
\\
We warn the reader that, although the overall behaviour should remain
unchanged, details might depend on the hydro scheme or the modeling adopted for feedback processes.
They are not expected to be crucial, though, as previous studies based both on SPH and grid codes with different implementations seem to suggest
\cite[e.g.][]{Maio2010, BiffiMaio2013, Wise2014, Pallottini2014}.
\\
Visual representations of the cosmic objects and filaments are displayed in Fig.~\ref{fig:maps}.
There, we project along the $z$ axis density, molecular fraction and total metallicity of a slice passing through the center of the box at various cosmic epochs.
At $z=6.594$ primordial gas condenses mainly via H$_2$ cooling which boosts runaway collapse and star formation episodes.
Then, massive stars with short lifetimes undergo supernova explosions and eject heavy elements in the surrounding medium.
As a consequence, the Universe can be enriched up to metallicities as high as $Z\sim 10^{-3}$ (i.e. $\sim$~10 per cent the solar value) already at such epochs.
The subsequent evolution (e.g. at $z=4.992$) is characterized by further collapsing phases which increases the amount of heavy elements and star formation activity, consistently with Fig.~\ref{fig:sfr}.
Such events are accompanied by thermal feedback which is responsible for local gas heating and partial destruction of existent molecules.
This can temporarily halt the SFR as long as H$_2$ reforms in shocks or in newly collapsing material and metals start cooling efficiently.
After a couple of Gyr, molecular-driven star formation and metal enrichment have provided remarkable metal spreading with metallicities reaching values of $Z\sim Z_\odot$, as evident from the panels corresponding to $z=2.995$.
At all times we recover a sort of typical \emph{inside-out} pattern for cosmic pollution, with heavy elements produced in cold dense star forming regions and ejected by feedback effects into the lower-density IGM or circum-galactic medium (CGM), which is usually characterized by larger entropy and temperature.
Environmental effects can play a relevant role, since they can dilute metallicities (via e.g. shocks by SN explosions or winds) and affect the gas metal content with respect to the stellar one.
These trends for metal content are consistent with previous studies \cite[e.g.][]{Salvaterra2013, MaioViel2014, Ma2015}.
In the following we will address these issues in more depth and we will investigate the consequences for the resulting elemental ratios.

\subsection{Selection criteria} \label{Sect:selection}

The key tool to shed light on stellar evolution and its effects on the chemical enrichment of the Universe is the analysis of the chemical abundance ratios.
In this respect, measurements from metal-poor objects at high redshift are available.
In particular, recent constraints \cite[][]{Cooke2015} on the thermal and physical properties of metal-poor DLAs suggest typical temperatures\footnote{
 The sample of metal-poor DLAs by \cite{Cooke2015} (their Table 4) shows a quite broad temperature distribution with an average of $9600^{+2500}_{-2600} (1\sigma)^{+12200}_{-5000} (2\sigma) \,\rm K $.
}
of some $10^4\,\rm K$ (with a $2\sigma$ scatter larger than ten thousand K).
This is consistent with the assumption that DLAs trace a galaxy population
with low masses and luminosities\footnote{
We note that, although the temperatures mentioned above are comparable to gas temperatures in the Milky Way, the typical properties of the $z\sim 3-7$ gas clouds considered here are significantly different from the ones of the Milky Way at $z=0$ (see Sect.~\ref{Sect:results}).
}.
\\
Thus, as a first approach in Sect.~\ref{Sect:results} we will present our theoretical predictions by paying attention to those cold/warm gaseous structures whose temperature is below $ 3\times 10^4\,\rm K$.
With this simple selection we consider simulated gaseous structures that are, in average, compatible with the thermal properties of metal-poor DLAs and that can reveal typical features of the galaxy formation process (Sect.~\ref{Sect:results}).
The metal content of these systems is characterized by $\rm [Fe/H]\lesssim -2 $ (see more later).
A full discussion about detailed distinctions between possible observational classifications (e.g. metal-rich DLAs vs metal-poor DLAs) is beyond the goals of this work. We will rely on metal-poor data simply because their chemical content is not affected by dust uncertainties \cite[][]{Molaro2006}
(for rough estimates see Appendix~\ref{appendix}).
Metal abundance ratios in these systems are therefore relatively reliable (see next Sect.~\ref{Sect:data}).
Of course, the one above is a simplified assumption that becomes useful when analizing large sets of data and might, in some cases, hide systems that could be locally classified as absorbing/Lyman-limit systems.
\\
In order to perform a theoretical analysis akin to observational techniques, in Sect.~\ref{Sect:synt_qso} we will generate more sophisticated simulated observations by extracting physical properties along lines of sight (LOSs) through the collapsed structures identified with the temperature cut.
This will ensure that these objects are actually DLAs.
We will find that the results derived from these two different techniques are generally consistent.
\\

\subsection{Observational data}\label{Sect:data}
\begin{table}
\centering
\caption[Fractions]{Reference solar values \cite[][]{Asplund2009}.}
\begin{tabular}{lcccccc}
\hline
\hline
Element: & H & C & N & O & Si & Fe \\
Solar abundance: & 12.00 & 8.43 & 7.83 & 8.69 & 7.51 & 7.50\\
\hline
\label{tab:solar_abundances}
\end{tabular}
\begin{flushleft}
\vspace{-0.5cm}
{\small
}
\end{flushleft}
\end{table}

DLAs are discovered along the line of sight to powerful background sources (such as quasars, active galactic nuclei, or gamma-ray bursts) on the basis of their strong damped \HI\ \Lya\ absorption line, together with associated metal-absorption lines.
\\
Since the gas in DLAs is largely self-shielded, most elements reside in a single dominant ionisation state in the \HI\ gas \cite[e.g.][]{Vladilo2001}, allowing the metal content to be derived out to redshift $z\sim 6$ \cite[e.g.][]{Prochaska2007, Pettini2008, Ellison2011, Becker2012}.
\\
Measurements of molecular fractions are often difficult and complicated by the dissociation effects of the ultraviolet background in the outer, unshielded zones \cite[][]{Petitjean2000,Ledoux2003, Heinmueller2006}.
Consequently, mostly upper limits are available up to $z\sim 4.2$ and range between $\sim 10^{-7}$ and $\sim 10^{-2}$ \cite[][]{Noterdaeme2008}.
Currently, only a dozen measurements have been reported, which are typically derived from the coldest densest regions of high-$z$ gas and are generally biased towards larger metallicities \cite[][]{Noterdaeme2008, Srianand2010, Albornoz2014}.
\\
In the following, abundance ratios between two arbitrary species X and Y are presented with the usual notation:
\begin{equation}
[{\rm X}/{\rm Y}] \equiv 
\log \frac{(n_{\rm X}/n_{\rm Y})}{(n_{\rm X}/n_{\rm Y})_\odot} = 
\log(n_{\rm X}/n_{\rm Y}) -\log(n_{\rm X}/n_{\rm Y})_\odot
\end{equation}
where logarithms are base-10, $n_{\rm X}$ and $n_{\rm Y}$ are the number fractions of the species considered and the subscript ${}_\odot$ refers to the corresponding solar values.
All the chemical abundances are scaled according to the solar values by \cite{Asplund2009}, with hydrogen normalized to 12 on a logarithmic scale.
See Table \ref{tab:solar_abundances} for the solar composition we have adopted in this paper.
\\
The majority of the data points used in our study derive from the metal-poor DLA abundances reported by \cite{Cooke2015} for redshifts $2.3 < z < 4.5$ and \cite{Becker2011, Becker2012} for redshifts $4.7 < z < 6.3$.
First determinations for high-$z$ N abundance (at $z=4.466$) date back to \cite{Dessauges2001, Dessauges2007}, while more recent ones between $z\simeq 2$ and $z\simeq 3$ can be found in \cite{Zafar2014}.
Observational data for molecular fractions are taken from \cite{Noterdaeme2008} for $1.8 < z < 4.2$, from \cite{Srianand2010} for $z \gtrsim 3$, and from \cite{Albornoz2014} for $z \simeq 2.66$. Further upper limits for H$_2$ fractions are also taken from \cite{Noterdaeme2008} for $1.8 < z < 4.2$.
\\
We stress that at high metallicities (as large as $\rm [Fe/H] > -2$) dust depletion can affect the gas-phase element abundance ratios, making a comparison between simulations and observations much more uncertain \cite[][]{Molaro2006}.
The advantage of metal-poor DLAs is that these concerns are irrelevant and they are a safe and robust benchmark to test our numerical predictions.
\\
The observational samples we adopt throughout this work have been collected from a variety of sources and may contain selection biases. For example, some surveys specifically attempt to find the lowest-metallicity DLAs or systems with the highest molecular fraction.
Such biases are very difficult to take into consideration and different abundance estimates can differ by more than 1~dex \cite[see recent discussions in e.g.][]{Bonifacio2015arXiv}.
For this reason, we will focus mainly on the global trends exhibited by the data.
Additional constraints on early chemical enrichment might be derived by abundance ratios inferred for long gamma-ray burst (GRB) host galaxies.
Unfortunately, such data sample is currently very poor to draw definitive conclusions \cite[see][]{Ma2015} and will not be considered here.


\section{Results}\label{Sect:results}


In this section we present the main results from our analysis and illustrate the evolutionary pathways of cosmic gaseous systems at $z\sim 2 - 7$.
Then, we will use measured elemental ratios from metal-poor DLAs to probe the chemical enrichment of the Universe in addition to the trends of several key abundance ratios over the first $\sim 2$~Gyr.

\subsection{Cosmic structures at intermediate redshift}
\begin{figure*}
\centering
$z = $ 6.594\\
\vspace{-0.2cm}
\includegraphics[width=0.85\textwidth]{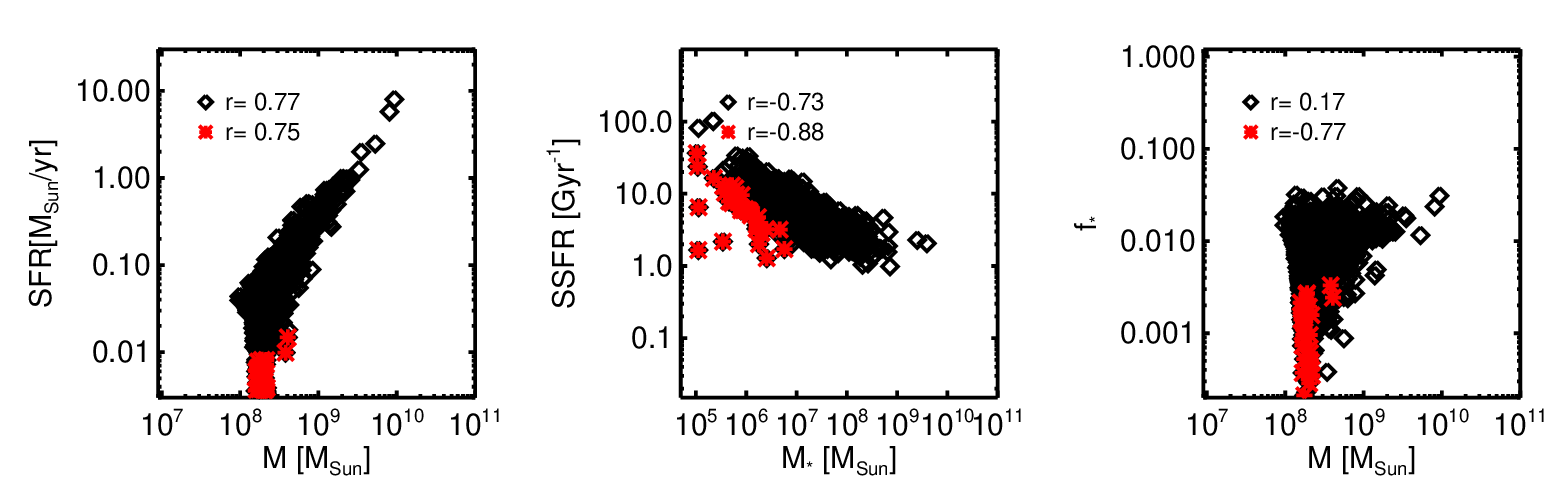}\\
$z = $ 4.992\\
\vspace{-0.2cm}
\includegraphics[width=0.85\textwidth]{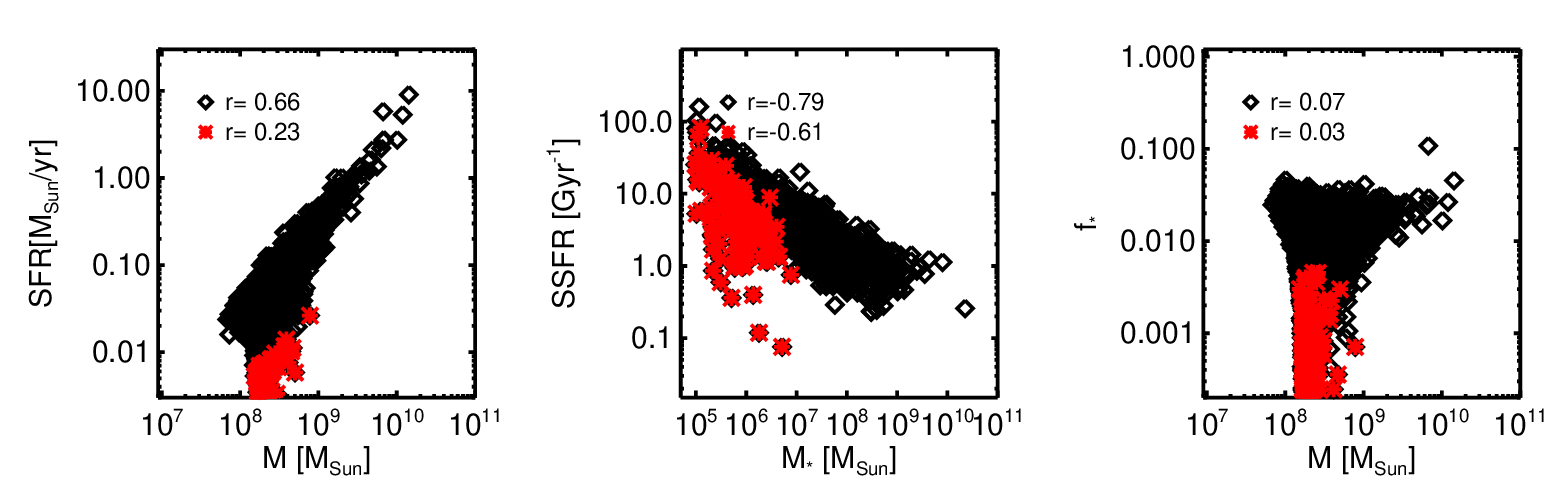}\\
$z = $ 4.186\\
\vspace{-0.2cm}
\includegraphics[width=0.85\textwidth]{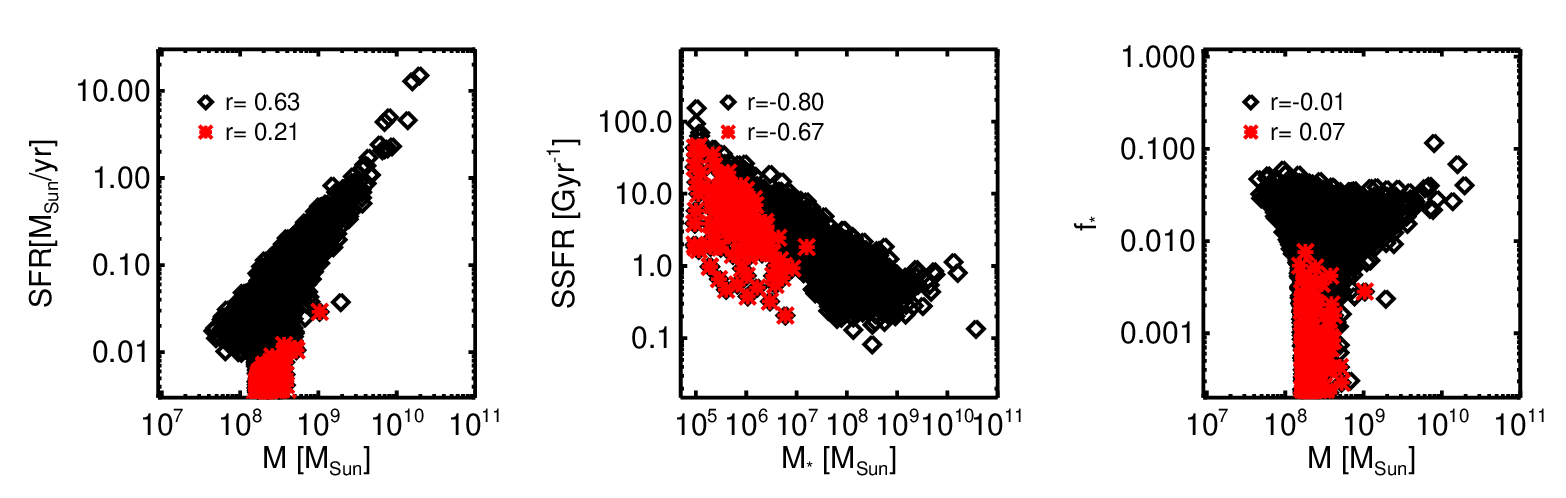}\\
$z = $ 2.995\\
\vspace{-0.2cm}
\includegraphics[width=0.85\textwidth]{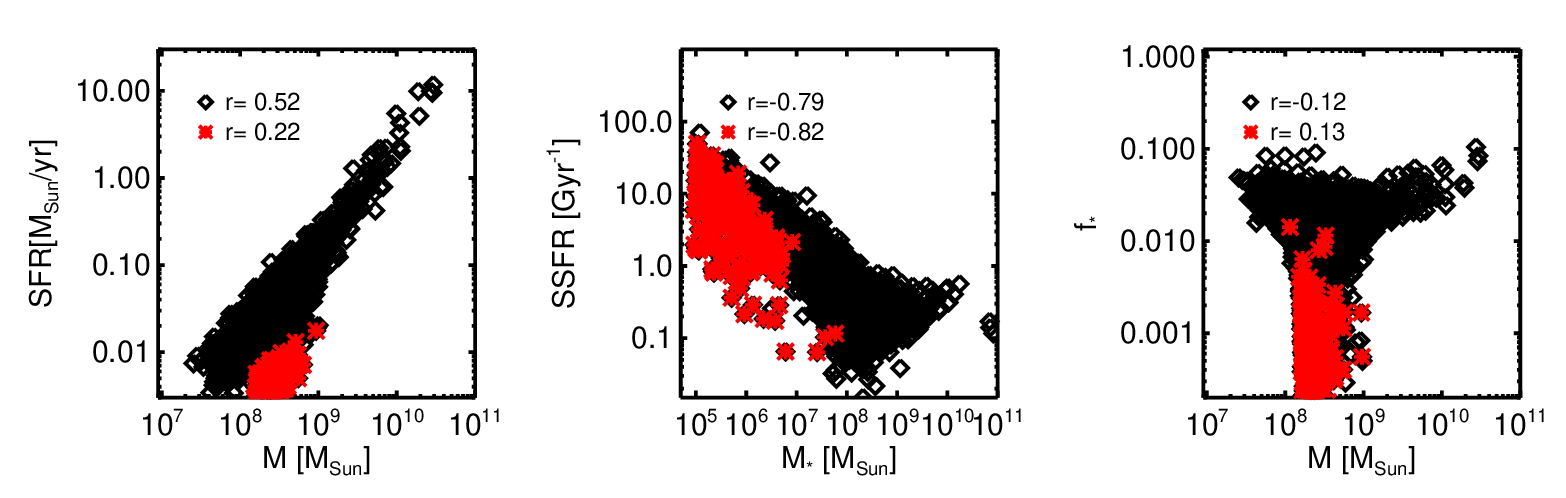}\\
\caption[]{\small
  The sequence shows the basic baryonic properties of our galaxy sample at redshift $z = $ 6.594, 4.992, 4.186, 2.995, respectively from top to bottom.
The three columns refer to star formation rate in $\rm M_\odot/yr$ as a function of gas mass in $\rm M_\odot$ (left), specific star formation rate in $\rm Gyr^{-1}$ as a function of stellar mass in $\rm M_\odot$ (center), and stellar fraction as a function of gas mass in $\rm M_\odot$ (right).
The Spearman correlation coefficient, $r$, is also quoted on the top-left corner of each panel for both the whole galaxy population (empty diamonds) and the metal-poor DLA candidates (asterisks).
}
\label{fig:sample}
\end{figure*}

Fig.~\ref{fig:sample} presents the basic baryonic properties of our simulated sample at redshift $z = $ 6.594, 4.992, 4.186, 2.995.
We show the star formation rate (SFR) as a function of gas mass, $M$, the specific SFR as a function of stellar mass, $M_\star$, and the stellar fraction, $f_\star$, as a function of $M$.
\\
Typical SFRs are roughly proportional to the gas mass and range between $\sim 10^{-3}$ and $\sim 10\, \rm M_\odot/yr$.
The gas content of our simulated galaxies is up to $\sim 10^{10}\,\rm M_\odot$ at redshift $z\simeq 6.594$, consistent with previous theoretical expectations at $z\sim 7$ \cite[][]{Maio2013}, and reaches $\sim 10^{11}\,\rm M_\odot$ at redshift $z\simeq 2.995$, growing by almost one order of magnitude in less than 1.5 Gyr. The galaxies that are associated with metal-poor DLAs (red asterisks) are normally characterised by low SFR values ($ \sim 10^{-2}\,\rm M_\odot/yr$) with a weak or null correlation with gas mass (having usually small Spearman coefficients, $r$).
We highlight that the estimated SFR for the Milky Way is $\rm\sim 1\, M_\odot/yr$ \cite[][]{Robitaille2010}.
\\
The specific SFR, $\rm SSFR \equiv SFR/M_\star$, is only slightly decreasing with stellar mass, $ M_\star$, as a consequence of the almost linear dependence of the SFR on $M_\star$ and in agreement with observational data at redshift $z\sim 2 $ \cite[e.g.][]{Daddi2007}.
The significance of such correlation is quantified by $r$ values that are always around $\sim -0.7$ and $-0.8$.
In general, SSFRs are larger at higher redshift; at $z\simeq 6.594$ there are structures with typical values of about $\rm \sim 10 \,Gyr^{-1}$ and in some cases reaching $\rm SSFR \sim 10^2\, Gyr^{-1}$.
This is consistent with estimates inferred from UV emission lines at high redshift \cite[][]{Stark2014} and is in line with recent high-$z$ theoretical \cite[][]{Salvaterra2013, Dayal2013} and statistical \cite[][]{deSouza2014,deSouza2015} studies supporting an evolution from a very bursty early Universe at $z\gtrsim 10$ to a more quiescent and evolved one at $z\lesssim 6$.
At later epochs, when the larger cosmic objects have consumed most of their gaseous fuel, the bulk of the SSFR distribution drops below a few Gyr$^{-1}$, with only a minor component above $\sim 10\,\rm Gyr^{-1}$, and broadly agrees with the low-$z$ estimates for $z\sim 2-4$ galaxies \cite[][]{Daddi2007,Michalowski2010,Reddy2012}.
By looking at DLA hosts one can see that some of them have large SSFR as a consequence of their extremely poor stellar content, that is often below 0.1 per cent.
This is consistent with the simple idea that such structures are probably small gas-rich objects that are susceptible to bursty episodes of star formation followed by periods of quiescence \cite[e.g.][]{Marcolini2006}.
As a comparison we mention that the Milky Way is instead a gas-poor galaxy with a stellar mass of $\rm\sim 10^{11}\, M_\odot$.
\\
This is more obvious from the behaviour of the stellar fraction, $f_\star$, as a function of $M$.
The former is not tightly linked to mass ($r \lesssim 0.2$) and, although larger masses can retain more gas and form more stars, feedback mechanisms dominate $\lesssim 10^9\,\rm M_\odot$ structures.
They cause a remarkable dispersion due to their more or less severe impacts in different environments and affect gas-to-star conversion efficiencies in a non-trivial way (see the following subsections).
\\
Baryon fractions for the Milky Way are difficult to measure.
Gas content in the solar neighborhood is limited and most of the visible
mass appears in stars.
By assuming a hosting dark halo with a mass of $\rm\sim 10^{12}\,M_\odot$ \cite[][]{Watkins2010} one can infer a value for $f_\star$ as high as $\sim 10$~per cent at $z=0$.

\subsection{Transient objects} \label{Sect:transient}

\begin{figure}
\centering
\includegraphics[width=0.23\textwidth]{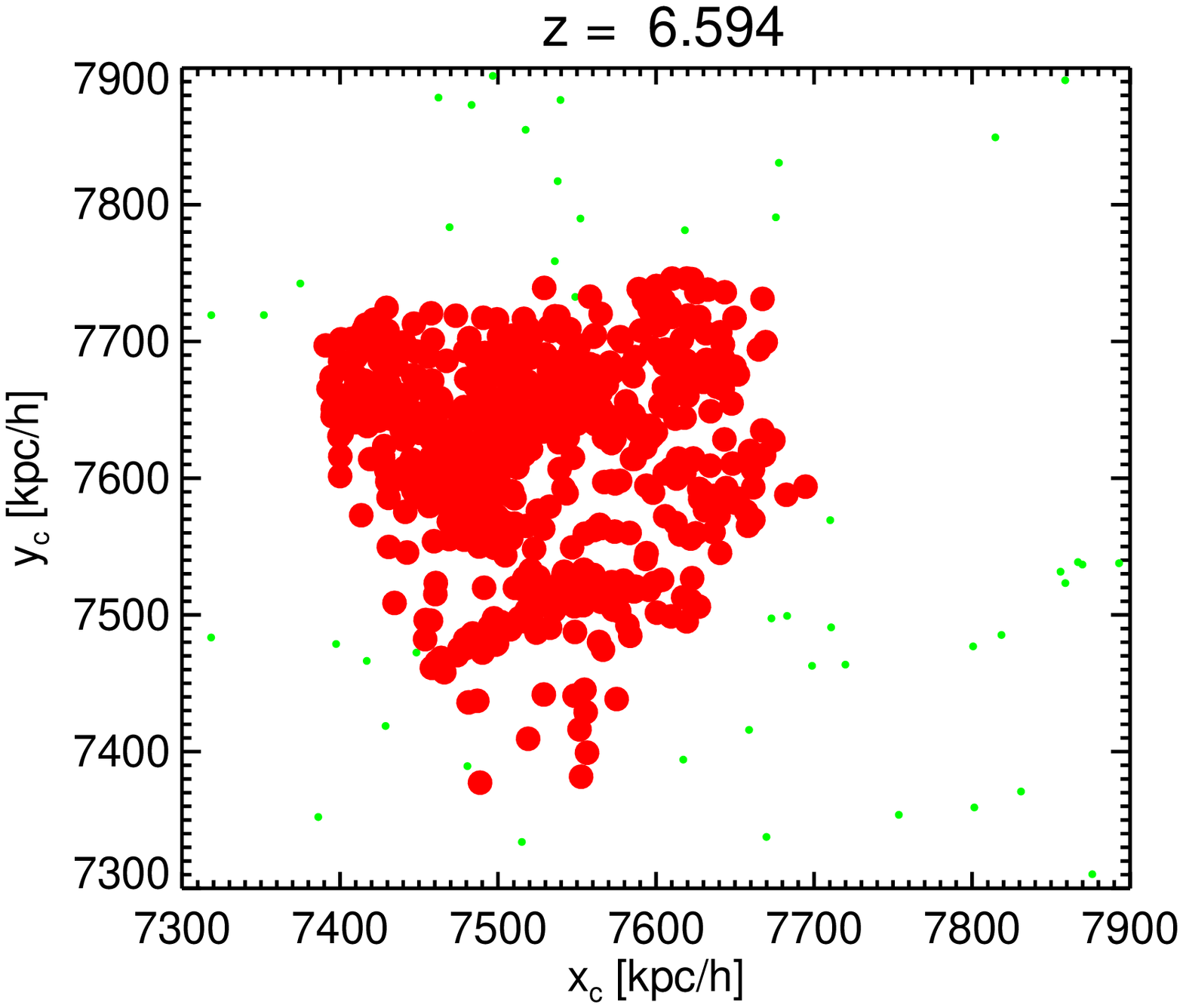}
\includegraphics[width=0.22\textwidth]{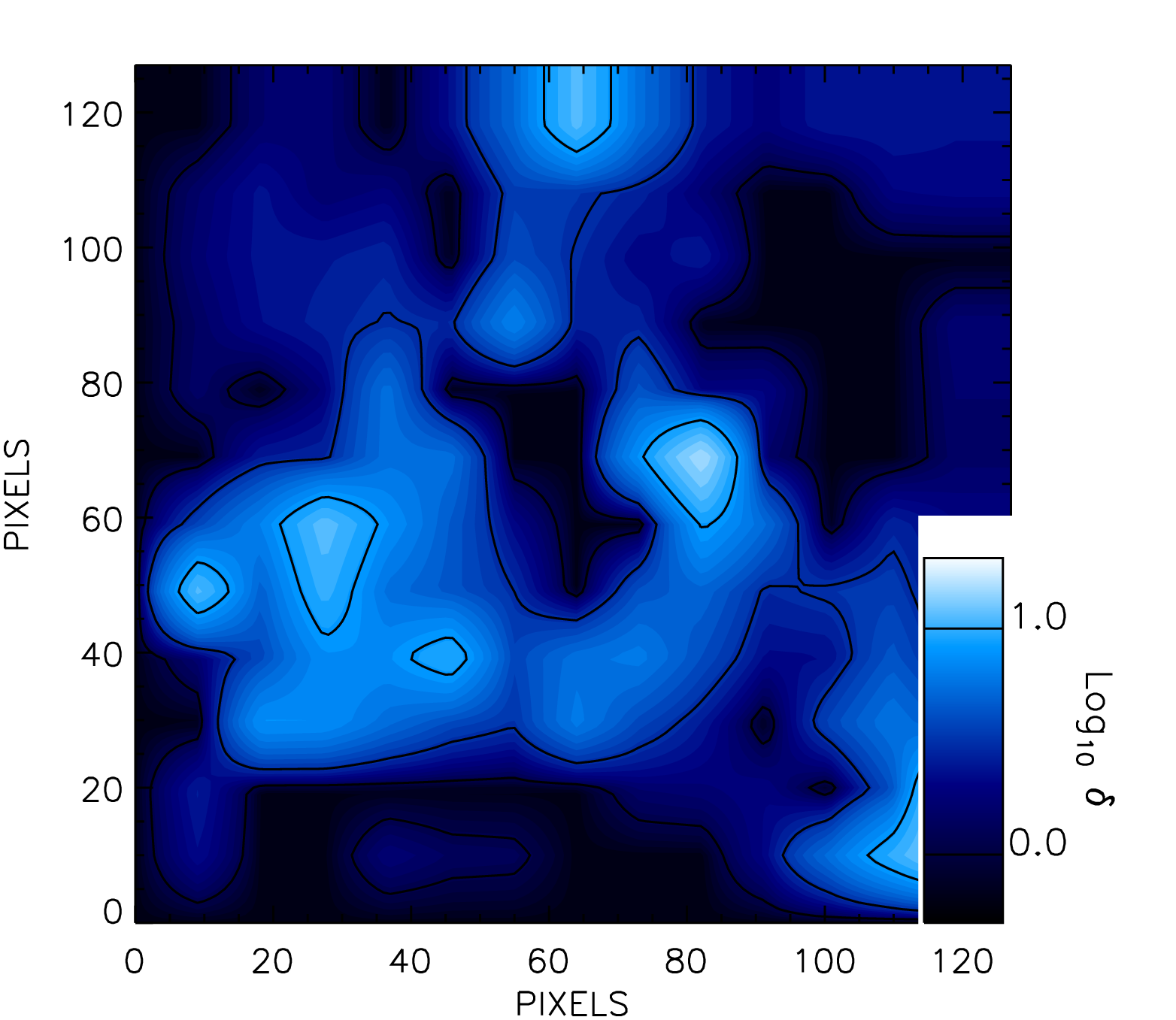}\\
\includegraphics[width=0.23\textwidth]{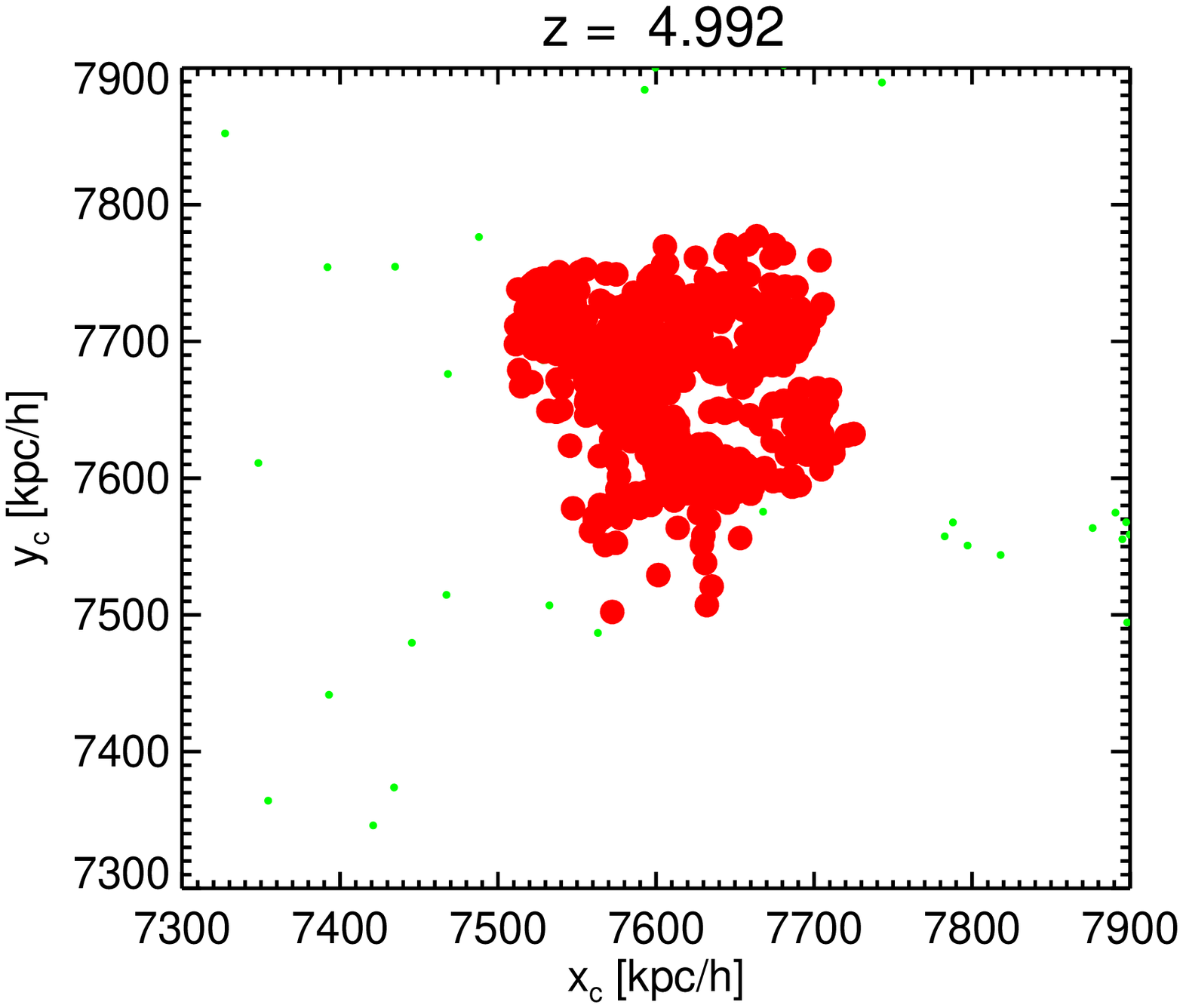}
\includegraphics[width=0.22\textwidth]{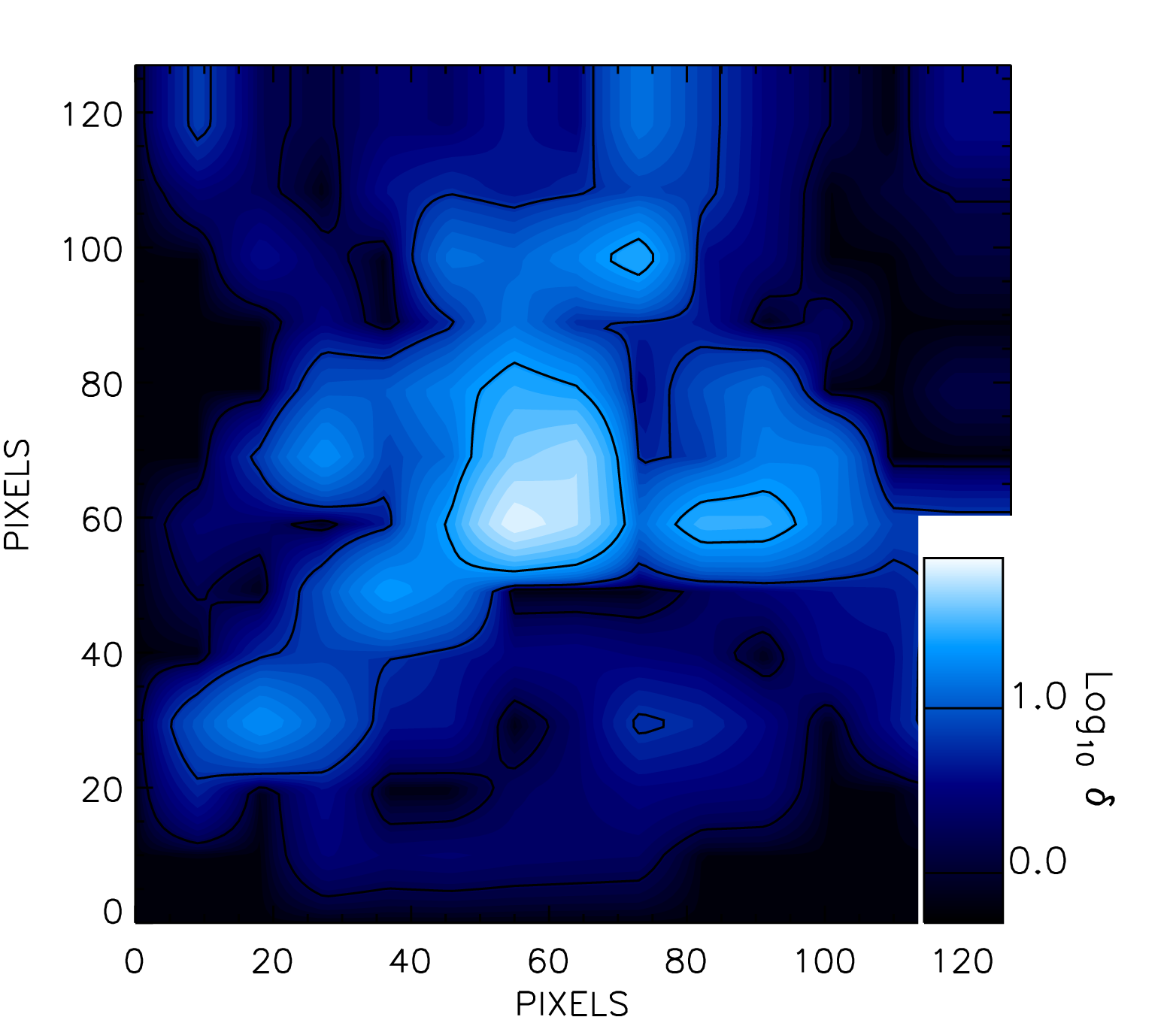}\\
\includegraphics[width=0.23\textwidth]{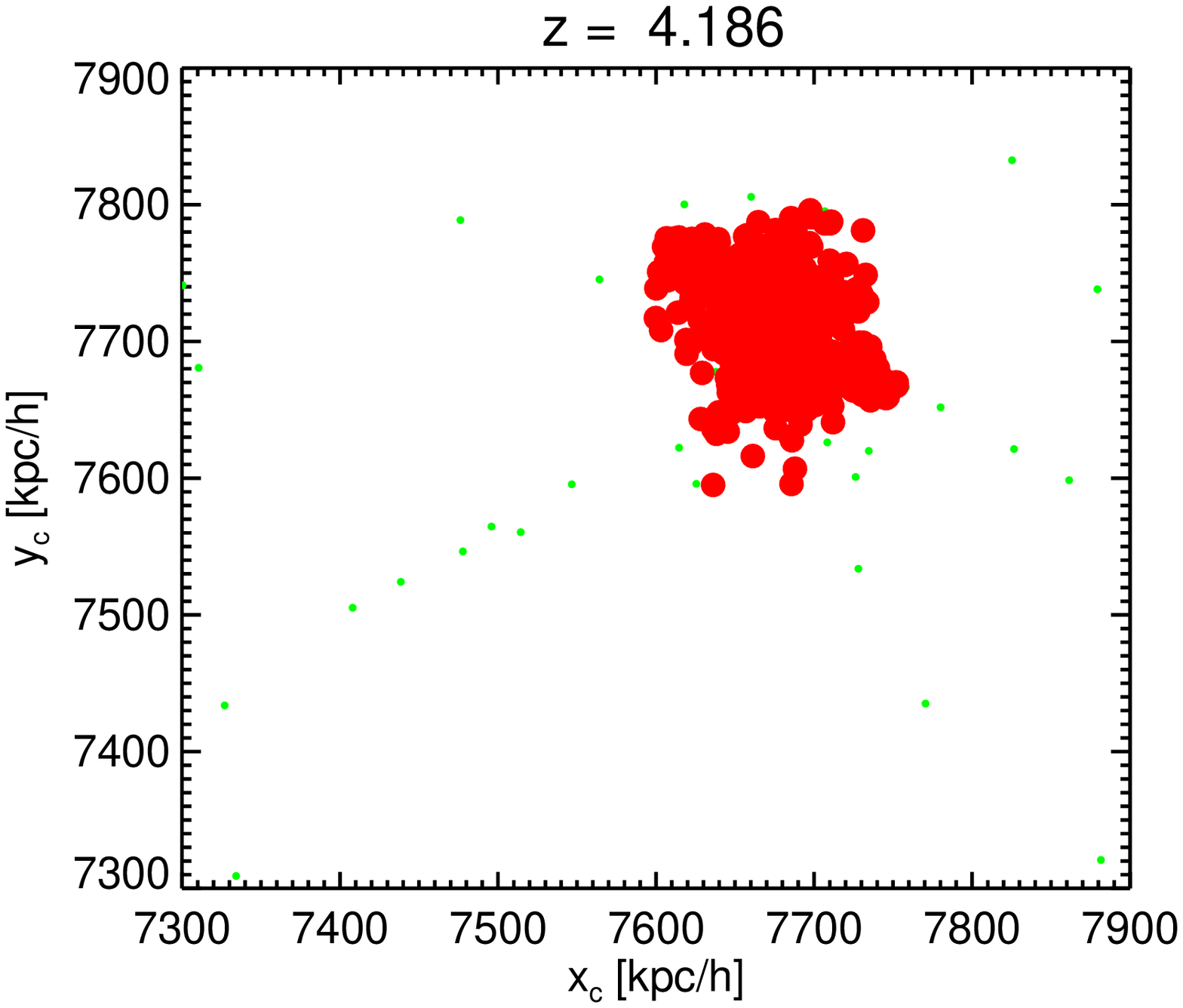}
\includegraphics[width=0.22\textwidth]{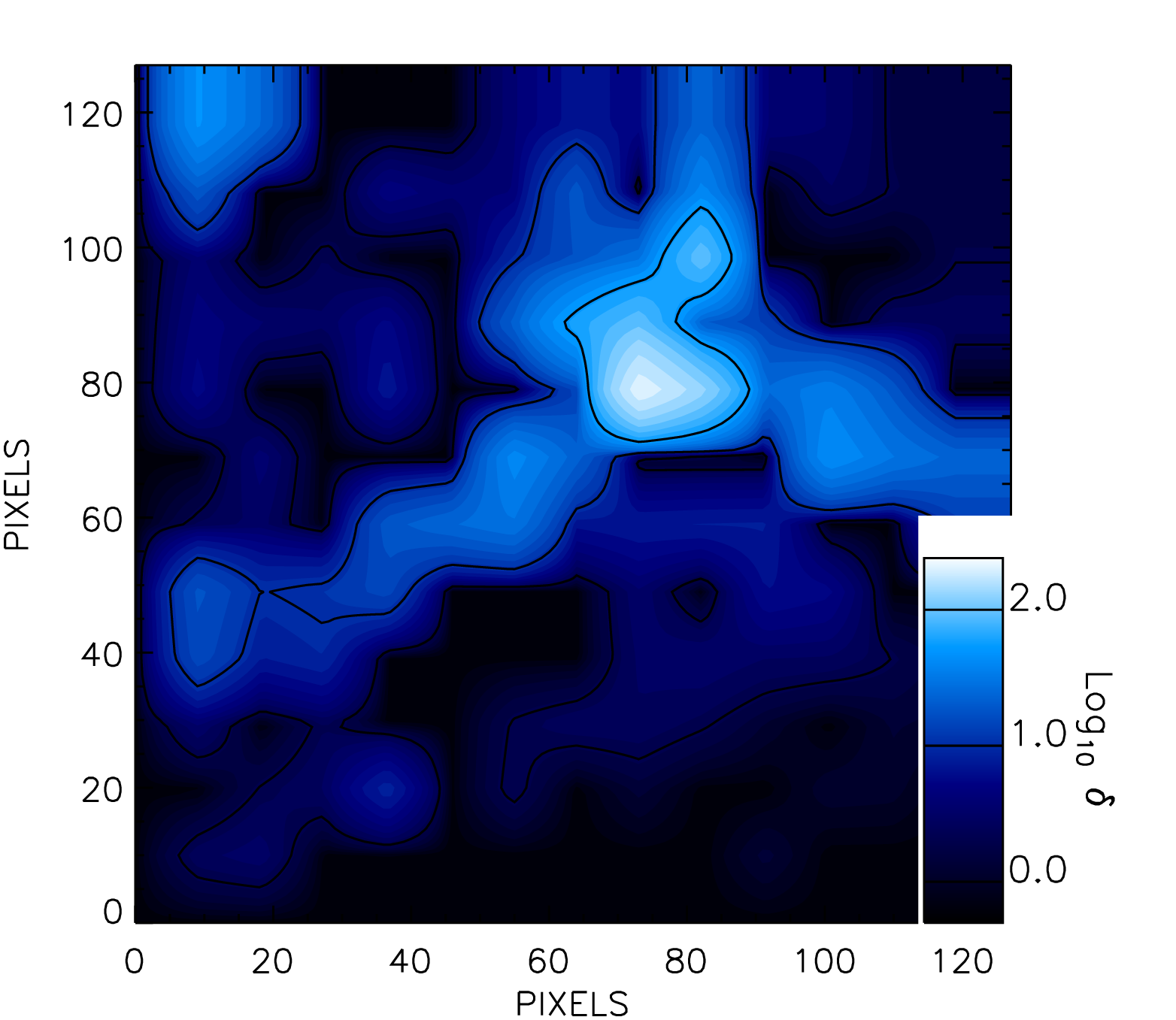}\\
\includegraphics[width=0.23\textwidth]{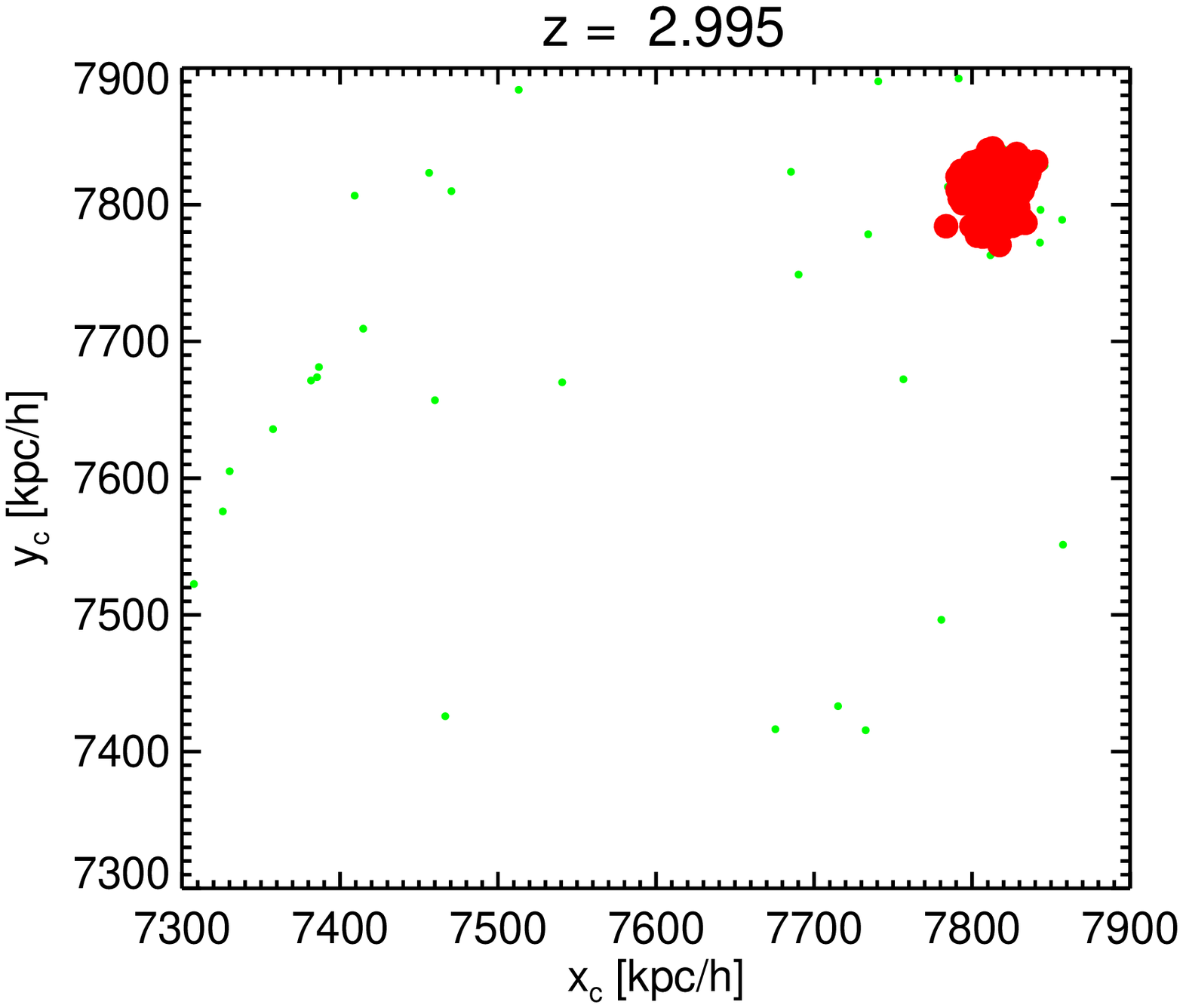}
\includegraphics[width=0.22\textwidth]{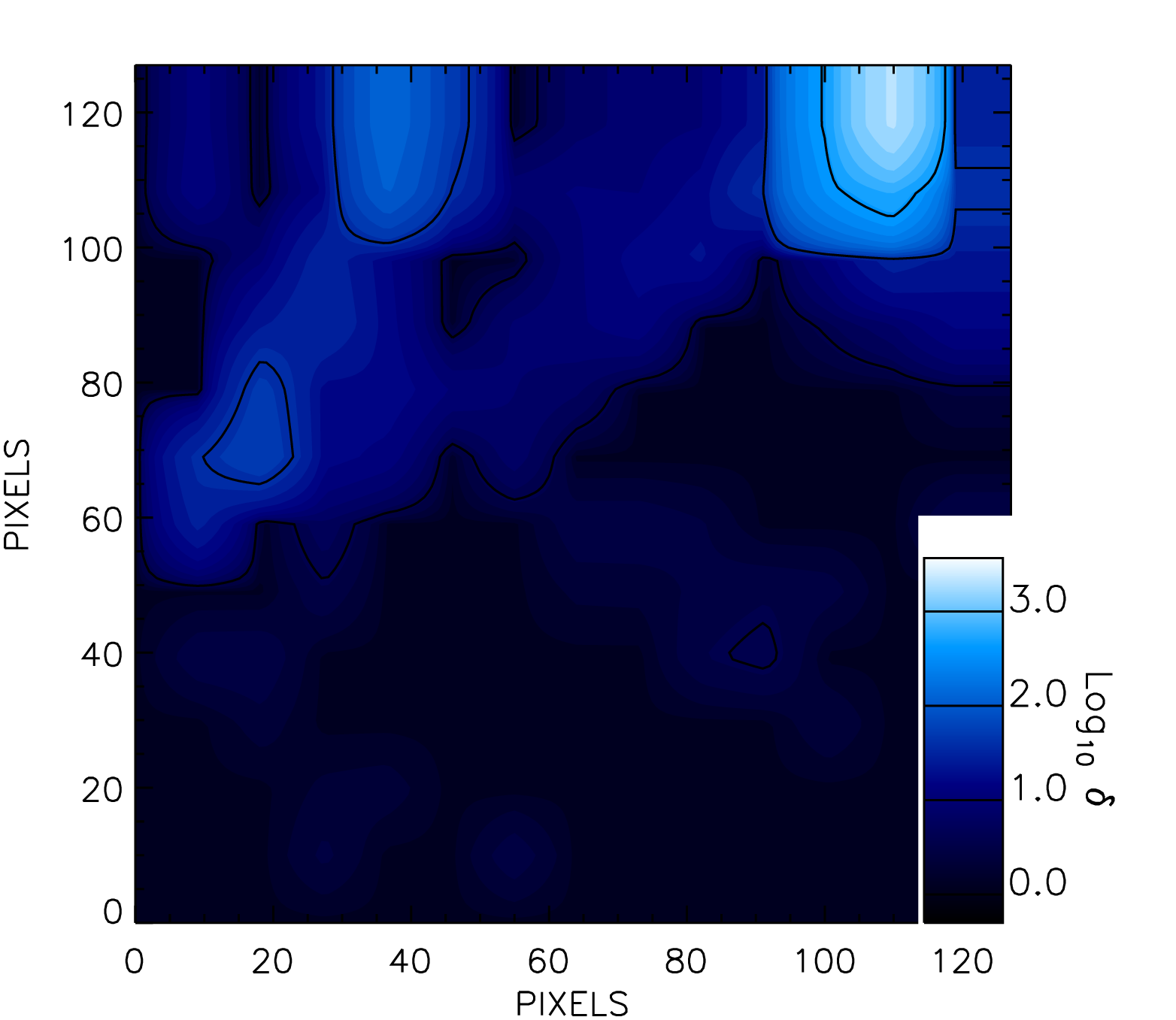}\\
\caption[]{\small
 Metal-poor DLA formation. The sequence shows the formation path of a metal-poor DLA candidate which could be observed at redshift $z\simeq 2.995$.
 On the left column we show the particle distributions of this metal-poor DLA candidate at different cosmological times ($z = 6.594$, 4.992, 4.186, 2.995, respectively from top to bottom) in comoving coordinates.
The right column depicts corresponding overdensity ($\delta$) distributions (within 10 comoving $\rm kpc/{\it h}$ thickness) in 128$\times$128 pixel maps marked by the three quartile contour levels.
Color ranges are adjusted to the scale of the individual maps to highlight the contrast for a better visualization.
}
\label{fig:DLAformation}
\end{figure}

Besides the general properties of the galaxy population at high and intermediate redshifts, it is interesting to focus on some particular case studies to investigate the formation of gaseous systems in more detail and also to shed light on the evolution properties of DLAs.
\\
Therefore, we identified a gas clump with a temperature $\lesssim 10^4\,\rm K$ and metallicity $\lesssim 10^{-2}\, Z_\odot$ at redshift $z\simeq 2.995$ and traced back the constituent gas particles to our higher redshift snapshots.
The resulting evolution is illustrated in Fig.~\ref{fig:DLAformation}.
On the left column we plot (comoving) positions at redshift $z=$6.594, 4.992, 4.186, 2.995 both for the clump particles (big red dots) and for the other particles (small green dots) found in the same plane of the center of mass of the clump.
In the right column we additionally show overdensity ($\delta = \rho/ \overline{\rho} $) maps with quartile contour levels, as extracted from a thin slice containing the reference gas clump.
Overall the condensing process of the initially diffuse slightly overdense gas is quite clear.
It proceeds from scales of some hundreds of comoving kpc (i.e. several tens of physical kpc) at $z\simeq 6.594$, cools and condenses into a denser object whose radius shrinks below a few physical kpc at $z\simeq 2.995$.
The increase of the gas overdensity is well highlighted by the color-coded pictures, which show a trend going from initial values of $\delta\sim 1-10$ up to $\delta > 10^2$ at $z=4.186$.
The correspondence between the red dots on the left panels and the peaks in $\delta$ on the right panels is striking and consistent with our previous description.
For example, the higher-redshift panels demonstrate only moderate density enhancements ($\delta\sim$ a few) locally accompanied by underdense or void regions in which the lack of material is visible already from the respective scatter plot.
At $z\lesssim 4$ the growth of the gas clump is in a quite advanced stage: $\delta$ reaches a few thousands and empty regions are found only in the outskirts (e.g. at $z=2.995$).
The entire evolution takes $\sim1.5$ Gyr, which is consistent with the cooling timescales expected from the thermal state of the gas.
The central densest regions would then proceed to collapse further to form stars and later suffer from destructive feedback or environmental processes.
\\
\begin{figure}
\centering
\includegraphics[width=0.23\textwidth]{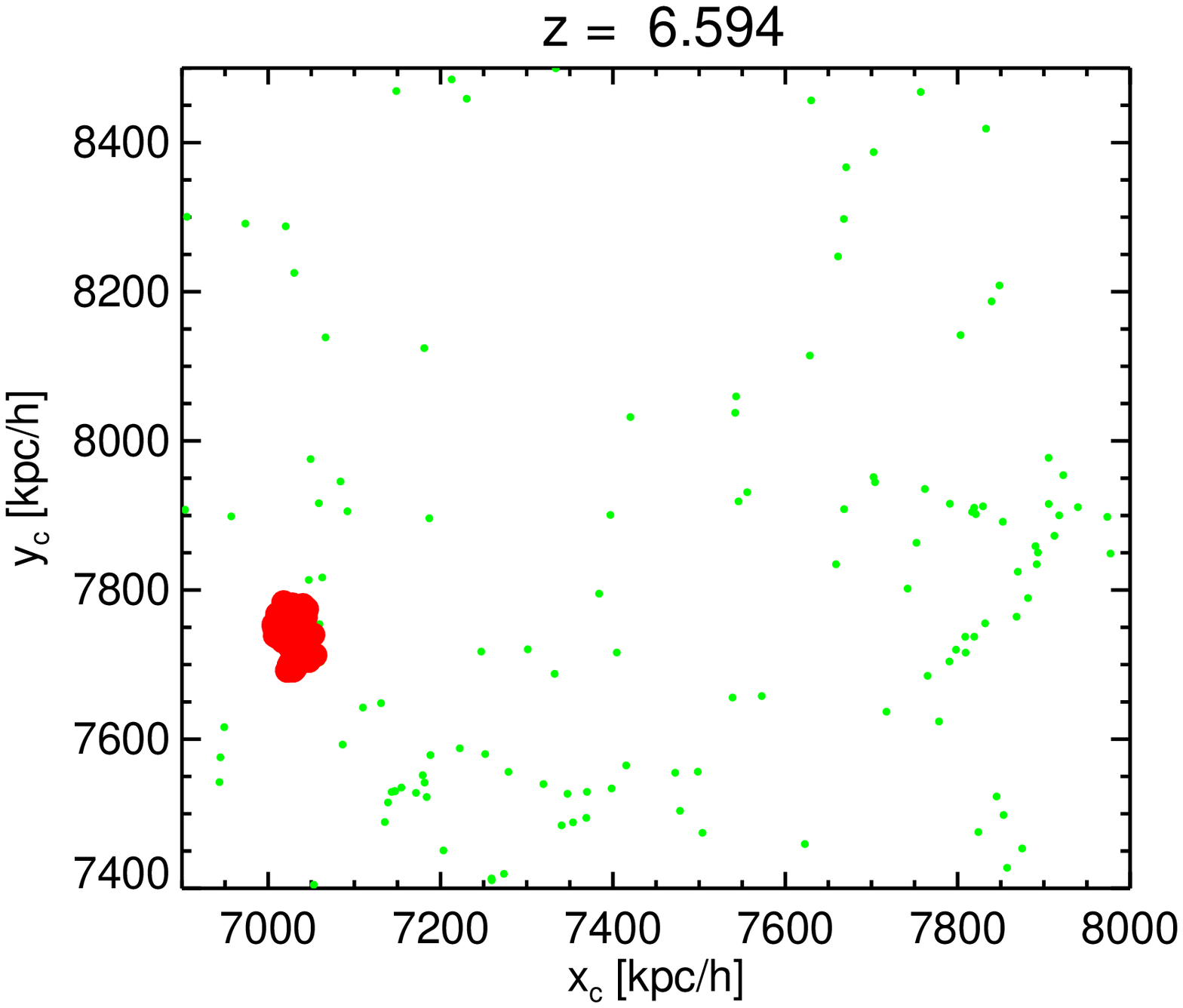}
\includegraphics[width=0.22\textwidth]{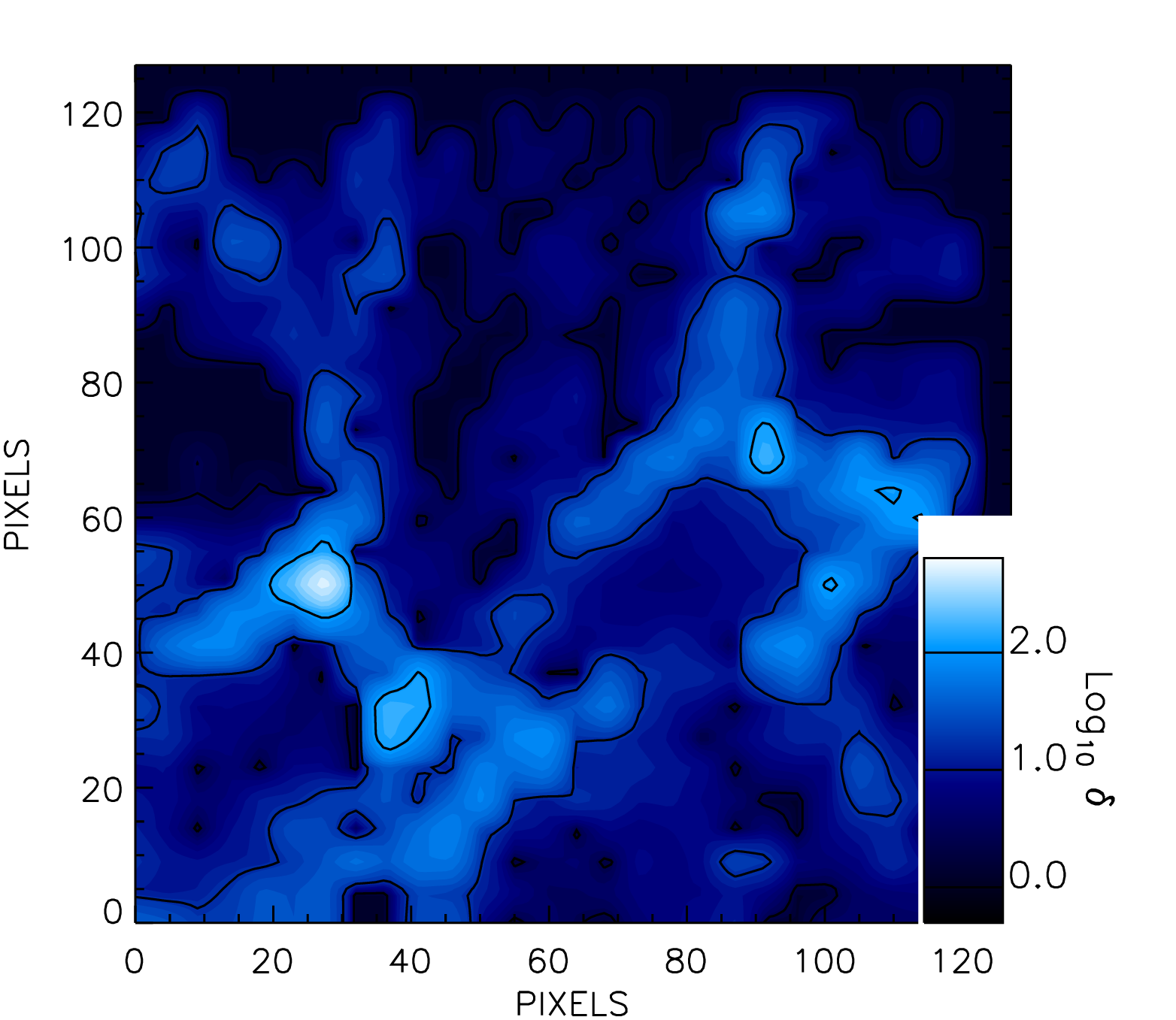}\\
\includegraphics[width=0.23\textwidth]{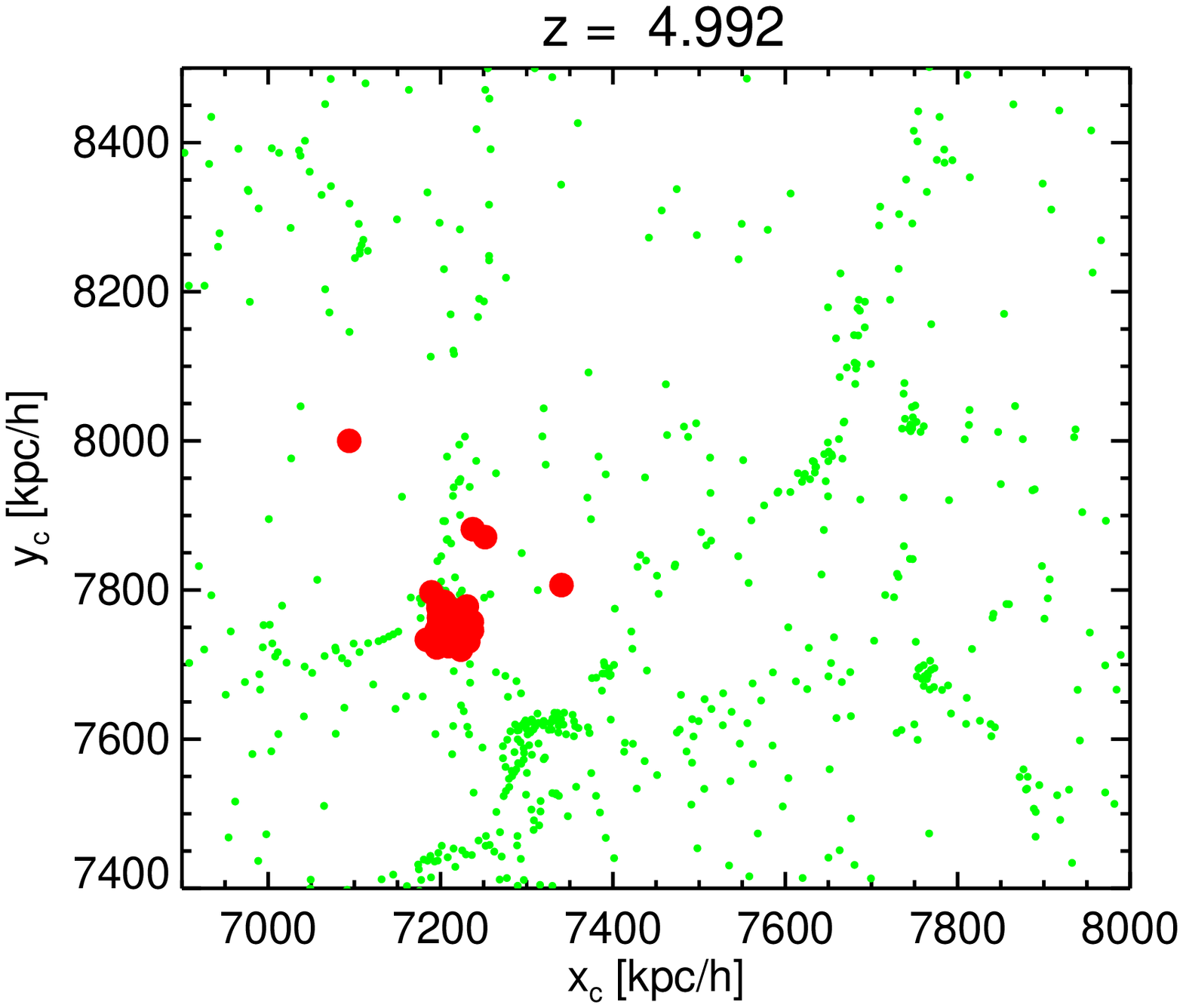}
\includegraphics[width=0.22\textwidth]{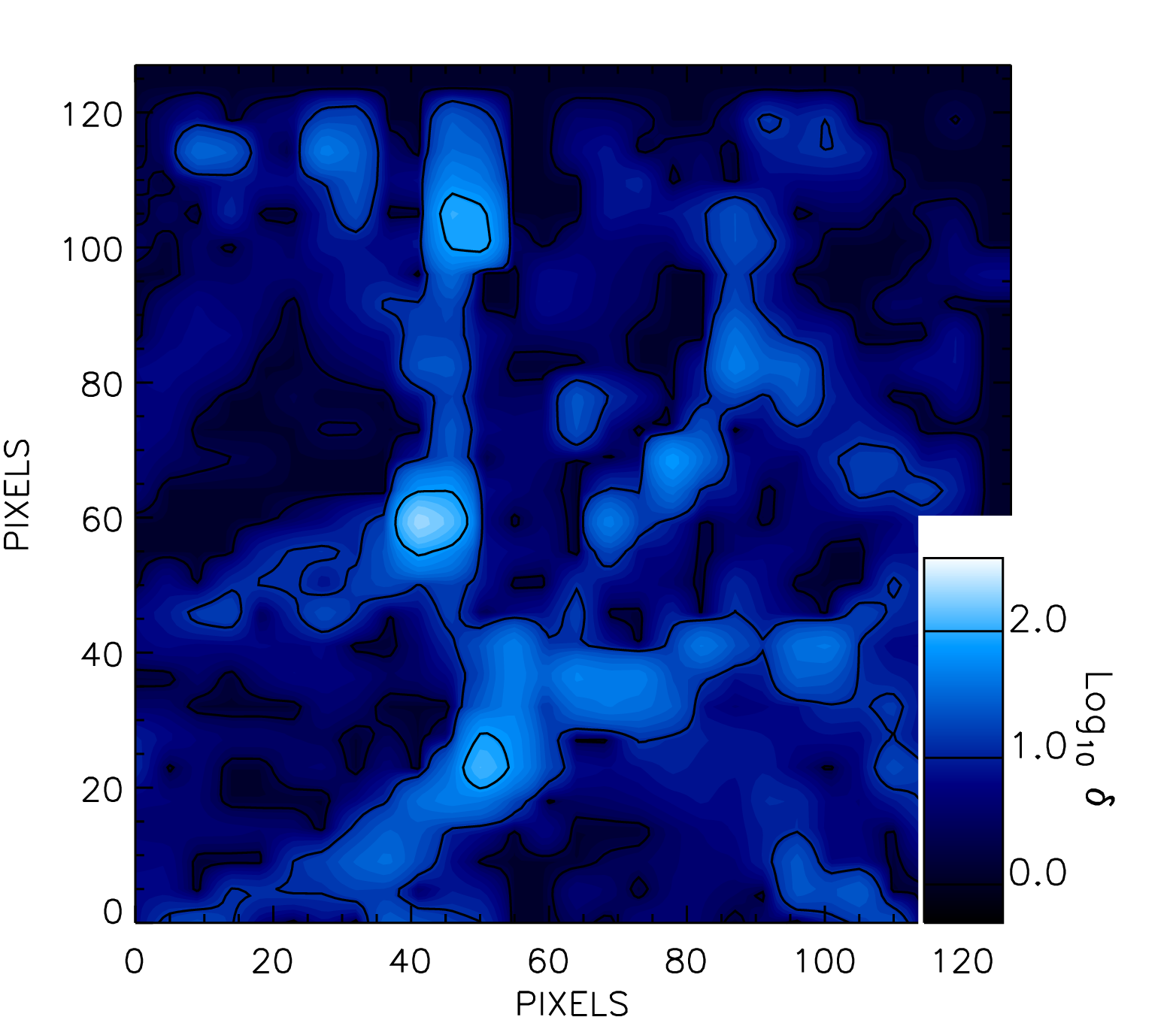}\\
\includegraphics[width=0.23\textwidth]{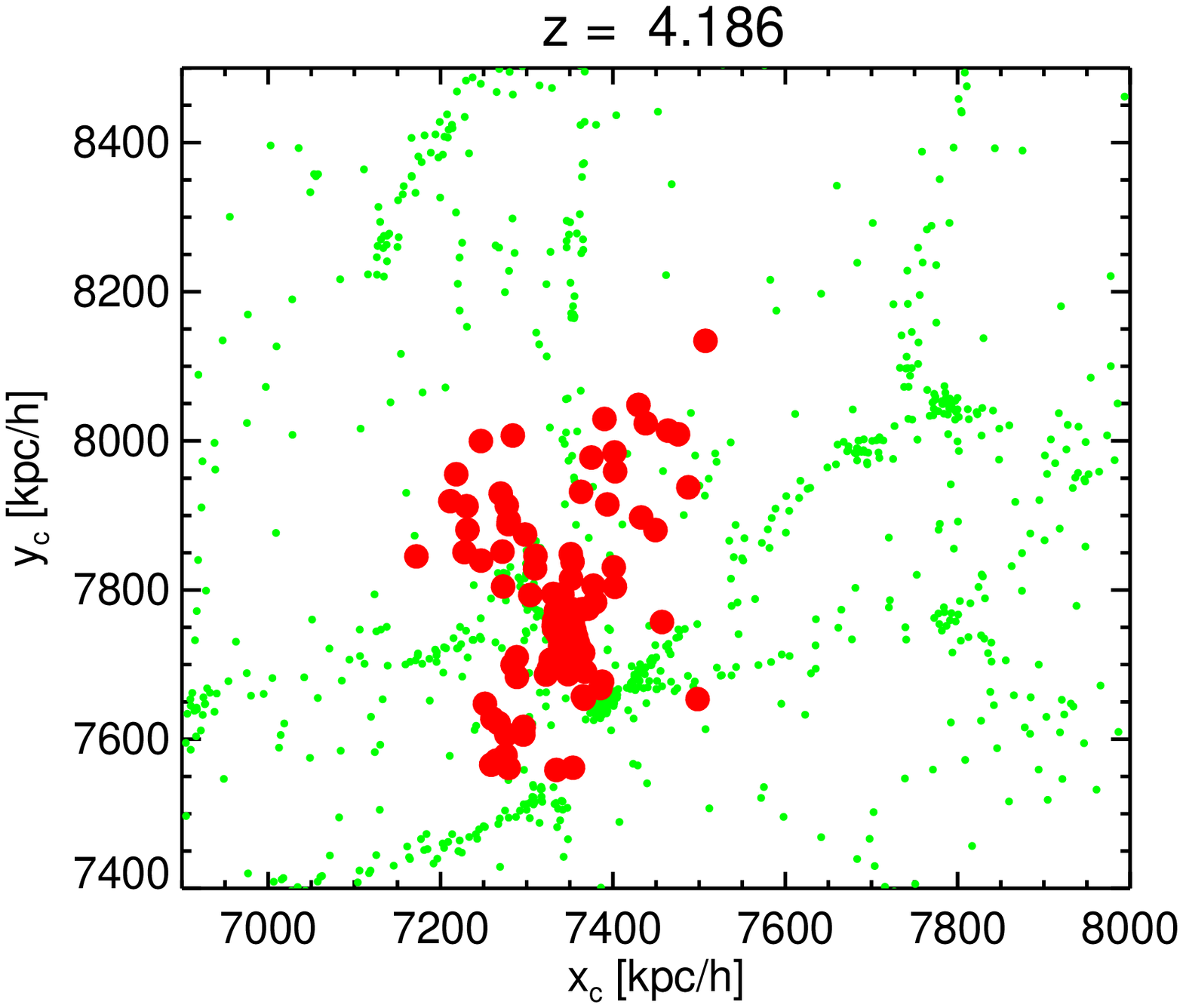}
\includegraphics[width=0.22\textwidth]{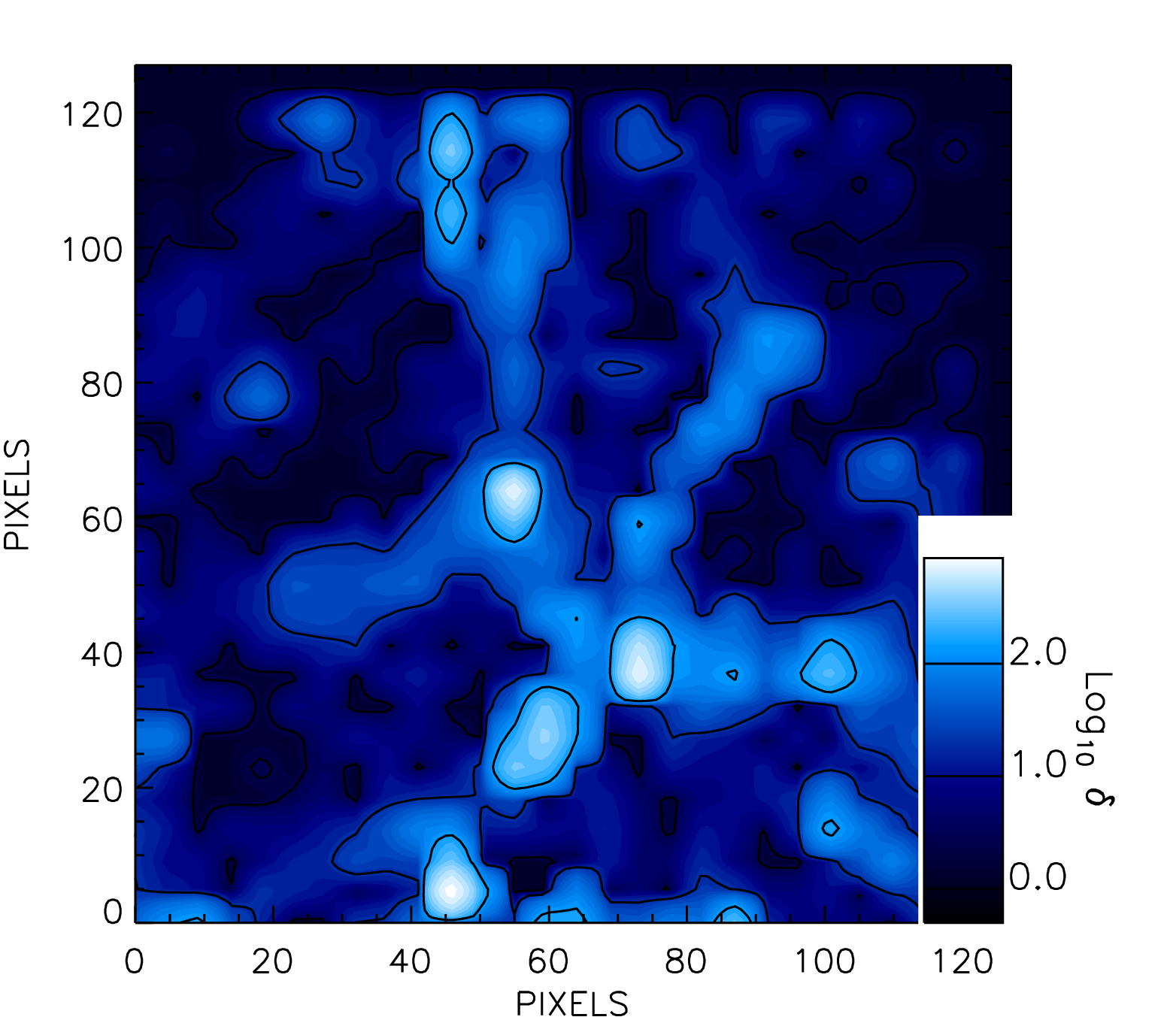}\\
\includegraphics[width=0.23\textwidth]{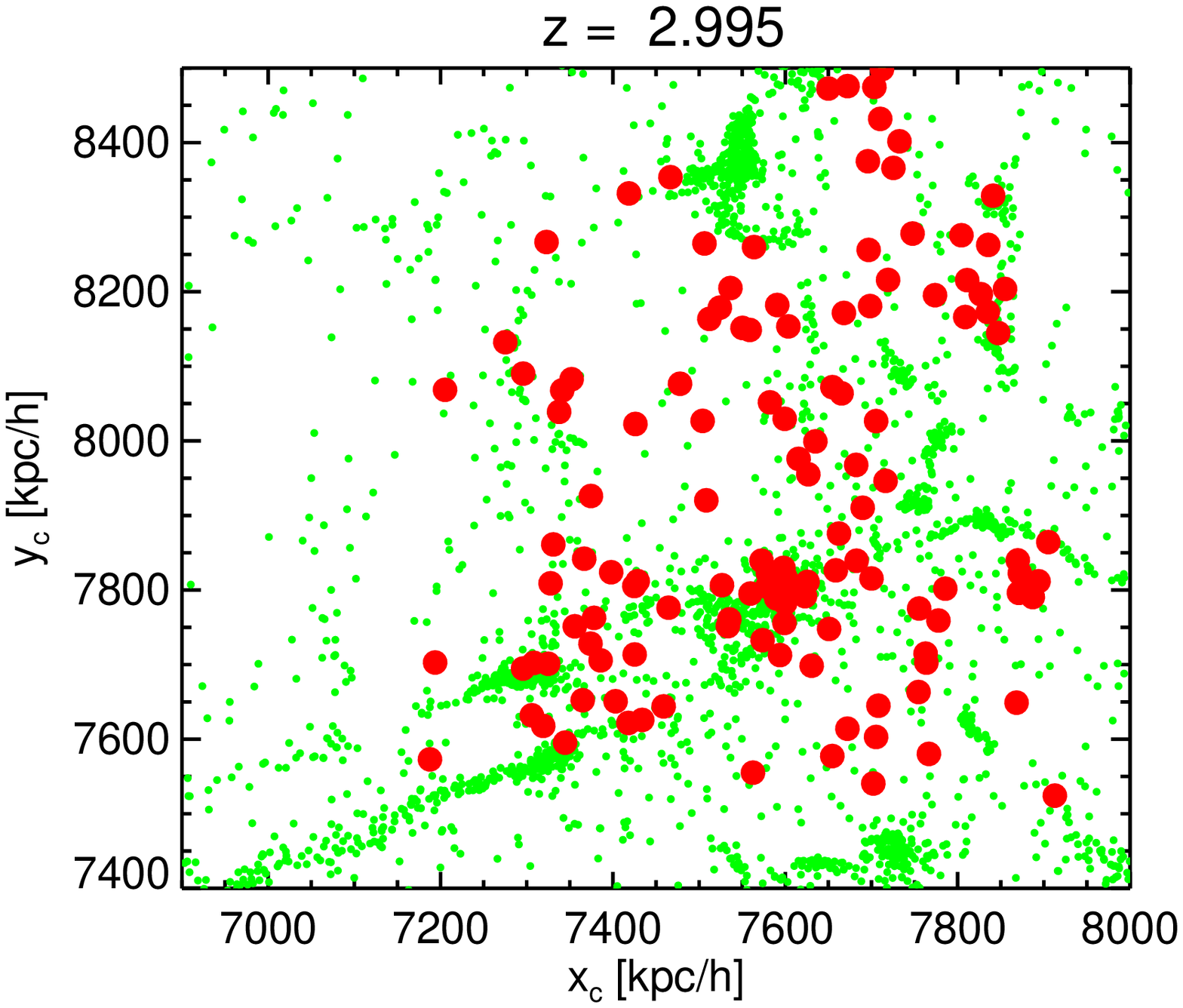}
\includegraphics[width=0.22\textwidth]{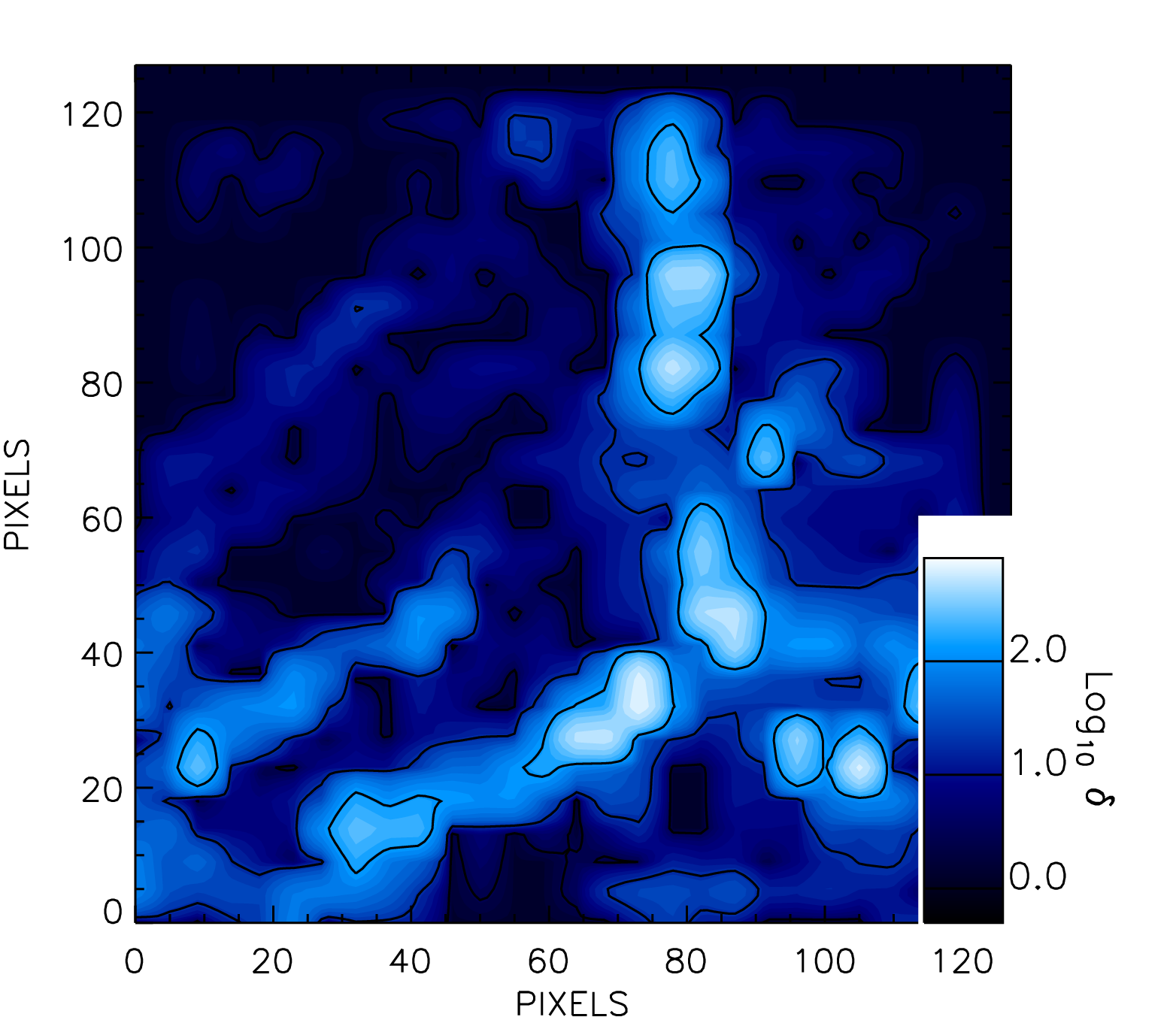}\\
\caption[]{\small
 Metal-poor DLA destruction. The sequence shows the evolution and final fate of a metal-poor DLA candidate which could be observed at redshift $z\simeq 6.594$. The left column shows the particle distribution of this metal-poor DLA candidate at different cosmological times ($z = $ 6.594, 4.992, 4.186, 2.995, respectively from top to bottom) in comoving coordinates. The right column depicts corresponding overdensity ($\delta$) distributions (within 10 comoving $\rm kpc/{\it h}$ thickness) in 128$\times$128 pixel maps marked by the three quartile contour levels.
}
\label{fig:DLAdestruction}
\end{figure}
This possibility can be verified by following the evolution of a pristine DLA which formed at higher redshift.
An example is provided in Fig.~\ref{fig:DLAdestruction}, where we show the fate of a cold damped system forming at $z\simeq 6.594$.
This structure hosts star formation at $z\sim 5$, and is significantly affected by stellar feedback processes at $z\sim 4$.
At $z\sim 2.995$ the initial system is almost completely disrupted by ongoing SN explosions and thermal heating, which spread the original cold gas far beyond the formation site.
For this case, the scales involved in the process range from some hundreds of comoving kpc (i.e. tens of physical kpc) at $z\simeq 2.995$ down to a few physical kpc at $z\simeq 6.594$.
The existence of such dense objects ($\delta > 10^2$) at these redshifts is a clear hint of primordial structure formation led by pristine molecular formation in the first Gyr.
Star formation feedback (e.g. SN heating, shocks), partly accompanied by environmental effects (gas stripping, interactions) at later times, plays a major role in shaping the gaseous systems.
\\
Evidence in support of this picture is provided by the panels in the second row ($z=4.992$), where typical $\delta$ values slightly decrease and gaseous material starts being ejected due to the first star formation episodes.
At later times, the amount of gas that is expelled is higher (see third row) and the density in the central region of the clump drops down to $\delta\sim 10^1-10^2$.
Simultaneously, structure formation in nearby sites sets in and further contributes disrupting the system by tidal interactions.
In the bottom panels, filamentary-like structures are recognizable, but the constituents of the initial clump (big red dots) are almost completely scattered around.
\\
These two examples suggest that the cold/warm structures seen as metal-poor DLAs can possibly be transient cosmological objects.
These would represent the final stages of gas collapse and fragmentation, captured when gas cooling is bringing temperatures to low values and locally the gas is condensing to high overdensities ($\delta \gtrsim 10^2$).
Their inner regions will eventually form stars and the associated feedback mechanisms will evacuate partially, if not completely, the system's gas reservoir.
\\
A more quantitative statement can be done by checking for the cooling capabilities of cosmic gas.
The cooling time, $t_{cool}$, is generally given by
\begin{equation}
t_{cool} = - \frac{E}{\dot E}
\end{equation}
where $E$ and $\dot E$ are the thermal energy and the energy loss rate.
This equation can be simplified into
\begin{equation}
t_{\rm cool} = \frac{3}{2} \frac{k_B T}{\Lambda~n_{\rm H}}
\end{equation}
where the prefactor $3/2$ holds strictly for mono-atomic gas, $k_B$ is the Boltzmann constant, $T$ is temperature, $\Lambda$ is the metal- and temperature-dependent cooling function, $n_H$ is the H number density.
For typical values in cgs units we can write
\begin{equation}
t_{\rm cool}
\simeq
2 \times 10^{10}~\frac{(T/10^4)}{ (\Lambda/10^{-22})~n_{\rm H}}~\rm s
\sim  10^3~\frac{(T/10^4)}{ (\Lambda/10^{-22})~n_{\rm H}}~\rm yr
\end{equation}
and easily get some useful estimates.
Given that the cooling times for diffuse gas ($\sim 10^{-6}\,\rm cm^{-3}$) are of order $\sim 1\,\rm Gyr$, much longer than catastrophic molecular-driven runaway cooling ($\ll 1\,\rm Gyr$), it is most likely to observe a metal-poor DLA prior to the onset of star formation (Fig.~\ref{fig:DLAformation}), before the effects of stellar feedback can heat or evacuate the hosted material (Fig.~\ref{fig:DLAdestruction}).

\subsection{Chemical patterns}

The most reliable probe of cosmic chemical evolution is the study of abundance ratios for different chemical elements.
Abundance ratios depend primarily on the stellar yields, timescales and IMF\footnote{
  Studies by \cite{ConroyVanDokkum2012} \cite[but see also][for different interpretations]{LaBarbera2015} suggest further IMF dependencies on [Mg/Fe], which is a proxy for $\alpha$ elements.
},
whereas the absolute gas-phase metallicity depends on the conditions of the environment, the star formation history, as well as feedback mechanisms.
We will test chemical evolution during the first Gyrs by means of molecular fractions and several important abundance ratios for the most common heavy elements.
In particular, among the tracked species, we will consider H$_2$ molecules and the main species produced by high- and low-mass stars, such as C, N, O, Si and Fe, for which reliable observational data exist.
Our reference indicators are [O/H] and [Fe/H], which are extremely useful to trace different stellar progenitors, since O (and the $\alpha$ elements) are mainly produced by short-lived massive stars, while Fe is more abundantly produced at later times by long-lived low-mass stars.
\\
Our findings are summarised in Figs.~
\ref{fig:DLAchemistry1},
\ref{fig:DLAchemistry2},
\ref{fig:DLAchemistry3} and
\ref{fig:DLAchemistry4},
where we plot the gas mass, $M$, average molecular fraction, $x_{mol}$, metallicity over H, $\rm Z/H$, and a number of element ratios for all of the simulated galaxies (empty diamond).
We highlight the metal-poor DLA host candidates with red asterisks.
As a comparison, we overplot observational measurements with error bars or experimental limits, as described in Sect.~\ref{Sect:data}.
We note that, since we are at relatively high or moderate redshifts ($z\gtrsim
3$), stellar evolution and metal enrichment are still in their infancy (the corresponding cosmic time is $\lesssim 2\,\rm Gyr$).
\\
Here we will largely focus on the theoretical gaseous chemical patterns at such epochs.
Previous results have shown that typical $Z$ values in the first Gyrs are distributed in a range of log$_{10}~(Z/Z_{\odot})$ between $-1$ and $-3.5$.
Rare outliers at lower $Z$ can be present although with probabilities below $0.1$ per cent.
We refer the interested reader to our previous works \cite[e.g.][]{Campisi2011, Salvaterra2013, Dayal2013, MaioViel2014, Ma2015} for more detailed analyses about probability distribution functions of metallicities at different redshifts.

\subsubsection{Molecular and metal content}
\begin{figure*}
{$\boldsymbol{z=6.594}$}\\
\includegraphics[width=\textwidth]{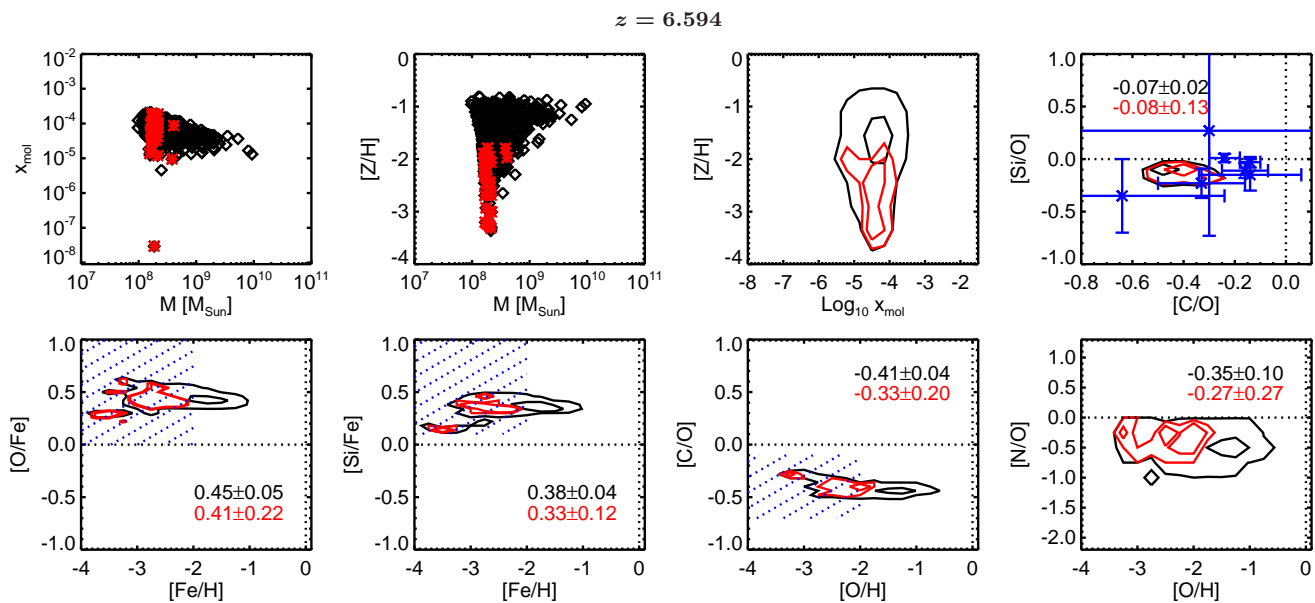}\\
\caption[]{\small
  Chemical abundances for the entire simulated population (black diamonds) and that hosting metal-poor DLAs (red asterisks) at redshift $z\simeq 6.594$.
The first row displays molecular (first panel) and metal content (second panel) as a function of gas mass, metallicity vs molecular fraction (third panel) and [Si/O] vs. [C/O] ratios with data points and error bars (fourth panel) from \cite{Becker2012}.
The second row collects important element ratios, such as 
[O/Fe] vs. [Fe/H] (first panel),
[Si/Fe] vs. [Fe/H] (second panel),
[C/O] vs. [O/H] (third panel),
[N/O] vs. [O/H] (fourth panel),
with observational data taken from \cite{Becker2012} (see text).
Contour levels refer to 1$\sigma$ and 3$\sigma$ confidence.
Shaded areas in the bottom row refer to the range allowed by observational constraints.
Reference solar data are taken from \cite{Asplund2009}.
}
\label{fig:DLAchemistry1}
\end{figure*}
\begin{figure*}
{$\boldsymbol{z=4.992}$}\\
\includegraphics[width=\textwidth]{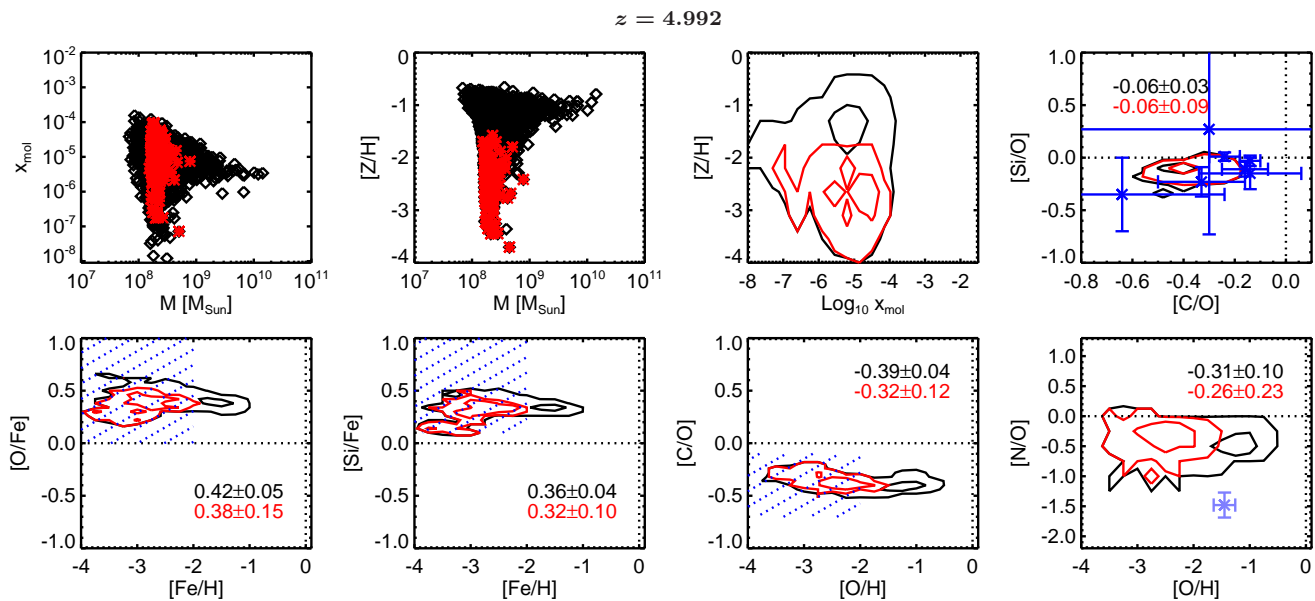}\\
\caption[]{\small
Same as Fig.~\ref{fig:DLAchemistry1}. Simulation snapshot refers to $z\simeq 4.992$.
[N/O] data point (light blue) is from \cite{Dessauges2001}.
}
\label{fig:DLAchemistry2}
\end{figure*}
\begin{figure*}
{$\boldsymbol{z= 4.186}$}\\
\includegraphics[width=\textwidth]{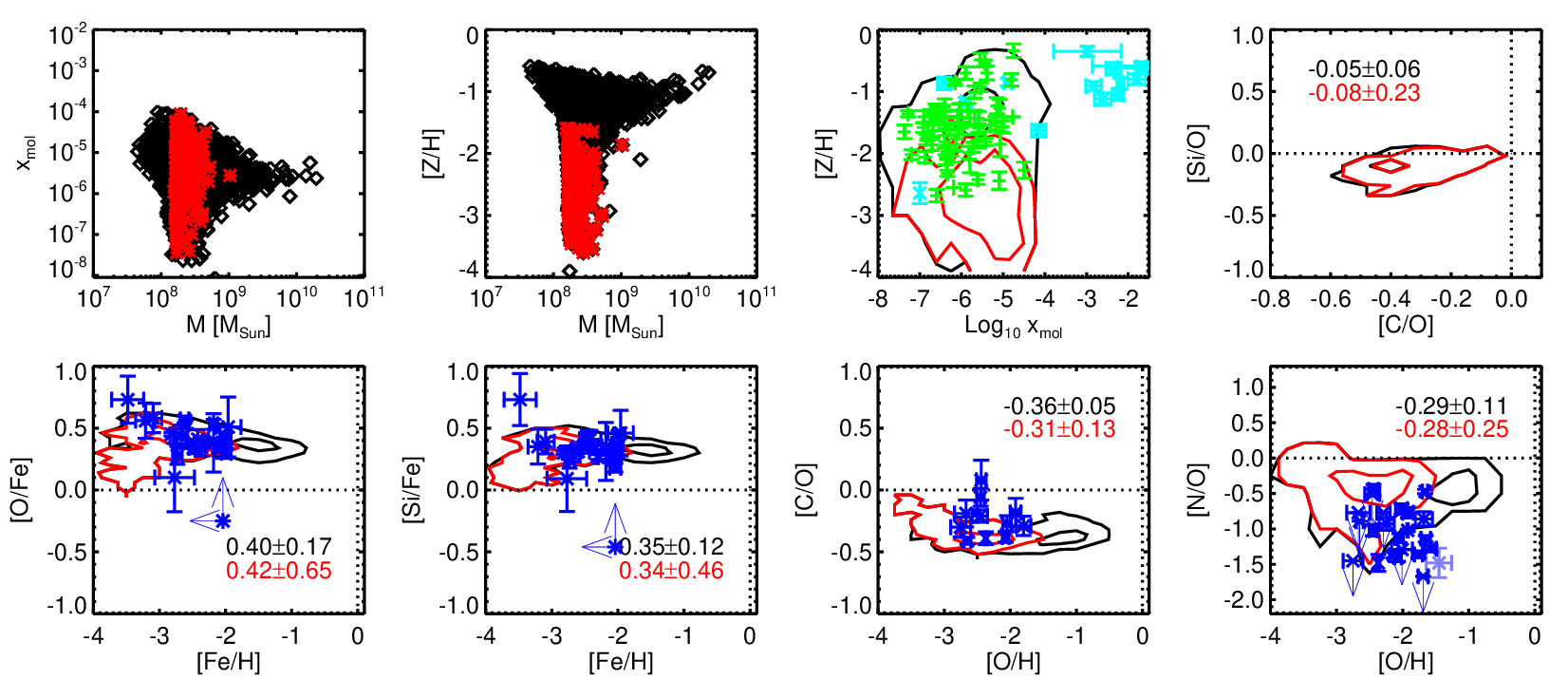}\\
\caption[]{\small
Same as Fig.~\ref{fig:DLAchemistry1}, except the simulation snapshot is taken at redshift $z\simeq 4.186$.
Data points for the metal ratios (blue) are taken from \cite{Cooke2015}.
[N/O] data point (light blue) is from \cite{Dessauges2001}.
Data points for molecular fractions (cyan) are taken from \cite{Noterdaeme2008} ($1.8 \lesssim z \lesssim 4.2$), \cite{Srianand2010} ($x_{\rm mol} \simeq 10^{-7}$ at $z\gtrsim 3$) and \cite{Albornoz2014} ($x_{\rm mol}\simeq 10^{-2.2}$ at $z \simeq 2.66$) and refer to the coldest molecular components within the hosting DLA.
Upper limits for H$_2$ fractions (green) are taken from \cite{Noterdaeme2008} ($1.8 \lesssim z \lesssim 4.2$).
}
\label{fig:DLAchemistry3}
\end{figure*}
\begin{figure*}
{$\boldsymbol{z=2.995}$}\\
\includegraphics[width=\textwidth]{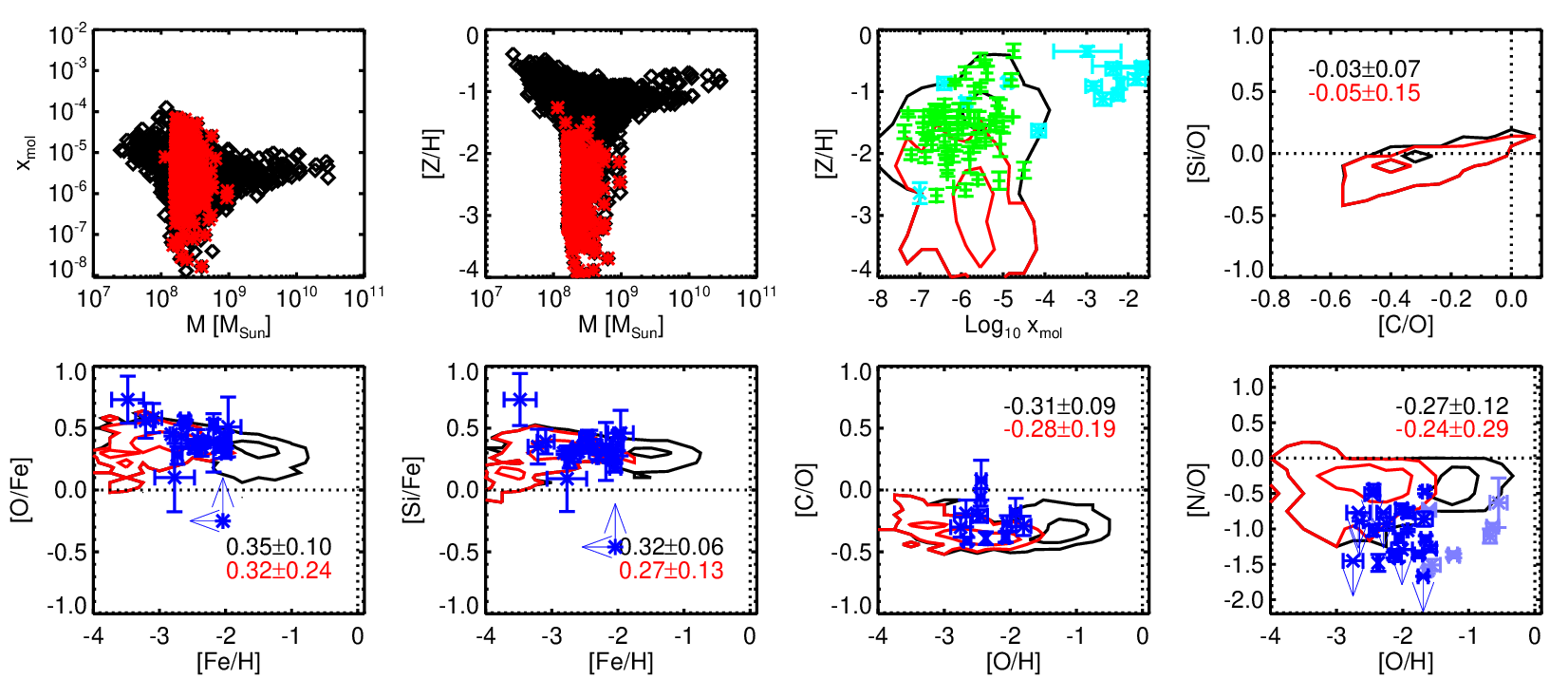}\\
\caption[]{\small
Same as Fig.~\ref{fig:DLAchemistry1}, except the simulation snapshot is taken at redshift $z\simeq 2.995$.
Data points for the metal ratios (blue) are taken from \cite{Cooke2015}, while data points for [N/O] ratios (light blue) are from \cite{Zafar2014}.
Data points for molecular fractions are the same as previous Fig.~\ref{fig:DLAchemistry3}.
}
\label{fig:DLAchemistry4}
\end{figure*}
The first two panels of Figs.~\ref{fig:DLAchemistry1}-\ref{fig:DLAchemistry4} show the metallicity and molecular content in our simulated galaxies for different gas masses (as in Fig.~\ref{fig:sample}), while the third panel displays the $x_{mol} - \rm [Z/H]$ trend.
\\
Mean molecular abundances at $z=6.594$ for the simulated galaxies span roughly two orders of magnitude, ranging between $\sim 10^{-5.5}$ and $\sim 10^{-3.5}$, while metallicities are between $\rm [Z/H]\sim -3.5 $ and $\rm [Z/H]\sim -1$ (Fig.~\ref{fig:DLAchemistry1}).
At later times the molecular content ranges between $ 10^{-8}\lesssim x_{mol} \lesssim 10^{-3} $ and the metal content between  $\rm -4\lesssim [Z/H] \lesssim -0.5$ (Fig.~\ref{fig:DLAchemistry4}).
The behaviour of chemical content with gas mass highlights the capability of larger objects to retain more metals and host molecule reformation thanks to the deeper matter potential wells.
Lower-mass structures, instead, are more strongly affected by star formation and feedback events and their gas gets easily heated by nearby star forming regions.
Consequently, molecules are dissociated and the resulting fractions cover a broad range around $x_{mol}\sim 10^{-8}-10^{-4}$ (Fig.~\ref{fig:DLAchemistry2}-\ref{fig:DLAchemistry4}) for gas masses below $\sim 10^{9}\,\rm M_\odot$.
A similar evolution in low-mass objects is recognizable for [Z/H] as well, while in the massive end [Z/H] tends to values of $\sim -1$ (Fig.~\ref{fig:DLAchemistry1}-\ref{fig:DLAchemistry2}) or even higher (Fig.~\ref{fig:DLAchemistry3}-\ref{fig:DLAchemistry4}).
High metallicities are additionally found in dense clustered regions, quite independently from the gas mass, where star formation takes place efficiently (thanks to the locally high molecular content) and spreads metals in the surroundings through an inside-out mode.
\\
As a consequence, objects with large $x_{mol}$ and high metal content (upper region in the $x_{mol} - \rm [Z/H]$ panel) are those that are hosting star formation and have already experienced heavy-element pollution, while objects with large $x_{mol}$ and small $Z$ values (lower region in the $x_{mol} - \rm [Z/H]$ panel) are those that are potentially star forming, but the onset of star formation has been quite recent, hence metal spreading has not yet been able to enrich the medium significantly.
We note that $x_{mol} \simeq 10^{-6}$ is roughly the amount of primordial H$_2$ that forms soon after the decoupling epoch (i.e. an approximate indicator of primordial  $x_{mol}$), therefore higher values actually testify the collapse of gas in the dark-matter potential wells that boosts H$_2$ and HD production.
On the other hand, values lower than $\sim 10^{-6}$ reflect environmental effects that determined molecule dissociation around $\sim 10^4\,\rm K$ (as clearly visible at lower redshifts).
\\
These considerations are useful to understand the features of molecular and metal content at later times, when star formation and feedback effects become progressively more important.
They are responsible for increasing the scatter of $x_{mol}$ and $Z$ values between $z=4.992$ and $z=2.995$ by several orders of magnitude, due to ongoing stellar evolution and feedback mechanisms that continue to enrich the surrounding medium and spread heavy elements over increasingly larger volumes.
As a result of the mechanical and thermal feedback from SN explosions, molecular fractions drop down to $\sim 10^{-8}$.
From the  $x_{mol} - \rm [Z/H]$ panels in Figs.~\ref{fig:DLAchemistry1}--\ref{fig:DLAchemistry2}, it is clear that DLAs with low molecular content and low metallicities are usually severely affected.
Such structures are likely to be externally polluted by nearby galaxies and suffer molecular dissociation from the UV background (compare the third panel of Fig.~\ref{fig:DLAchemistry1} and the one of Fig.~\ref{fig:DLAchemistry4}).
Available data points at $z=4.186$ and at $z=2.995$ show agreement for the $x_{mol} - \rm [Z/H]$ distribution, with the caveat that observed values at $x_{mol}\sim 10^{-3}-10^{-2}$ refer to data that probe the cold neutral medium of high-redshift galaxies (with temperatures $\sim 80\,\rm K$) and do not refer to the average content of the whole hosting structures, for which the molecular fraction would be significantly lower.
The range of values are nevertheless in broad agreement.
In fact, most of the DLA gas is diffuse and warm ($T \gtrsim 3000\,\rm K$), consequently, H$_2$ is scarce and detected only in $\sim 10$ per cent of the cases \cite[e.g.][]{Albornoz2014}.
Additional upper limits for $x_{mol}$ with available metallicity estimates \cite[][]{Noterdaeme2008} shown in Figs.~\ref{fig:DLAchemistry3}--\ref{fig:DLAchemistry4} lie at $x_{mol}<10^{-4}$ and are in agreement with the simulations.
We note that high-$z$ observational samples \cite[e.g.][]{Rafelski2012,Rafelski2014, Jorgenson2013, Jorgenson2014} usually consist of objects with $\rm [Z/H]\gtrsim -3$.
This floor is quite interesting, because when compared to stellar metallicities it is found that values inferred from individual stars do not feature such a plateau and attain values lower than $ -4$ \cite[e.g.][]{Cayrel2004, Caffau2011, Roederer2014}.
Considering that it is quite difficult to measure signals from low-metallicity gas, the $x_{mol} - \rm [Z/H]$ panels suggest a non-zero, albeit small (below $\sim 0.1$~per cent), probability of finding them as a result of the metal enrichment process from star forming regions towards the outskirts of galaxies and beyond.

\subsubsection{Abundance ratios}

We continue our discussion on metal enrichment from stellar evolution by showing selected relevant abundance ratios as functions of both [Fe/H] and [O/H].
In fact, [Fe/H] and [O/H] can be used to track the temporal evolution from early times, when massive SN~II progenitors eject a lot of O, to later times, when long-lived low-mass stars eject large amounts of Fe.
This means that going from low to high [Fe/H] is like following stellar evolution from early to later stages or, equivalently, from short-lived larger stellar masses to long-lived smaller stellar masses.
We cross-check our results by looking at
[O/Fe] vs. [Fe/H],
[Si/Fe] vs. [Fe/H],
[C/O] vs. [O/H],
[N/O] vs. [O/H]
and 
[Si/O] versus [C/O]
(see panels in Figs.~\ref{fig:DLAchemistry1}-\ref{fig:DLAchemistry4}).
For the various distributions we quote the mean and 1$\sigma$ dispersion in the corresponding legends.
We now describe our results on gas metallicities\footnote{
  In the following analysis we will refer to gas metallicities, but it should be noted that the majority of metal poor stars are substantially $\alpha$-enhanced and metallicities measured via $\alpha$ elements in DLAs and via Fe in stars are not totally consistent.}
and discuss more in depth their physical origin.
\\
In general, there is some redshift evolution in the various ratios with variable spreads due to contaminations from chemical feedback by different stellar phases.
This is quite remarkable, because the scatter in the ratios highlight the limits of the closed-box model and possible deviations from a single stellar population scenario or simple semi-analytical expectations.
\\
More in detail, at early regimes C, O and $\alpha$ elements are abundantly produced by massive stars and their trends can be constrained by higher-redshift ($z\gtrsim 5$) data on [Si/O] versus [C/O].
Values of [Si/O] are nearly solar already at $z = 6.594$ and lie within the error bars suggested by the observations \cite[][]{Becker2012}.
This is not very surprising, because $\alpha$ elements are produced in SN~II by the same nuclear process ($\alpha$ capture) starting from O nuclides, hence their abundance is usually proportional to the one of O.
As a result, mean [Si/O] values for the whole (metal-poor DLA) population evolve only from $-0.07$ ($-0.08$) at $z=6.594$ to $-0.03$ ($-0.05$) at $z= 2.995$ reaching $-0.06$ ($-0.06$) at $z=4.992$ and $-0.05$ ($-0.08$) at $z=4.186$.
The mean values for [C/O] instead increase from $\sim -0.4$ up to $\sim -0.3$, being more sensitive to the ongoing stellar phases -- see also the next discussion.
\\
We then check the behaviour of [O/Fe] as a function of [Fe/H], which represents a very important diagnostic of SN~II and SN~Ia evolution.
[O/Fe] predictions lie in a range around mean values of $\sim 0.4$ at $z=6.594$ and $z=4.992$ and they are led by short-lived massive SNe that eject large amounts of O.
At lower redshift a knee around [Fe/H]$\sim -2$ is recognizable and marks the transition from short-lived massive SNe~II to long-lived low-mass SNe~Ia that produce huge amounts of iron and cause a decline of [O/Fe] values at $z=4.186$ and $z=2.995$.
Simulated data are in agreement with data by \cite{Becker2012} and \cite{Cooke2015} at $z=6.594$ and $z=4.992$ (despite the lack of observational information on [Fe/H]) and with \cite{Cooke2015} at $z=4.186$ and $z=2.995$.
Also in this case we note a large spread of the simulated data which is a consequence of the mixing of different stellar phases due to ongoing star formation and feedback mechanisms.
\\
Similar considerations can be drawn about [Si/Fe] ratios, that demonstrate super-solar abundances and are driven by O evolution during their early phases and by Fe production at later times.
\\
The [C/O] ratios for the whole galaxy (metal-poor DLA) population present typical mean values that are slightly sub-solar, between $-0.41$ ($-0.33$) at $z=6.594$ and $-0.31$ ($-0.28$) at $z=2.995$ \cite[consistent with e.g.][]{Pettini2008, Fabbian2009, Becker2012} and feature a U-shaped trend with a minimum of $\sim -0.5$.
Hints of this trend can be already seen at $z=6.594$ and $z=4.992$, when primordial pollution episodes by massive SNe eject the first metals and pollute the medium.
At later times, [C/O] can reach the solar value (i.e. at $z=4.186$) or even overcome it ($z=2.995$).
In general, the relative increase of carbon abundance [C/O] is due to the metallicity-dependent carbon yields of massive stars and the delayed release of C from low- and intermediate-mass stars \cite[e.g.][]{Akerman2004, Cescutti2009}.
Therefore, due to early SN enrichment\footnote{
  Pristine or low-$Z$ SN~II have yields that account for values of [C/O] typically larger than the ones derived by high-metallicity SN~II.
},
[O/H] increases up to about $-2$ or $-1$ while [C/O] decreases down to $\sim -0.5$ (Fig.~\ref{fig:DLAchemistry1} and \ref{fig:DLAchemistry2}).
After a few $10^8\,\rm yr$, AGB stars expel further heavy elements and make the [C/O] trend at $\rm [O/H]>-2$ increase (Fig.~\ref{fig:DLAchemistry3} and \ref{fig:DLAchemistry4}).
At low metallicities ongoing explosions of massive SNe determine the increase of [C/O] ratios.
Observations agree with theoretical expectations quite independently from $z$ or from unknown [O/H] values.
\\
Uncertainties appear for N abundances, both theoretically and observationally.
Theoretical limitations are not due to major failures in the calculations, but to the specific input yield tables.
N evolution is in fact a very problematic issue for which there is still a lack of understanding and a missing agreement about its production processes.
These are due to non-trivial dependencies on several parameters of nucleosynthesis and stellar-structure models, such as metallicity, magnetic fields, rotation, winds, binary systems, etc. \cite[see][for further details]{Francois2004, Pettini2008, CescuttiChiappini2010, Cescutti2013, Zafar2014}.
Furthermore, most of the available data give only upper limits to [N/O] that qualitatively point towards an increasing (almost bimodal) spread at low metallicity.
Overall, the broad [N/O] values suggest that metal pollution derives from stars forming in quite different metallicity environments, from very metal-poor (as testified by low [N/O] ratios) up to very metal-rich ones (as in the cases with high [N/O]).
Considering that N is poorly produced by massive stars and is mostly expelled by intermediate-mass stars during their AGB\footnote{
  Model uncertainties for AGB yields are still large, though. See e.g. \cite{Karakas2010} for a comprehensive discussion.
}
phases it is possible that when intermediate- or low-mass star formation took place (at $z\lesssim 7$) the cosmic medium could be locally enriched up to solar levels with a spread in the ratios of a few orders of magnitude.
Such large dispersion in [N/O] ranges from solar values down to $\rm [N/O] \sim -1.5$ and is consistent with the delayed delivery of primary N into the surrounding medium with respect to O \cite[][]{Pettini2008, Zafar2014}.
In addition to the few data points available for high-$z$ meal-poor DLAs, several determinations of extra-galactic \HII\ regions \cite[e.g.][and references therein]{Izotov2004, vanZee2006, Pettini2008} give [N/O] ratios between $\sim -2$ and $\sim 0$ and feature a decreasing trend from values in the interval $\sim [-1, 0]$ at $\rm [O/H]\sim 0$ down to $\rm [N/O]\sim -2$ at $\rm [O/H]\sim -1.5$.
At very low metallicities observations seem to suggest a sort of plateau for [N/O] (albeit very scattered), due to the primary production of N that basically traces O abundance \cite[][]{Pettini2008, Zafar2014}.
Given the observational difficulties and the current theoretical limitations, more studies and improved yield calculations are needed in this respect in order to reach a general consensus.
\\
In conclusion, we notice that theoretical ratios are generally within the observational range and, despite the relatively large error bars of the available data, this is extremely encouraging.

\subsubsection{Cosmic chemical evolution} \label{Sect:cosmic-chemical-evolution}

\begin{figure}
\centering
\includegraphics[width=0.47\textwidth]{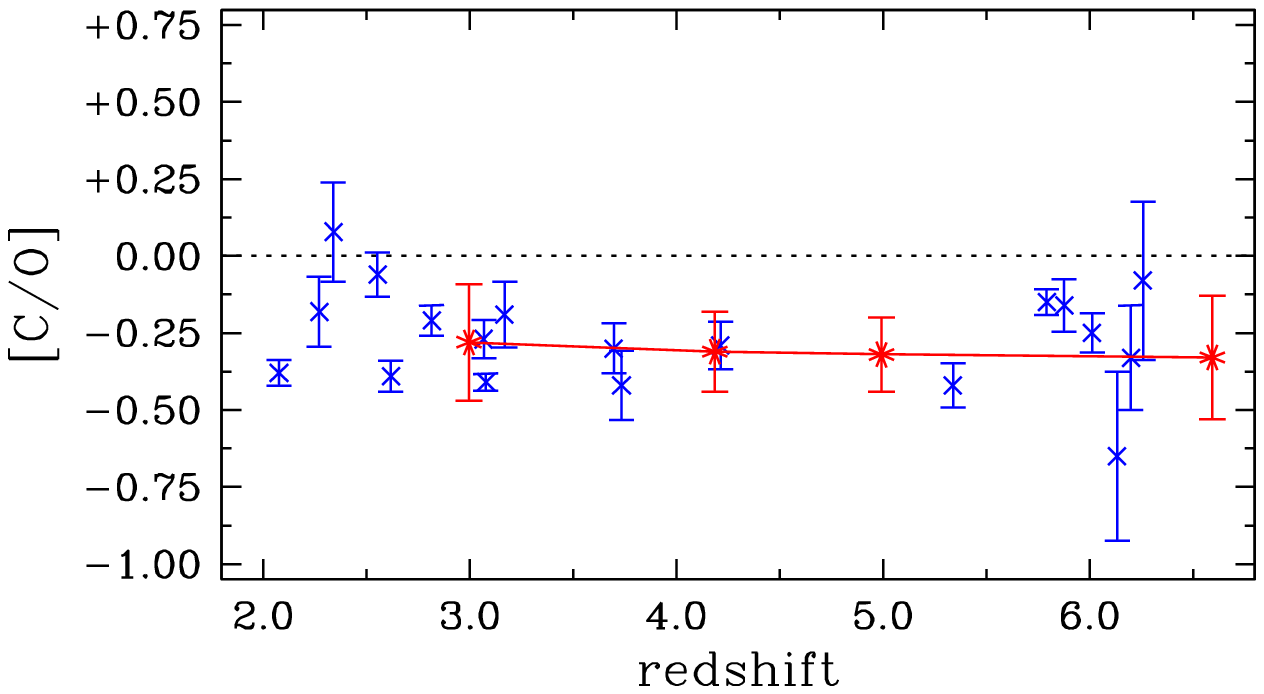}\\
\includegraphics[width=0.47\textwidth]{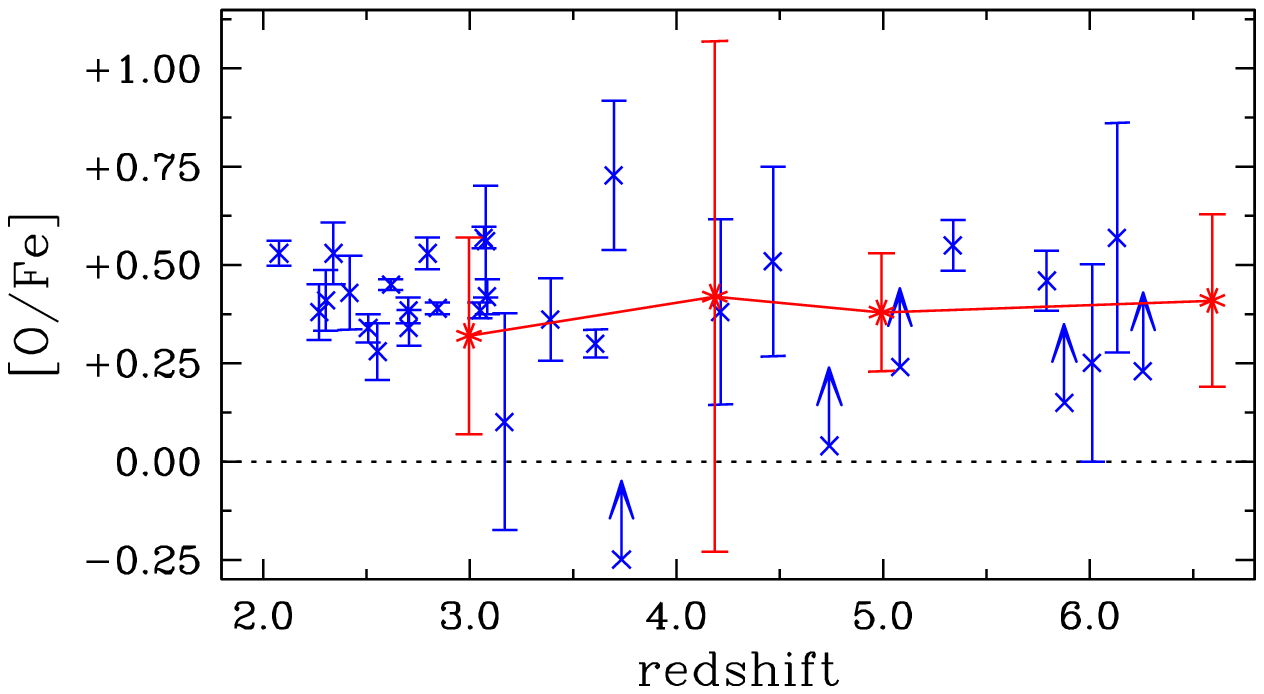}\\
\includegraphics[width=0.47\textwidth]{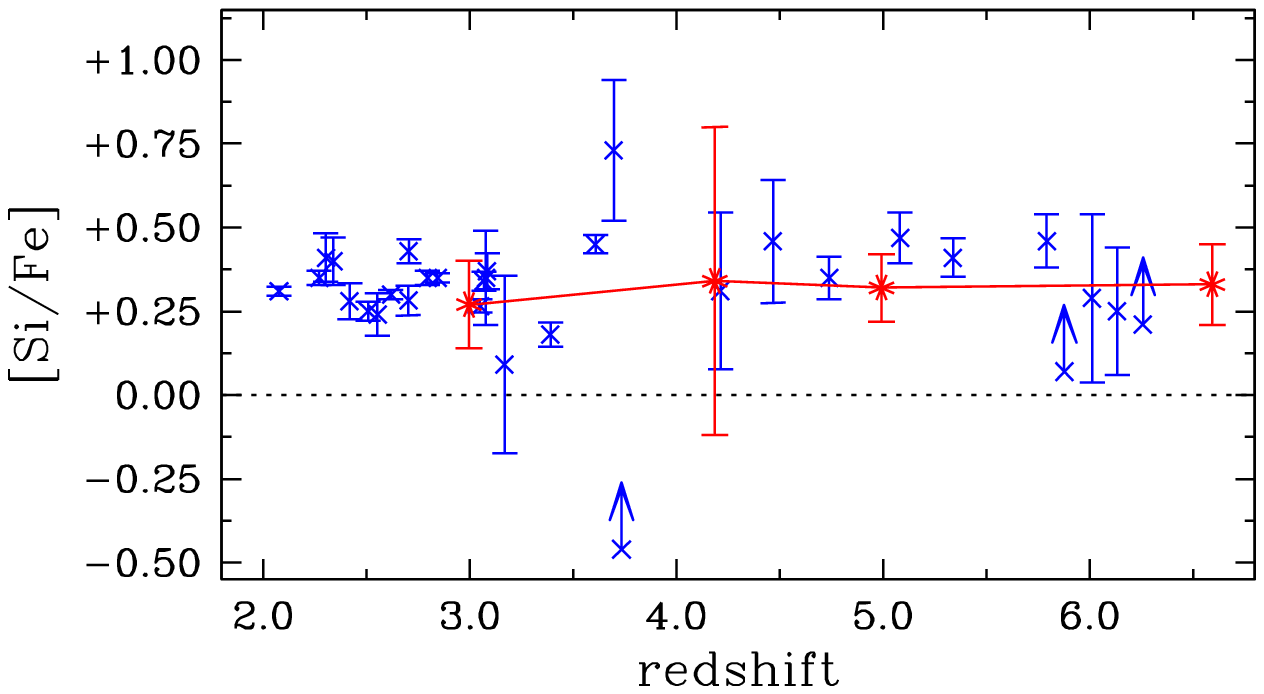}\\
\caption[]{\small
  Observational determinations (blue crosses) of abundance ratios at different redshifts for [C/O] (upper panel), [O/Fe] (central panel) and [Si/Fe] (lower panel) with 1$\sigma$ error bars compared to our theoretical predictions (red solid line with asterisks) with corresponding 1$\sigma$ dispersion.
}
\label{fig:abundances_redshift}
\end{figure}
We now explore the cosmological evolution of metal enrichment by comparing the elemental ratios of the DLA population at different redshifts.
We note that, at the epochs considered here ($z\simeq 2 - 7 $), the Universe is already a-few-Gyr old and hosts numerous SN~II explosions and AGB pollution events.
Long-lived SN~Ia are still rare at $z\sim 5-6$, but become progressively important at later times.
To address the implications of different stellar evolutionary phases we consider some indicative elements, such as C and O for SN~II and AGB, Fe for low-mass SN~Ia and Si as a proxy for $\alpha$-element production by massive stars.
\\
In Fig.~\ref{fig:abundances_redshift} we collect the expected values of [C/O] (upper panel), [O/Fe] (central panel), [Si/Fe] (lower panel) with 1$\sigma$ error bars for redshifts $z\gtrsim 2$ and compare these to the observational data points.
In general, our numerical calculations agree with observations, since the theoretical predictions are within the error bars or consistent with the experimental limits for all redshifts.
Values referring to the low-temperature population and the whole galaxy population at different redshifts are quite similar (further details are in Sect.~\ref{app}).
\\
Theoretical [C/O] ratios take values of $-0.28$, $-0.31$, $-0.32$ and $-0.33$ for the metal-poor DLA population and $-0.31$, $-0.36$, $-0.39$ and $-0.41$ for the whole galaxy population, at redshift 2.995, 4.186, 4.992, 6.594, respectively.
These estimates are in agreement with SN~II yields for different initial stellar metallicities \cite[][]{WW1995, ChieffiLimongi2004, HegerWoosley2010} and with most of the data.
Interestingly, we see a spread in the simulated data that spans a factor of a few.
For the metal-poor DLA sample, the theoretical dispersions are $\sim 0.20$ at $z\simeq 6.594$ and $\sim 0.19$ dex at $z\simeq 2.995$, over a timescale of $\sim 1.5$ Gyr.
However, values of $0.13$ and $0.12$ are found at $z=4.186$ and $z=4.992$.
This spread is a sign of enrichment by multiple star formation episodes which take place in different environments and which pollute the medium through stars in different evolutionary stages and with different metallicities.
Indeed, the ratio between C and O can easily deviate from the single-stellar-population expectations for a number of reasons.
For example, enrichment by SN~II belonging to the same {\it simple stellar population} (i.e. born in a medium with similar conditions, with similar initial metallicities and at similar epochs) would feature a given [C/O] at any given time.
However, if SN progenitors were born simultaneously in media with {\it different initial metallicities} then the SN~II yields would differ as well and the resulting abundance ratios would change accordingly\footnote{
  Large deviations would be expected for several elemental ratios, such as [N/O], [Al/O], [Fe/O], etc..
}.
In the same spirit, if, c.~p., stars were {\it not born simultaneously} the various stellar evolution phases would proceed in an asynchronous way and SN~II explosions could pollute the surroundings when AGB stars are also spreading heavy elements.
Consequently, abundance ratios would deviate from the predictions for a simple stellar population, because of the presence of multiple populations.
Such deviations would be more important mostly at later times, when the evolved stars that coexist with short-lived stars become effective at polluting the surrounding medium.
In fact, although at $z\sim 2-4$ SN~II are still dominant\footnote{
  We remind the reader that at these epochs the cosmic star formation rate density reaches its peak values.
},
SN~Ia and AGB stars can contribute already at $z\lesssim 6$, when the cosmic time elapsed is around 1~Gyr or larger.
\\
Obviously, these considerations apply to all the elemental ratios and to [O/Fe] in particular.
Given the larger error bars, the [O/Fe] evolution is essentially flat with typical values around $0.4$ at all redshifts.
The galaxy population features [O/Fe] values of 0.35, 0.40, 0.42 and 0.45 at $z = $ 2.995, 4.186, 4.992, and 6.594, respectively, while the metal-poor DLA candidates have corresponding ratios of 0.32, 0.42, 0.38 and 0.41.
Such results are consistent with recent observational determinations of metal-poor DLAs \cite[][]{Cooke2015,Becker2012} and the error bars are consistent with the whole data sample at any $z$.
The slight decrement in [O/Fe] at $z\simeq 3-4$, when the cosmic time is $t(z)\simeq 2.1 - 1.5 \,\rm Gyr$, reflects the additional Fe contributions from low-mass stars that formed roughly $\sim 1\,\rm Gyr$ earlier, at $z\sim 6-10$, and that explode as SN~Ia enriching the medium at these very epochs.
The interplay with coexisting short-lived stars, just forming in recently collapsed regions, induces the large dispersion in the abundance ratio at $z \sim 3$.
The trend in the expected spread is not regular, but rather stochastic and ranges within a factor of $\sim 4$ so that the standard deviation, $\sigma$, results to be about $0.15$ dex at $z\simeq 4.992$ and $0.65$ dex at $z\simeq 4.186$.
\\
Finally, the whole galaxy population at $z = $ 2.995, 4.186, 4.992, and 6.594, is characterized by [Si/Fe] values of 0.32, 0.35, 0.36, 0.38, respectively.
There is very little difference from the metal-poor DLA population, whose corresponding [Si/Fe] ratios are 0.27, 0.34, 0.32, 0.33.
Similarly to [O/Fe] also [Si/Fe] features roughly constant values\footnote{
  Here we should keep in mind that a given metal-poor DLA might have had a different star formation history than a typical DLA. Similarly, a typical DLA may have had a different chemical evolution to the Milky Way. Based on studies of the local-group dwarf galaxies, the Milky Way and the dwarf spheroidals seem to deviate from this plateau in $\rm [\alpha/Fe] $ when $\rm [Fe/H] \sim -2 $ roughly. Thus, it is interesting that the red contours in the ratio distributions (Figs.~\ref{fig:DLAchemistry1}-\ref{fig:DLAchemistry4}) tend to move to lower values than the black points with decreasing redshift, which could be exactly this effect.
}
of $\sim 0.3$ \cite[consistently with e.g.][]{Vladilo2002, Wolfe2005} with a decrease at $z\simeq 3-4$ and a larger scatter at $z\simeq 4.186$.
In fact, standard deviations vary between $\sim 0.10$ ($z=4.992$) and $\sim 0.46$ ($z=4.186$).
Also in this case, as for [O/Fe], the drop at $z\sim 3-4$ is the result of SN~Ia enrichment that boosts Fe content with respect to O and $\alpha$ elements.
\\
We note that data in Fig.~\ref{fig:abundances_redshift} can be easily explained by metal spreading and feedback effects from regular stellar populations at $2\lesssim z \lesssim 7$ without invoking exotic popIII generations.
This is consistent with several studies in the literature predicting a rather small popIII contribution to cosmic star formation.
\\
We finish by stressing that evolution and abundances of different heavy elements are strictly linked to timescales and yields of different stellar phases, which also depend on the patchy distribution of metals in the original star forming sites and on the effects of feedback mechanisms (such as inflows and outflows).
This poses serious limits on homogeneous models or on the closed-box model and highlights the need of detailed 3D modelling, as we reported herein, to predict both the trend and the dispersion that are observed for elemental ratios at various redshifts.

\subsubsection{Direct comparisons}\label{app}
\begin{figure}
  \includegraphics[width=0.5\textwidth]{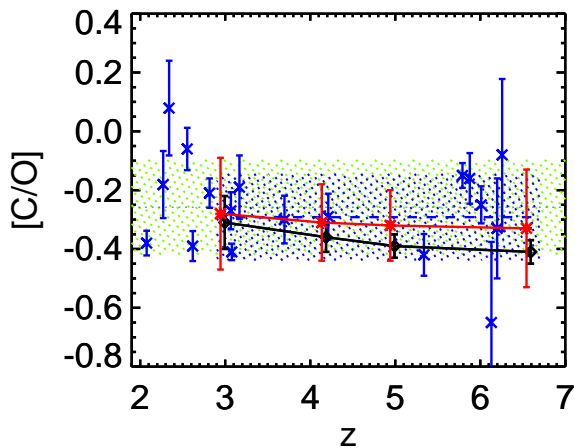}\\
  \vspace{-0.5cm}
  \caption[]{\small
    Mean abundance evolution for the whole simulated [C/O] sample (black solid line with diamonds) and for the metal-poor DLA candidates (red solid line with asterisks) compared to available observational data by \cite{Cooke2015} and \cite{Becker2012}. The blue dotted line marks the observational average over the whole redshift range with corresponding 1$\sigma$ dispersion (green shaded area). The blue dashed line marks the observational average referring to the redshift range $2.9\lesssim z \lesssim 6.7$, that is directly comparable to our theoretical expectations, with corresponding 1$\sigma$ dispersion marked by the blue shaded area.
  }
  \label{fig:direct}
\end{figure}
In the following, we briefly check the implications of outliers in a more direct comparison between observations and theory for the case of [C/O] ratios.
Fig.~\ref{fig:direct} shows mean abundance evolution for the whole simulated [C/O] sample (solid black line) and for the DLA candidates (solid red line) compared to available observational data.
The horizontal dotted line and the light-shaded area mark the mean value of the whole observational data sample and corresponding 1$\sigma$ deviation.
The horizontal blue dashed line and the blue-shaded area mark mean value and 1$\sigma$ deviation for data in the redshift range $2.9 \lesssim z\lesssim 6.7$.
This latter is more directly comparable to the theoretical expectations for our galaxy population (black solid line and diamond symbols) and for the metal-poor DLA sample (red solid line and asterisks).
The theoretical expectations for the whole galaxy population are typically below the ones referring to metal-poor DLAs, because these latter sample mainly the low-metallicity end, hence the resulting average ratios are a bit larger.
Slightly lower mean values are obtained when restricting the observation range from $z \gtrsim 2$ to $2.9\lesssim z\lesssim 6.7$.
This is due to the [C/O] values observed at $z\sim 2.4$ that increase the mean of $\sim 0.05 $ dex, i.e. about 12 per cent.
We note that all the data are compatible with metal spreading from regular population~II stellar generations.
The effects of SN~Ia going off at $z\lesssim 4-5$ should result in larger line widths as a consequence of the establishment of a more and more caothic environment.


\section{Observational signatures} \label{Sect:synt_qso}


%
\begin{figure*}
\centering
\includegraphics[scale=0.7]{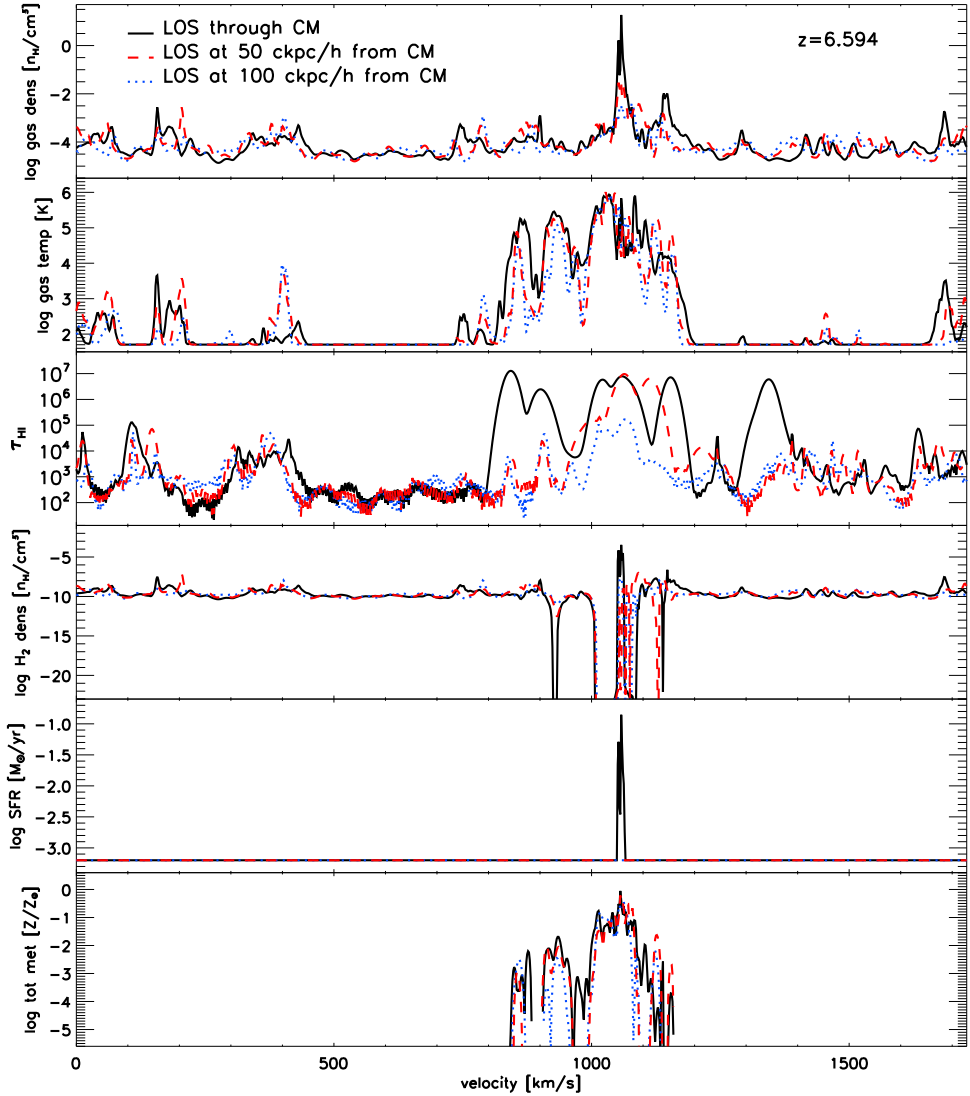}
\caption{\small
  Projected physical quantities extracted from three lines of sight close to the center of mass of the most massive halo at $z=6.594$. Black solid line: LOS through the CM. Red dashed line: LOS at 50 comoving kpc$/h$ from the CM. Blue dotted line: LOS at 100 comoving kpc$/h$ from the CM. From top to bottom the panels show:
total gas density ($n_{\rm H}/$cm$^3$),
gas temperature (K),
\HI\ optical depth ($\tau_{\rm \HI}$),
H$_2$ density ($n_{\rm H}/$cm$^3$),
SFR ($\rm M_\odot/yr$)
and total metallicity (solar units).
}
\label{fig:sim_qso}
\end{figure*}
As a final task, we compute simulated spectra (Sect.~\ref{Sect:spectra}) in order to retrieve further information that are directly comparable against observations.
We generate simulated observations by extracting spectra of \HI, H$_2$ and metals along lines of sight (LOSs) through the cosmological box and following the procedure outlined in \cite{Theuns1998,Tescari2009,Tescari2011}.
We model self-shielding by following \cite{Rahmati2013} (their fitting formula is valid up to $z\simeq 5$, even though we assumed its validity up to $z=6.594$).
We then discuss the column density distribution functions (Sect.~\ref{Sect:cddf}) and the statistical trends derived for different redshifts (Sect.~\ref{Sect:statistics}).

\subsection{Spectral analysis} \label{Sect:spectra}

Fig.~\ref{fig:sim_qso} shows some physical quantities interpolated along three LOSs around the most massive halo at $z=6.594$.
The black solid line is a LOS through the center of mass (CM) of the halo.
The red dashed and blue dotted lines are two LOSs at, respectively, 50 comoving kpc$/h$ and 100 comoving kpc$/h$ from the CM in the same direction.
\\
In the top panel the density of the gas (in $n_{\rm H}/$cm$^3$) is plotted. A sharp spike corresponding to the position of the center of the halo is clearly visible at $v\approx 1000$ km$/$s.
Apart from the central spike, the gas density profile does not change remarkably when considering LOSs far from the CM.
\\
The same is true for the temperature profile (second panel).
We note that the temperature profile is characterized both by hot material with temperatures above $\sim 10^4\,\rm K$ and by cold material with values below $\sim 10^4\,\rm K$.
The former one represents turbulent gas that has been heated mainly by stellar evolution and star formation feedback, while the latter is material that is still largely unpolluted (see metallicity profile below) and is condensing into gas clumps.
Such gaseous clumps are found both at 50 and at $100\,\rm kpc/{\it h}$ (comoving) from the CM and represent counterparts of DLA systems.
\\
This is confirmed by the optical depth, $\tau_{\rm \HI}$, profile (third panel) which is boosted by high-density cold material.
The optical depth markedly increases in correspondence to the center of mass of the halo (black solid line), reaching values of $\sim 10^7$.
Additional peaks with $\tau_{\rm \HI}>10^3$ represent intervening cold material that gives rise to sub-DLAs or Lyman-limit systems.
This is even more clear when looking at the off-center LOSs (red dashed and blue dotted lines) which are less impacted by the star formation episodes in the center of the halo and feature integrated optical depths up to $\tau_{\rm \HI} \sim 10^5-10^7$.
We note here that simpler \HI\ modelings based on density cuts only \cite[such as the ones in][]{ Nagamine2004, Nagamine2007, Tescari2009}, reproduce similar trends for these off-center LOSs, although tend to overproduce $\tau_{\rm \HI}$ values for particles near the central star forming region.
\\
In the fourth panel of Fig.~\ref{fig:sim_qso} we plot the H$_2$ density profile (resulting from our non-equilibrium treatment) which is particularly interesting because of its enhancement corresponding to dense gaseous environments.
Indeed, at the center of the halo the H$_2$ fraction experiences an exponential boost due to non-equilibrium catastrophic runaway cooling\footnote{
  We stress that this is an {\it out-of-equilibrium} process taking place when densities increase and the gas passes from a loitering regime to an unstable regime without pressure support.
}
in dense collapsing sites and reaches peak values $\sim 10^{-2}$ with subsequent molecular-driven star formation episodes.
Outside the halo's zone of influence, molecular hydrogen does not experience dramatic formation processes and hence the dotted and dashed lines never reach values as high as the black line.
Nevertheless, local H$_2$ peaks can be easily identified with the dense material (see also first panel) at low temperatures (second panel) and with large  $\tau_{\rm \HI} $ (third panel) that has not yet undergone star formation episodes.
\\
This is clearly visible in the fifth panel where a spike in the SFR profile is present corresponding to dense H$_2$-rich gas in the center of the halo, where stars are being formed actively (solid black line).
Away from the central region (dashed and dotted lines), instead, SFR contribution is null -- for the sake of clarity in this latter cases the SFR is arbitrarily set to $\rm log_{10}~SFR[M_{\odot}/yr] = -3.2$ in the panel (horizontal lines).
\\
The main effects of star formation are a global heating of the local environments and metal spreading into the surrounding medium.
Stellar feedback is responsible both for H$_2$ destruction (see e.g. troughs of the H$_2$ profile in the fourth panel) and for the build-up of metallicity gradients in the neighbourhood of star forming sites.
In fact, the resulting  metallicity profiles (last panel) are quite noisy and display significant amounts of heavy elements (up to $\sim Z_\odot$) in correspondence with the broad temperature distribution around $v\approx 1000\,\rm km/s$ shown in the second panel.
Cold dense clumps far away from the central star forming regions are basically pristine.
\\
We will not go into the details of the individual elemental abundances, because they demonstrate shapes that are quite similar to $Z$, but we will perform a more statistical investigation.
The discussion presented above has been based on three LOSs only and, although useful to shed light on the complex interplay between chemical and stellar feedback, it might suffer from low number statistics.

\subsection{Column density distribution function} \label{Sect:cddf}

\begin{figure}
\centering
\includegraphics[scale=0.43]{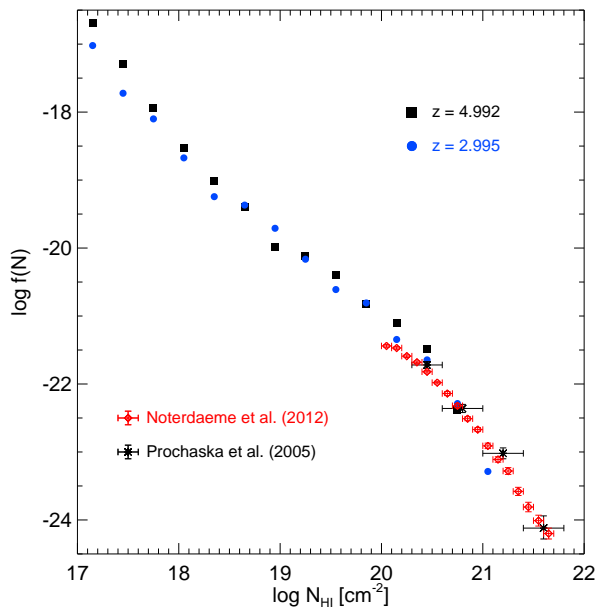}
\caption{\small
Simulated column density distribution functions, $f(N)$, at $z=2.995$ (filled circles) and $z=4.992$ (filled squares) and observational determinations at $z\simeq 3$ from \citet{Prochaska2005} (diamonds) and \citet{Noterdaeme2012} (asterisks) with error bars.
}
\label{fig:cddf}
\end{figure}
To have a statistically significant observable we compute the column density distribution function, $f(N)$, and make sure it is consistent with available observational determinations.
This is a common sanity check that we perform to be confident on how global properties of absorption systems are reproduced by our numerical calculations.
\\
We shoot 9000 random LOSs through the simulation box at each redshift and store the corresponding \HI\ column densities.
The column density distribution function $f(N)$ is then obtained, following previous works \cite[e.g.][]{Tytler1987}, as the fraction of LOSs in the
intervals $(N_{\rm\HI}, N_{\rm \HI} + \d N_{\rm \HI})$ and $(X, X + \d X)$,
where $N_{\rm \HI}$ denotes the \HI\ column density and $X$ the absoprption distance:
\begin{equation}
\d X = \frac{H_0}{H(z)} (1+z)^2 \d z .
\end{equation}
With these definitions, $f(N) \d N \d X$ is the number of LOSs in the considered $X$ and \HI\ column density bins.
\\
In Fig.~\ref{fig:cddf} we show $f(N)$ extracted from our simulation box at redshift $z=2.995$ (filled circles) and $z=4.992$ (filled squares) with observational determinations at $z\simeq 3$ from \cite{Prochaska2005} and \cite{Noterdaeme2012}.
We find a generally decreasing behaviour consistent with available data for both DLAs and Lyman-limit systems and no evidence for strong redshift evolution in the epochs probed here.
The drop at large $\rm N_{\HI}$ is rather steep, with a slope $\lesssim -2$.
The trends for both $z$ are essentially caused by the shorter cooling times at larger densities (see previous discussion in Sect.~\ref{Sect:transient}).

\subsection{Statistics of thermal and chemical properties}\label{Sect:statistics}

\begin{figure*}
\centering
\includegraphics[scale=0.6]{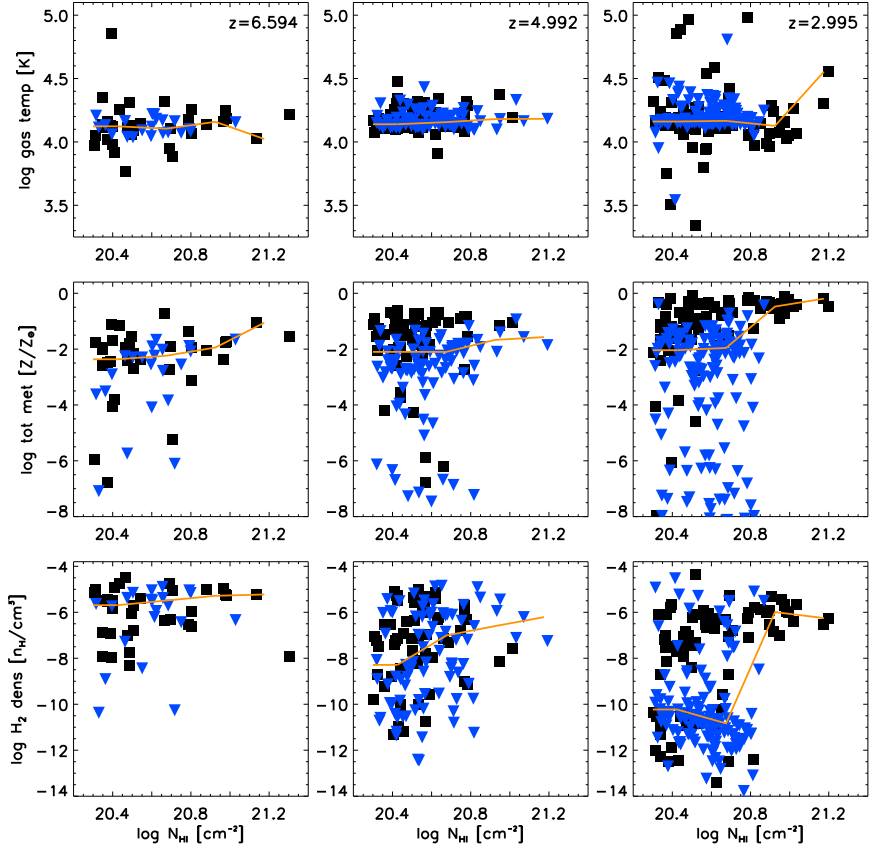}
\caption{\small
  Physical quantities of simulated DLAs. Gas mean temperature (top row), total metallicity (middle row) and H$_2$ mean density (bottom row) as a function of \HI\ column density. We consider three different redshifts: $z=6.594$ (left column), $z=4.992$ (middle column) and $z=2.995$ (right column). In each panel, the solid orange line shows the median of the corresponding physical quantity. Blue triangles refer to LOSs passing through the center of mass, while black squares refer to off-center LOSs.
}
\label{fig:dla_stats}
\end{figure*}
\begin{figure*}
\centering
\includegraphics[scale=0.6]{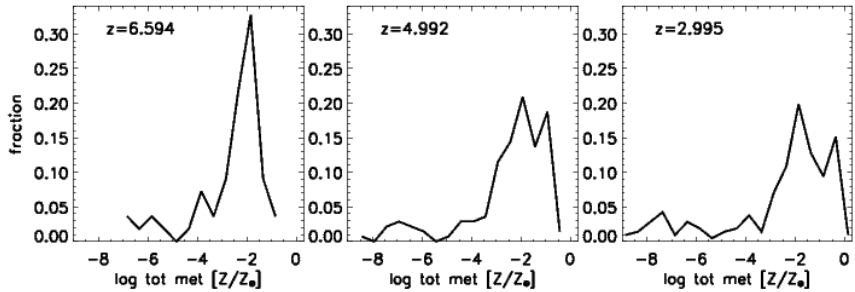}
\caption{\small
\HI-weigthed metallicity distributions at redshift $z=6.594$ (left panel), $z=4.992$ (central panel) and $z=2.995$ (right panel).
}
\label{fig:dla_met}
\end{figure*}
We extend the previous analysis to the cold gas clouds defined in Sect.~\ref{Sect:selection} and studied in Sect.~\ref{Sect:results}.
For each object, we fix the origin of an orthogonal reference system on the CM, such that the $z$ direction points towards the observer.
Then, for each object 17 LOSs are considered with displacements between 0 and $\pm 15\,\rm kpc/{\it h}$ (physical) from the center:
 1 LOS along the $z$ direction through the CM;
 8 LOSs radially displaced on the $x$ axis (4 LOSs positively displaced and 4 LOSs negatively displaced);
 8 LOSs radially displaced on the $y$ axis (4 LOSs positively displaced and 4 LOSs negatively displaced).
To randomize the viewing angle for each object a permutation of the $x$, $y$ and $z$ axes is performed.
We take into account only absorbing systems with N$_{\rm \HI}\ge 2\times 10^{20}$ atoms cm$^{-2}$ to have more directly comparable quantities with measurements.
Since our implementation follows non-equilibrium species calculations and holds also in absence of pressure support it is possible to obtain reliable information on the gas thermal and chemical state and to explore the transition from \HI\ atoms to H$_2$ molecules.
\\
In Fig. \ref{fig:dla_stats} we show the mean temperature (top row), metallicity (middle row) and H$_2$ content (bottom row), all weighted by \HI\ volume density, as a function of \HI\ column density.
We consider the evolution with redshift from 
$z=6.594$ (left column) to 
$z=4.992$ (middle column) and 
$z=2.995$ (right column) 
for the resulting $ 12587 $ LOSs 
(680 at $z=6.594$, 
2907 at $z=4.992$ and 
9000 at $z=2.995$).
In each panel of the figure the solid orange line shows the median values.
The blue triangular points refer to LOSs passing through the CM, while the black squares refer to off-center LOSs.
\\
At a fixed \HI\ column density we notice some scatter in the temperatures
(especially at low $\rm N_{\rm \HI}$), even if the median remains constant around $ 10^4\,\rm K$ across the whole column density range.
This scatter reflects the different environmental conditions of the absorbing systems and the effects of intervening hot material between the observer and the targeted structures.
Moving from $z = 6.594$ to $z = 2.995$, the majority of LOSs report values of the \HI-weighted median temperature below $\sim 3\times 10^4 \,\rm K$ (i.e. 4.5~dex), consistently with our discussion in Sect.~\ref{Sect:results}.
The rare LOSs at $\rm log_{10} [T/\rm K] > 4.5$ (1 at $z = 6.594$ and 8 at $z = 2.995$) intercept parcels of gas just experiencing bursts of star formation.
The LOSs with lower temperatures are instead passing through dense gas dominated by molecular (H$_2$ and HD) cooling and undergoing run-away collapse.
\\
In the second row of Fig. \ref{fig:dla_stats} we present the redshift evolution of the \HI-weighted total metallicity (in solar units) along the different LOSs.
At a fixed $\rm N_{\rm \HI}$ the metallicity value is severely scattered, with a small fraction (of the order of $\sim 10$~per cent) of values at $Z<10^{-4}Z_\odot$.
We note that, given these numbers and the fact that only a bunch of metal-poor DLAs are currently known, it is statistically difficult to draw conclusions on such low-metallicity objects and further observational data are definitely needed.
It is also difficult to measure metallicities at redshift $z\simeq 6.6$, because of the lack of information about the \HI\ column density due to the \cite{GP1965} effect.
The median metallicity is roughly constant at $Z\lesssim 10^{-2}Z_\odot$ and
increases by one order of magnitude ($\gtrsim 10^{-1}Z_\odot$) in high-density regions due to ongoing metal pollution between $z=6.594$ and $z=2.995$.
Already at earlier epochs ($z=6.594$), the median $Z$ shows a large metallicity spread as a result of the initial stages of metal pollution:
the first phases of this process in the denser cores of primordial structures cause very inhomogeneous enrichment patterns.
At $z=4.992$ the LOS metallicities remain around $\rm 10^{-2}\,Z_\odot$, despite some outliers below $10^{-4}\,Z_\odot$.
\\
At $z=2.995$ the metallicity distribution of the LOSs peaks still at $\sim 10^{-2}\, Z_\odot$, while $Z<10^{-4}\, Z_\odot$ data account for $\sim 10$~per cent of the cases.
Interestingly, we observe a steep increase in the high-N$_{\HI}$ regions caused by ongoing pollution from the numerous low-mass stars that can now die as SN~Ia (see discussion in Sect.~\ref{Sect:cosmic-chemical-evolution}).
\\
For sake of completness, in Fig.~\ref{fig:dla_met} we show the corresponding \HI-weighted metal distribution functions at $z=6.594$, $z=4.992$ and $z=2.995$.
As previously mentioned, there is little redshift evolution, although at earlier times $Z$ values are more sharply peaked around $0.01\, Z_\odot$ ($z=6.594$) and later on between $0.01-0.1\, Z_\odot$.
\\
Finally, in the bottom row of Fig.~\ref{fig:dla_stats} we plot the H$_2$ content as a function of \HI\ column density.
Larger-$\rm N_{\rm \HI}$ structures sample more gas and hence have more chance to host H$_2$ formation, however, it is not simple to state whether H$_2$ correlates with $\rm N_{\rm \HI}$ along the LOS.
The trends feature evident temporal and local variations of H$_2$ abundances accompanied by a broad spread.
The denser regions host the first star formation events and metal spreading episodes.
Star formation feedback is then responsible for molecule dissociation in the surrounding environment.
The resulting behaviour is therefore a consequence of the increasing star formation activity with cosmic time.
In fact, its feedback dissociate molecules and affect the slope of the $\rm \HI$-$ \rm H_2$ relation.
\\
As a final remark, we stress that both shielding corrections and weighting scheme could have impacts on the final results.
We have checked that with the alternative shielding prescriptions by e.g. \cite{Nagamine2004, Nagamine2007} and \cite {Tescari2009} we get typically lower scatter in the $\rm \HI$-$ \rm H_2$ relation, due to the lack of temperature effects, which are instead accounted for by \cite{Rahmati2013}.
Different weighting schemes can introduce further systematics.
Ongoing studies \cite[][]{Rubin2014arXiv} or future observational campaigns might be able to test LOS modeling.
\\
Thermal properties for off-center LOSs (black squares) and for LOSs passing through the CM (blue triangles) do not differ significantly at any redshift.
By looking at the latter case we can confirm that, in the central regions, high H$_2$ values drive non-equilibrium runaway gas collapse and star formation.
Subsequent feedback causes gas evacuation, metal spreading and molecule dissociation with a remarkable decrement in the H$_2$ fractions.
Very recent observational analyses of $z\sim 2 - 3$ data by 
\cite{Noterdaeme2015arXiv}
seem to agree with the expected large scatter in H$_2$ abundances.
%


\section{Conclusions}\label{Sect:conclusions}


We have employed cosmological N-body hydrodynamical chemistry calculations following gas atomic and molecular collapse and heavy-element production from stars with different masses and metallicities during the first few Gyr of the Universe, in the redshift range $2\lesssim z \lesssim 7$. Our detailed non-equilibrium chemistry integration of e$^-$, H, H$^+$, H$^-$, He, He$^+$, He$^{++}$, D, D$^+$, H$_2$, H$_2^+$, HD, HeH$^+$ abundances and stellar evolution treatment with individual yields for different species (He, C, N, O, Ne, Mg, Si, S, Ca, Fe) allowed us to investigate the chemical patterns of intermediate-redshift galaxies and compare the predictions of our calculations to observational measurements.
\\
Throughout this work we have assumed a $\Lambda$CDM scenario.
We warn the reader, though, that in some particular cases \cite[e.g. warm dark matter,][]{MaioViel2014} the background cosmological model can influence the baryonic-structure evolution, mostly at higher redshift.
This can play a role during the very beginning of the onset of star formation, but we checked that typically star formation and feedback mechanisms tend to dominate and alleviate possible discrepancies in the epochs of interest here \cite[see][]{Maio2006, MaioIannuzzi2011, PaceMaio2014, MaioViel2014}.
Similarly, the exact numerical parameters adopted for the initial mass functions, wind prescriptions, initial-condition gas velocities and related issues are likely to induce some changes, which are expected to be minor
\cite[see more in][]{Maio2011, Campisi2011, Tescari2014, Ma2015}.
Uncertainties might derive from the lack of a definitive treatment of diffusion mechanisms, which, despite several attempts, still remain an unsolved problem in astrophysics.
\\
Theoretical yields are affected by a plethora of uncertain physical processes in stars (such as explosion mechanisms, differential rotation, the initial composition, magnetic fields, nuclear reaction rates, etc.) all of which can influence the final ratios.
Nevertheless, we find agreement between our theoretical expectations and observational determinations.
We note that the detailed values of metal yields for individual elements are not supposed to change gas hydrodynamics significantly, because metal content and cooling in the regimes explored here are usually dominated by oxygen, for which there are well determined metallicity-dependent yields \cite[e.g.][]{Francois2004}.
\\
Physical processes in the inter-stellar medium are modelled by means of subgrid effective models, which have also been tested through simulations of isolated objects \cite[][]{Maio2013b}.
The effects of changing particular model parameters have been shown not to be dramatic for star formation and enrichment processes, as long as gas cooling capabilities (which have been treated in a quite detailed fashion in this work) are properly addressed for both atomic and molecular phases.
\\
We do not expect statistically significant changes when considering different IMFs at early times due to e.g. pristine (popIII) generations.
Indeed, previous works \cite[][]{Tornatore2007,Maio2010,Wise2012,Muratov2013, Pallottini2014, Hirano2015} have shown that these should be cosmologically subdominant at any redshift, almost irrespectively of their mass spectrum.
Isolated regions with pristine pockets of popIII star forming gas can survive down to very low redshift, however, they are not supposed to contribute significantly to cosmic SFR densities (Fig.~\ref{fig:sfr}), neither in the case of massive popIII stars nor in the case of low-mass popIII stars \cite[][]{Maio2011}.
Furthermore, recent analyses based on measured GRB host metallicities \cite[][]{Ma2015} seem to exclude massive popIII star formation episodes for current detections at $z\gtrsim 4.7$.
\\
We find that redshift $\sim 2-7$ galaxies  have baryonic masses up to $ 10^{11}\,\rm M_\odot $ with typical stellar masses as low as $\sim 10^{5}\,\rm M_\odot $.
Structures at intermediate redshifts have typical SFRs around $\sim 10^{-2}$ and $\sim 20\, \rm M_\odot/yr$ and a very broad SSFR range of $\sim 10^{-2} - 10^2 \,\rm Gyr^{-1}$ with typical values of $\sim 10\,\rm Gyr^{-1}$ at $z\gtrsim 6.5$ and more quiescent ones around $\sim 0.1-1\,\rm Gyr^{-1}$ at $z\lesssim 3$ (Fig~\ref{fig:sample}).
The combined effects of feedback and environment can cause large excursions in the stellar fraction with typical values extending from $\lesssim 10^{-3}$ to $\sim 0.1$.
Cold gas-rich damped clumps usually have low SFRs ($\lesssim 10^{-2}\,\rm M_\odot/yr$) and stellar fractions ($\lesssim 10^{-2}$) and represent objects that are just building up their stellar mass \cite[consistent with observational hints;][]{Fynbo2008}.
Our results suggest that metal-poor DLAs can be transient objects formed by condensation of warm material ($\sim 10^4\,\rm K$) and captured in the evolutionary stage when cooling processes are lowering the kinetic temperature, allowing the gas to condense to higher overdensities (Figs.~\ref{fig:DLAformation}-\ref{fig:DLAdestruction}).
This explains why, despite molecular (H$_2$) fractions might be locally large, their average values are usually very scattered, consistently with data by e.g. \cite{Noterdaeme2008, Srianand2010, Albornoz2014}.
To explicitly demonstrate this evolutionary phase of metal-poor DLAs, we have traced a few simulated gaseous clumps over the redshift interval $z\simeq 2.995 -6.594$ and have shown their large-scale contraction.
On the other hand, we have additionally checked the possible fate of metal-poor DLA candidates formed at high redshift and found that they are likely to be partially or totally destroyed due to feedback from {\it in situ} star formation or tidal interactions.
\\
In the comparisons with available data, the metallicities and molecular fractions derived from our numerical calculations are in agreement with the observations.
\\
By tracking the cosmic chemical evolution for some of the most common elements and by testing the resulting abundance ratios (such as [Fe/H], [O/H], [C/O], [Si/O], [O/Fe], [Si/Fe], [N/O]) against observational measurements (Figs.~\ref{fig:DLAchemistry1}-\ref{fig:DLAchemistry4}), we find that average trends and dispersion of simulated systems agree with observed systems from $z\sim 2.995$ to $z\sim 6.594$. The ratios we consider do not present strong redshift evolution, since metal enrichment is dominated by massive stars at these epochs with some contribution from intermediate-mass stars and SN~Ia. In all cases, mean values for e.g. [C/O], [O/Fe], [Si/Fe] behave within the error bars of observational estimates \cite[][]{Becker2012,Cooke2015}. The late-time Fe enhancement from long-lived SNe Ia decreases [O/Fe] and [$\alpha$/Fe] ratios consistently with stellar evolution time-scales (Figs.~\ref{fig:abundances_redshift}-\ref{fig:direct}).
\\
An interesting feature of our analysis is the existence of considerable scatter along the theoretical trends over the redshift intervals we probe: typical standard deviations vary from a fraction of one dex up to more than half of a dex.
We interpret these spreads as due to simultaneous enrichment by stars at different stellar stages or belonging to different stellar populations.
This is a remarkable deviation from the simple-stellar-population scenario and highlights the limits of homogeneous models.
\\
From an observational point of view, the effects of SN~Ia going off at $z\lesssim 4-5 $ should appear in the broadening of measured line widths at later times as a consequence of the establishment of a more and more turbulent regime.
\\
Simulated spectral observations (Fig.~\ref{fig:sim_qso}) allow us to stress the connection between gas, H$_2$, star formation and the complex interplay between chemical and stellar feedback in the presence of cold optically thick molecular-rich clumps.
Our synthetic spectra statistics at $z\sim 2-7$ (Fig.~\ref{fig:dla_stats}) suggest that the median intervening gas temperature evolves little during structure growth after reionization.
The increase in metallicity from high to lower $z$ and the large scatter observed in H$_2$ abundances are mainly due to the  increasing star formation activity (Figs.~\ref{fig:sfr}-\ref{fig:sample}) in increasingly larger objects, after the  primordial bursty phases.
\\
These findings demonstrate the importance of baryon chemistry to understand the properties of cosmic structures at early and intermediate epochs and its power to unveil the origins (SN~II, AGB and SN~Ia) of heavy elements and of their trends during cosmological evolution.
\\
Despite disentangling theoretical expectations and observational features might be challenging, enrichment of metal-poor gas could hide indirect signatures of previous stellar generations.
Future studies of elemental ratios as expected from different stellar populations within 3-dimensional numerical simulations will shed light on the properties of low-$Z$ star formation and on the existence of primordial popIII stars.


\section*{acknowledgements}
We thank the anonymous referee for detailed and constructive comments which helped us extend and improve significantly the original manuscript.
We acknowledge useful discussions with A.~Arino-i-Prats, S.~Borgani, J.~Cooke, R.~Cooke, S.~Cristiani, G.~De~Lucia, K.~Dolag, F.~Matteucci, P.~Molaro, M.~Viel, F.~A.~Villaescusa~Navarro, F.~Vincenzo.
U.~M. was supported through a Marie Curie Fellowship by the European Union Seventh Framework Programme (FP7/2007-2013) under grant agreement n. 267251.
E.~T. is supported by the Australian Research Council Centre of Excellence for All-sky Astrophysics (CAASTRO) through project number CE110001020.
This research was undertaken with the assistance of resources from the National Computational Infrastructure (NCI)\footnote{http://nf.nci.org.au}, which is supported by the Australian Government.
We would like to thank Volker Springel for making available to us the non-public version of the {\small{GADGET-3}} code.
For the simulations we used the {\textit{raijin}}, {\textit {vayu}} and {\textit{xe}} clusters at the NCI National Facility.
For the post-processing we also used the {\textit {edward}} High Performance Computing (HPC) cluster at the University of Melbourne\footnote{https://edward-web.hpc.unimelb.edu.au/users/}.
Mass functions have been computed by means of the online calculator
http://hmf.icrar.org by \cite{Murray2013}.
We acknowledge the NASA Astrophysics Data System and the JSTOR archives for their bibliographic tools.


\begin{appendix}

\section{Dust estimates} \label{appendix}

\begin{figure}
\includegraphics[width=0.5\textwidth]{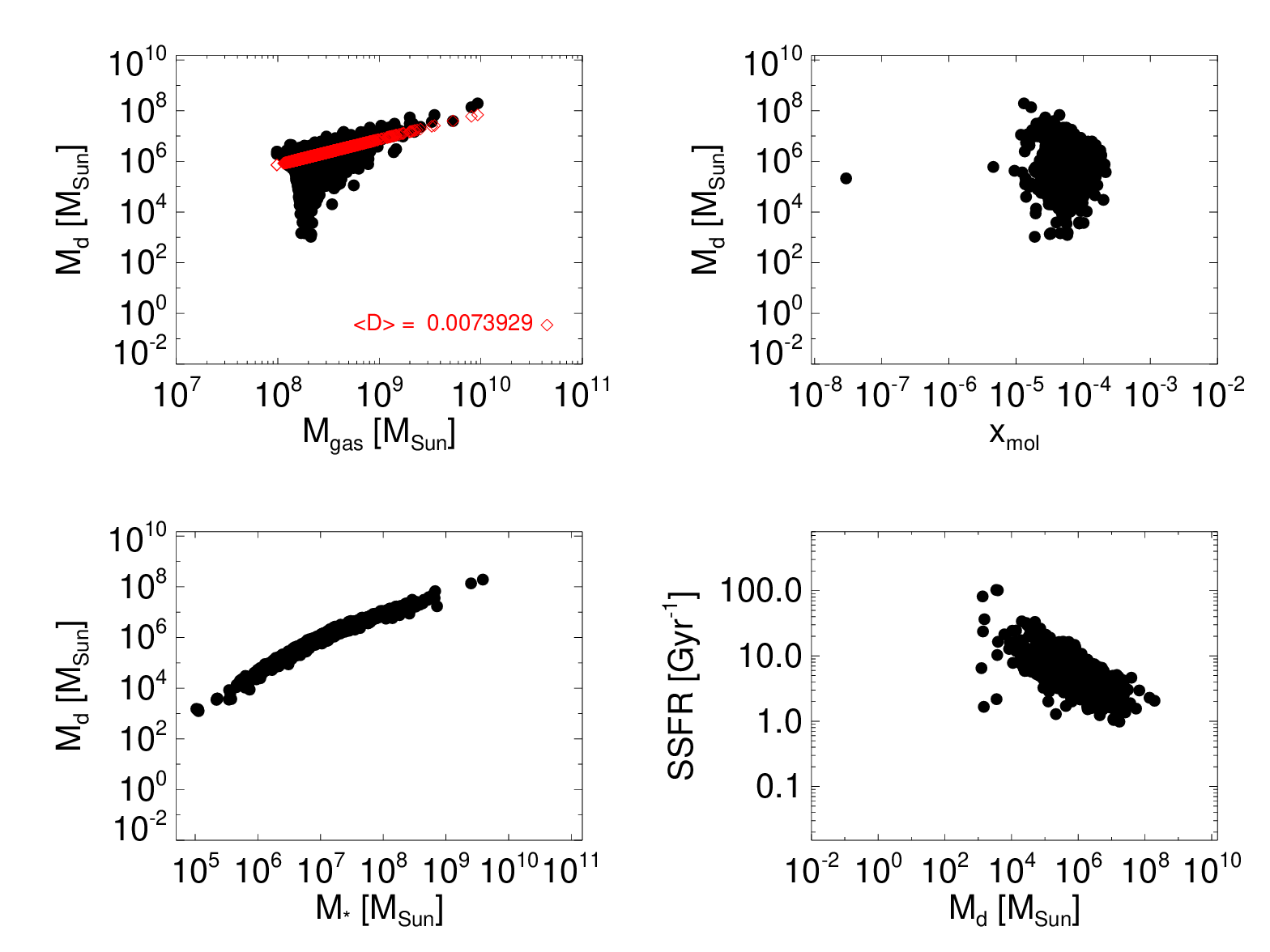}
\vspace{-0.25cm}
\caption[]{\small
Dust mass estimates as function of 
gas mass (top left), 
average molecular fraction (top right), 
stellar mass (bottom left) and 
specific SFR (bottom right)
at $z=6.594$.
The average dust-to-gas ratio, $D$, is quoted in the top left panel, with
corresponding expectations (red empty diamond).
}
\label{fig:DustProperties6}
\end{figure}

\begin{figure}
\includegraphics[width=0.5\textwidth]{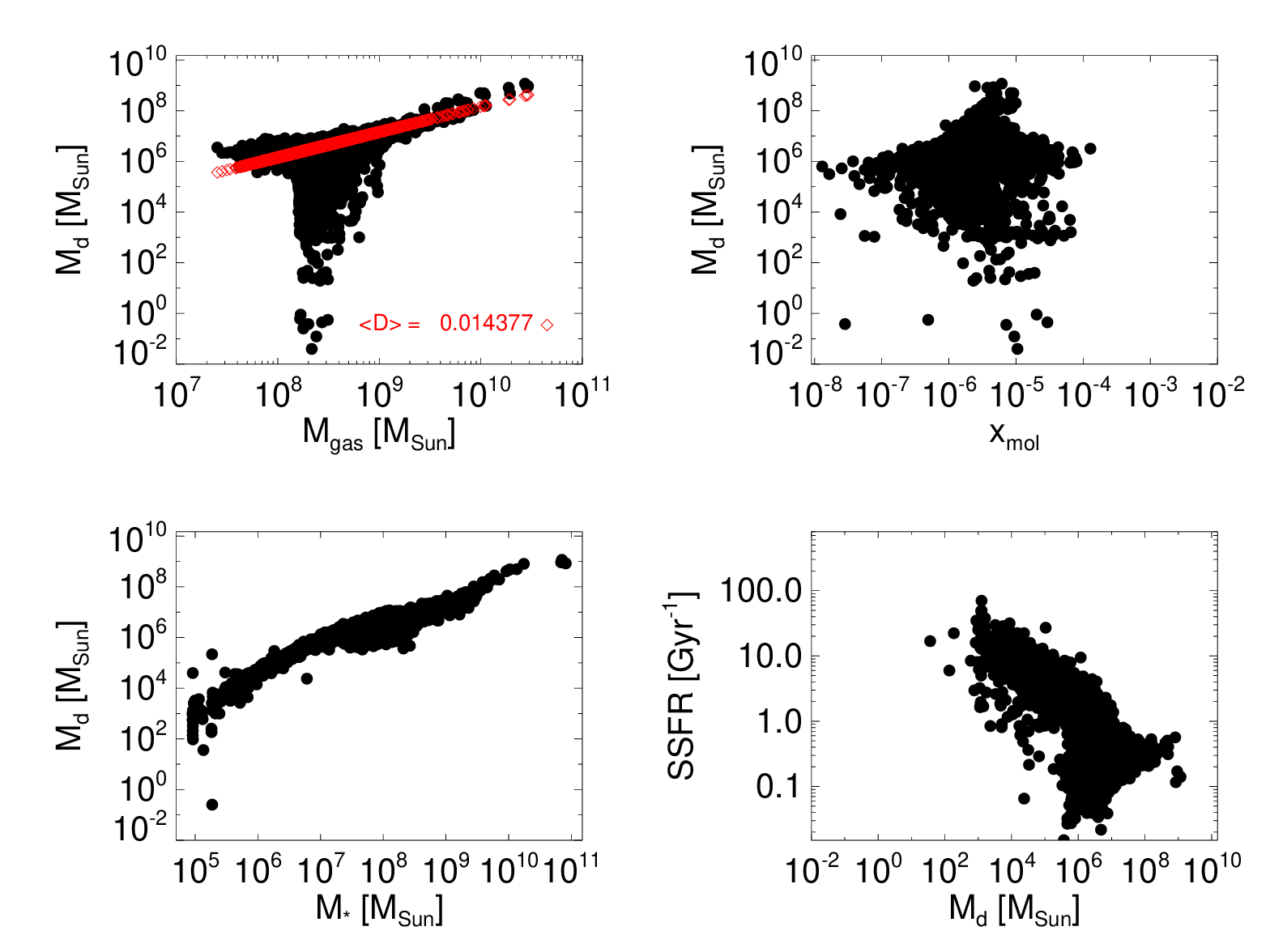}\\
\vspace{-0.25cm}
\caption[]{\small
Dust mass estimates as function of
gas mass (top left), 
average molecular fraction (top right),
stellar mass (bottom left) and 
specific SFR (bottom right)
at $z=2.995$.
The average dust-to-gas ratio, $D$, is quoted in the top left panel, with
corresponding expectations (red empty diamonds).
}
\label{fig:DustProperties3}
\end{figure}

The sudden appearance of large amounts of dust in early epochs can be explained by cooling and condensation of heavy elements \cite[][]{Feder1966} in massive core-collapse SN ejecta.
These, thanks to the short SN lifetimes and the high melting temperatures of the principal refractory elements (Si, Mg, Fe, Ca, Ti, Al, etc.), act as major sources of dust production that can be additionally reprocessed in the inter-stellar medium 
\cite[][]{YozinBekki2014,Mancini2015,Mattsson2015}.
Investigations of external galaxies and Magellanic Clouds have suggested a tight dependence of the dust-to-gas ratio, $D$, on the local metallicity, $Z$.
A stringent correlation has been obtained via sub-millimeter analyses by e.g. \cite{Galametz2011} and has been supported by analytical models of dust grain evolution \cite[][]{Feder1966}.
The resulting trend shows an increasing $D$ for increasing $Z$, $D \propto Z^{1.5}$, over a broad range of $Z$.
Although averaged over the many complex and still poorly understood processes that rule formation and destruction of dust grains
(e.g. the role of different stellar sources, metal yields, local star
formation, radiative transfer, sticking coefficients, grain size distribution, grain destruction/growth timescales in the inter-stellar medium, reverse-shock effects, underlying models for grain evolution, etc.),
the $D(Z)$ behaviour can be very precious to get hints on the typical fraction of metals condensed onto dust in different metallicity regimes.
Environmental effects can be surely important, as also testified by the large error bars for the data regression \cite[][]{Galametz2011}, but the general trend is fairly established.
Thus, the empirical $D(Z)$ relation represents a cheap and fruitful tool when applied to simulated cosmic structures, since it implicitly includes and constrains all the relevant unknowns.
Here, we apply the relation fitted by \cite{Galametz2011} to infer the dust mass implied by the metal content of each object in our data sample.
\\
In Figs.~\ref{fig:DustProperties6} and \ref{fig:DustProperties3} we show the expected dust mass as a function of gas mass, average molecular fraction, stellar mass and specific SFR for the whole data sample at redshift $z=6.594$ and $z=2.995$, respectively.\\
We observe already at higher $z$ a well established increasing trend of dust mass, $M_{\rm d}$, with gas mass.
The amounts of dust can be as large as $\sim 10^7-10^8\,\rm M_\odot$ already in the first billion years and the corresponding average $D$ value can reach $\sim 0.007$.
The effects of metal spreading is reflected in the broad $M_{\rm d}$ distribution which departs from the average behaviour (empty diamonds), mostly in the low-mass end.
When looking at the trend with molecular content we do not see any particular correlation, due to the dominant role of metallicity.
The behaviour with stellar mass, is instead, more clear with peak values reached for $M_\star\sim 10^9-10^{10}\,\rm M_\odot$.
Interestingly, this is consistent with recent observations 
at $z\sim 6.5-7.5$ \cite[e.g.][]{Schaerer2015,Watson2015,Maiolino2015}
whose data suggest normal galaxies with stellar masses $M_\star \sim 10^9-10^{10}\,\rm M_\odot$ and dust masses $M_{\rm d} \sim 10^7-10^8\,\rm M_\odot$.
In terms of SSFR, more quiescent objects host typically larger dust content, as a hint of their previous activity.
\\
At later times ($z=2.995$) the picture shown in Fig.~\ref{fig:DustProperties3} is similar, although the typical dust-to-gas ratio increases by a factor of 2, to $D\sim 0.014$, and the dust fraction is comparable to metallicities (see main text).
\\
From a qualitative point of view the over-linear dependence of $D\propto Z^{1.5}$ helps understand why dust issues should be more relevant in more enriched gas.
Finally, we note that, despite the crude approximations, our approach based on the dust-metallicity relation performs relatively well when compared to data or to more sophisticated models that have to include unknown free parameters.


\end{appendix}


\bibliographystyle{mn2e}
\bibliography{bibl.bib}

\label{lastpage}
\end{document}